\documentclass[aps,prl,reprint,superscriptaddress,showpacs,twocolumn]{revtex4-2}

\usepackage{color}
\usepackage{graphicx}
\usepackage{tabularx}
\usepackage[american]{babel}
\usepackage[T1]{fontenc}
\usepackage{amsmath}
\usepackage{amssymb}
\usepackage{amstext}
\usepackage{amsthm}
\usepackage{latexsym}
\usepackage{verbatim}
\usepackage{natbib}
\usepackage{array}
\usepackage{color} 
\usepackage{bm} 

\usepackage{ulem}   

\def\G{\Gamma}

\def\Ef{$E_{F}$}
\def\kz{$k_{z}$}
\def\kf{$k_{F}$}
\def\dxz{$d_{xz}$}
\def\dyz{$d_{yz}$}
\def\dxy{$d_{xy}$}
\def\dx2my2{$d_{x^2-y^2}$}
\def\dz2{$d_{z^2}$}

\def\In25{CoSn$_{0.75}$In$_{0.25}$}
\def\Fe23{Co$_{0.77}$Fe$_{0.23}$Sn}

\usepackage[draft]{todonotes}   

\usepackage{xspace}

\definecolor{vbgreen}{rgb}{0.00,0.45,0.20}

\begin{document}

\title {Contrasting fate of holes introduced by Fe or In in kagome metal CoSn}

\author{V\'{e}ronique Brouet}
\affiliation {Universit\'{e} Paris-Saclay, CNRS, Laboratoire de Physique des Solides, 91405, Orsay, France.}

\author{Laure Thomarat}
\affiliation {Universit\'{e} Paris-Saclay, CNRS, Laboratoire de Physique des Solides, 91405, Orsay, France.}

\author{Patrick Le F\`evre}
\affiliation {Univ Rennes, CNRS, IPR - UMR 6251, F-35000 Rennes, France}

\author{Ethan Saiman}
\affiliation {Universit\'{e} Paris-Saclay, CNRS, Laboratoire de Physique des Solides, 91405, Orsay, France.}
\affiliation {Synchrotron SOLEIL, L'Orme des Merisiers, D\'epartementale 128, 91190 Saint-Aubin, France}

\author{Jean-Pascal Rueff}
\affiliation {Synchrotron SOLEIL, L'Orme des Merisiers, D\'epartementale 128, 91190 Saint-Aubin, France}

\author{Oleg Rubel}
\affiliation{Department of Materials Science and Engineering, McMaster University, 1280 Main Street West, Hamilton, Ontario L8S 4L8, Canada}

\author{Victor Bal\'{e}dent}
\affiliation {Universit\'{e} Paris-Saclay, CNRS, Laboratoire de Physique des Solides, 91405, Orsay, France.}
\affiliation{Institut universitaire de France (IUF), Paris, France}

\author{François Bertran}
\affiliation {Synchrotron SOLEIL, L'Orme des Merisiers, D\'epartementale 128, 91190 Saint-Aubin, France}

\begin{abstract}

CoSn is a kagome metal with well-defined flat bands 0.2~eV below the Fermi level \Ef. Moving them across \Ef~would offer a rare opportunity to investigate their impact on electronic properties. Using angle-resolved photoemission spectroscopy (ARPES) and X-ray absorption at the Co K-edge, we examine two routes to introduce holes in this compound, through Co/Fe or Sn/In substitutions. Substituting In for Sn produces an almost rigid upward shift of the bands toward \Ef, as expected for hole doping, whereas substituting Fe for Co leads to an unexpected downward shift. Density functional theory calculations reproduce these opposite trends and trace the unusual behavior of Fe to its local magnetic moment. The moment keeps the hole on the Fe impurity and imposes an attractive electrostatic potential on neighboring Co atoms. Overall, both substitutions bring weakly dispersing states close to \Ef, the flat bands of the Co kagome lattice upon In substitution and partly localized states around magnetic Fe impurities upon Fe substitution.
	
\end{abstract}

\date{\today}

\maketitle

One of the most exciting features of kagome metals is their ability to host flat bands (FB) resulting from interference between equivalent paths in the kagome lattice \cite{YinHasanReview22}. For a partially filled FB, one could expect to stabilize exotic states, such as the fractional quantum Hall effect \cite{JiangPRB19}. More conventional instabilities, such as ferromagnetism, are also promoted, both by the high density of states that facilitates fulfilling the Stoner criterion and by the underlying quantum geometry \cite{HuPRL26}. Moreover, the coexistence between carriers in narrow FB and more dispersive bands is reminiscent of heavy-fermion systems and opens new avenues to engineer correlated systems \cite{ChenCondMat25}. The perfect FB is however obtained with a simple tight-binding (TB) model for a perfect two-dimensional (2D) kagome plane with only nearest-neighbor interactions \cite{GuoPRB09,MazinNatCom14}. In real materials, it inevitably acquires finite dispersion and often coexists with other bands that may obscure its specific role, as in the strongly three-dimensional Co$_3$Sn$_2$S$_2$ \cite{YinNatPhys19} or Ni$_3$In \cite{HanPRB24,YeNatPhys24}. In other cases, like Nb$_3$X$_8$ \cite{GaoPRX23} or Rb$_2$Pd$_3$S$_8$ \cite{LiPRB22}, there is a clean FB at the top of the electronic structure, but it seems difficult to dope. Recently, Kondo-like behaviors were observed in Cr-based systems, which may realize this long sought purely d-electron Kondo effect \cite{LiuNature24,ZhangCondMat2026}.  

\begin{figure*}[tbp]
\includegraphics[width=1\linewidth]{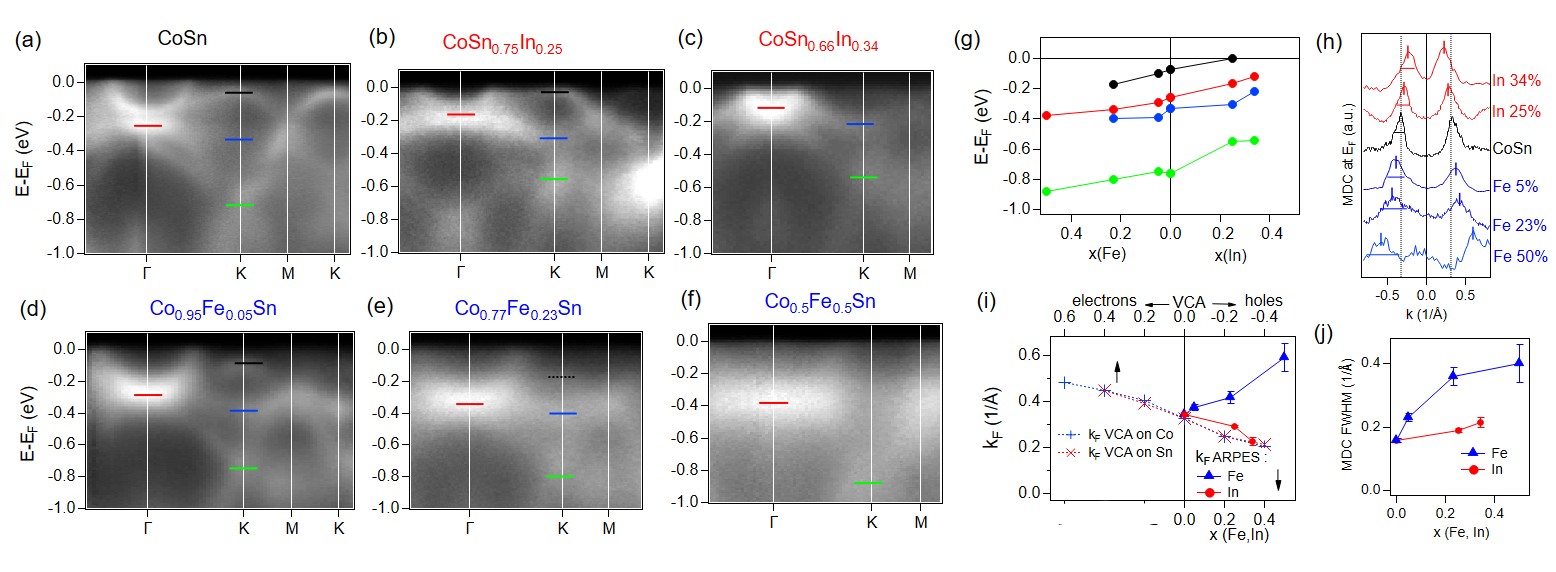}
	\caption{(a-f) ARPES energy-momentum images measured at \kz=0 in the indicated compounds [photon energy is 70~eV for (a,f) and 126~eV for (b,c,d,e)]. To show all bands, the images are reconstructed from four different $\Gamma$KM measurements in the first and second BZ and with LH and LV polarization (see supplementary \cite{sup} for more details). The colored markers emphasize the evolution of salient features of the band structure. (g) Evolution of the markers' energy positions as a function of Fe (left) and In (right) substitutions. (h) Momentum Distribution Curves (MDC) at \Ef~extracted from (a-f). The Fe (blue) or In (red) dopings are indicated on the side. (i) Evolution of k$_F$ extracted by MDC fits \cite{sup} as a function of substitution level $x$ (bottom axis), and as calculated by VCA approximation (top axis) on Co or Sn (see text). (j) Full MDC widths at half maximum (FWHM) as a function of the substitution level $x$.}
	\label{Fig_ARPES}
\end{figure*}

As first emphasized by Meier et al. \cite{MeierPRB20}, CoSn-type compounds are a good platform to investigate FB in kagome lattices. Their relatively simple structure preserves the remarkable features of the TB model, e.g., flat bands and Dirac cones. In CoSn, density functional theory (DFT) calculations predict Dirac cones around -1~eV and FB around -0.5~eV at $\Gamma$ for both \dxz/\dyz~and \dxy/\dx2my2~orbitals, with dispersions of respectively 1.5~eV and 0.2~eV (\cite{MeierPRB20,LiuNatCom20} or Fig. \ref{Fig_Bands}). They were observed by ARPES \cite{KangNatCom20_CoSn,LiuNatCom20}, which even resolved the expected FB topology around $\Gamma$ using circular polarization \cite{KangNatPhys25}. ARPES further showed that the FB were even closer to \Ef~than expected in DFT, around -0.2~eV at $\Gamma$ and as close as -0.05~eV at K, possibly due to selective band renormalization \cite{BrouetPRB25}.  

Moving these bands progressively toward the Fermi level would make it possible to study their impact on the electronic properties. There are in principle two ways to hole-dope these systems, either by substituting Fe for Co or by substituting In for Sn \cite{SalesPRM21}. In Co$_{1-x}$Fe$_x$Sn, an A-type antiferromagnetic ground state emerges for $x>0.4$, and its N\'eel temperature increases with $x$ up to $T_N=365$~K in FeSn, where the magnetic moment is $1.85\mu_B$ \cite{SalesPRM21,SalesPRM19}. The respective roles of itinerant and local interactions in this transition can be questioned, as the magnetic splitting was observed by ARPES to persist above $T_N$, as if local moments were still present on Fe \cite{RenNatCom24}. Compositions with low Fe content have received little attention beyond their initial characterization \cite{SalesPRM21}, but a recent scanning tunneling microscopy (STM) study \cite{ChenPRX26} reported signatures of correlated electronic states, a pseudogap and nematic order, at low temperature for $0.05<x<0.25$, which motivates a closer look at this regime. For CoSn$_{1-x}$In$_x$, there is a solubility limit imposing $x\leq0.4$. An antiferromagnetic state was detected below 24~K at $x=0.4$ \cite{SalesChemMat22}. The magnetic tendencies however appear weaker, judging from the smaller Curie-like contribution to the susceptibility induced by In doping compared with Fe doping \cite{SalesPRM21}.

In this letter, we reveal with ARPES and X-ray absorption spectroscopy at the Co K-edge a markedly different evolution of the CoSn electronic structure in the two cases. While the In case is consistent with the idea of hole doping, the Fe case is not. We compare these findings with DFT calculations, which reproduce the opposite trends. They show that the carriers are not distributed homogeneously between Fe and Co, which stabilizes a local magnetic moment around Fe. As a consequence, there could be two different ways to induce magnetic correlations in CoSn, either directly in the Co band by moving the FB to \Ef~with In doping or as localized magnetic impurities around Fe dopants. 

Single crystals of Co$_{1-x}$Fe$_{x}$Sn and CoSn$_{1-x}$In$_{x}$ were grown from a Sn/In flux \cite{SalesPRM21}. The exact composition was measured with energy-dispersive X-ray analysis and found homogeneous with a precision of about 2\%. The crystals form needles along the $c$ direction with 0.1-0.5~mm cross sections and cleave easily in the perpendicular kagome plane \cite{sup}. They were characterized by X-ray diffraction, transport and magnetic measurements with results similar to Ref.~\citenum{SalesPRM21}. ARPES experiments were carried out on UHV-cleaved single crystals at the CASSIOPEE beamline of the SOLEIL synchrotron, at low temperatures (T$\sim$15~K) and with an overall resolution better than 15~meV.  High-energy-resolution fluorescence-detected X-ray absorption spectroscopy (HERFD-XAS) \cite{Hamalainen1991, Glatzel2013, Lafuerza2020} was performed at the Co K-edge (7710 eV) on the GALAXIES beamline \cite{Rueff2015, Ablett2019} of the SOLEIL synchrotron. The measurements were conducted using a Ge(444) crystal analyzer with a Bragg angle of 84° to select the Co K$_\beta$ emission line (7649 eV). DFT calculations were done using the full-potential linear augmented plane wave method implemented in the WIEN2k software \cite{blaha2001} and the projector augmented-wave method implemented in VASP \cite{Kresse_PRB_47_1993, Kresse_CMS_6_1996, Kresse_PRB_54_1996, Kresse_PRB_59_1999}, the latter being used for the large supercells \cite{sup}. We used the \texttt{fold2Bloch} module \cite{RubelPRB14, Rubel23} to unfold the supercell band structures onto the primitive Brillouin zone (BZ) of CoSn. 

\vspace{0.5cm}

In Fig.~\ref{Fig_ARPES}, we compare ARPES data taken in compounds with different amounts of Fe and In, at photon energies corresponding to \kz=0. To visualize all bands at once, we have added images taken in different BZs and with different polarization (raw data and details are given in supplementary \cite{sup}). It is clear that the evolution of the data is markedly different and in fact opposite. For a similar hole count of $\sim$ 25\%, the bands move towards \Ef~for In [Fig. \ref{Fig_ARPES}(c)] but away from \Ef~for Fe [Fig. \ref{Fig_ARPES}(e)]. We indicate with markers the position of important features of the electronic structure. In red, the bottom of an electron band at $\Gamma$, nearly degenerate with the two FB; in black, the top of the \dxy/\dx2my2~FB at K and in blue that of \dxz/\dyz; in green, the two Dirac cones that cannot be clearly distinguished from each other at this \kz~\cite{LiuNatCom20}. We report their evolution as a function of doping in Fig. \ref{Fig_ARPES}(g). They trace a roughly linear evolution, but with opposite slope for In (right) and Fe (left). In a rigid band model, one would expect bands to move towards \Ef~for hole doping, as observed for In, but not for Fe. Our data contradict the intuitive expectation of a similar behavior for both dopants often used in previous works \cite{SalesPRM21,ChenPRX26}. Doped CoSn was indeed never systematically studied before by ARPES, but only for 40\%~In \cite{SalesChemMat22} and 2\%~Fe \cite{ZhangCondMat21}, both reporting an upward shift. While the results at 40\%~In is consistent with our data, the Fe shift 2\%~was probably too small to establish a reliable trend. 

Band shifts below \Ef~do not necessarily imply a change in band filling. For example, a renormalization that varies with substitution could shift the bottom of a band, and such renormalization is significant in CoSn \cite{BrouetPRB25}. However, \kf, hence the number of electrons this band contains, would remain unchanged in this case. In contrast, a clear evolution of \kf~for the band crossing \Ef~at \kz=0 is visible in the MDC spectra of Fig. \ref{Fig_ARPES}(h) and again opposite for Fe and In [Fig. \ref{Fig_ARPES}(i)]. As other bands cross \Ef~as a function of \kz~\cite{sup}, we cannot deduce from this evolution alone the number of carriers. Assuming a rigid shift of the electronic structure, we can nevertheless estimate what number of electrons corresponds to the observed $k_F$, by calculating the band structure in the virtual crystal approximation (VCA) \cite{BellaichePRB00}, where either Co or Sn is replaced with fictitious atoms having Z+$x_{VCA}$ electrons (the corresponding band structures are given in supplementary \cite{sup}). We report on the top axis of Fig. \ref{Fig_ARPES}(i) the evolution of \kf~at \kz=0 as a function of the number of added holes ($x_{VCA}$<0) or electrons ($x_{VCA}$>0). First, we observe a similar evolution if carriers are introduced on Co or Sn showing that no difference would be expected at this level of approximation. Second, the evolution is roughly linear and corresponds quite well on the hole side to the evolution observed with In, as if each In were effectively adding one hole to the band structure. On the contrary and unexpectedly, the evolution of Fe is much closer to that obtained in VCA for electrons, as if each Fe were inducing an effective electron doping. Alternatively, electrons may have been transferred to this band from other more three-dimensional bands.

An independent marker of the strength of the perturbation induced by Fe is the significant broadening observed in the MDC linewidths [Fig. \ref{Fig_ARPES}(j)], twice as large for Fe as In, which finally obscures the electronic bands at strong Fe contents [e.g. in Fig. \ref{Fig_ARPES}(f)]. 


\vspace{0.2cm}
\begin{figure}[tbp]
	\center
	\includegraphics[width=0.99\linewidth]{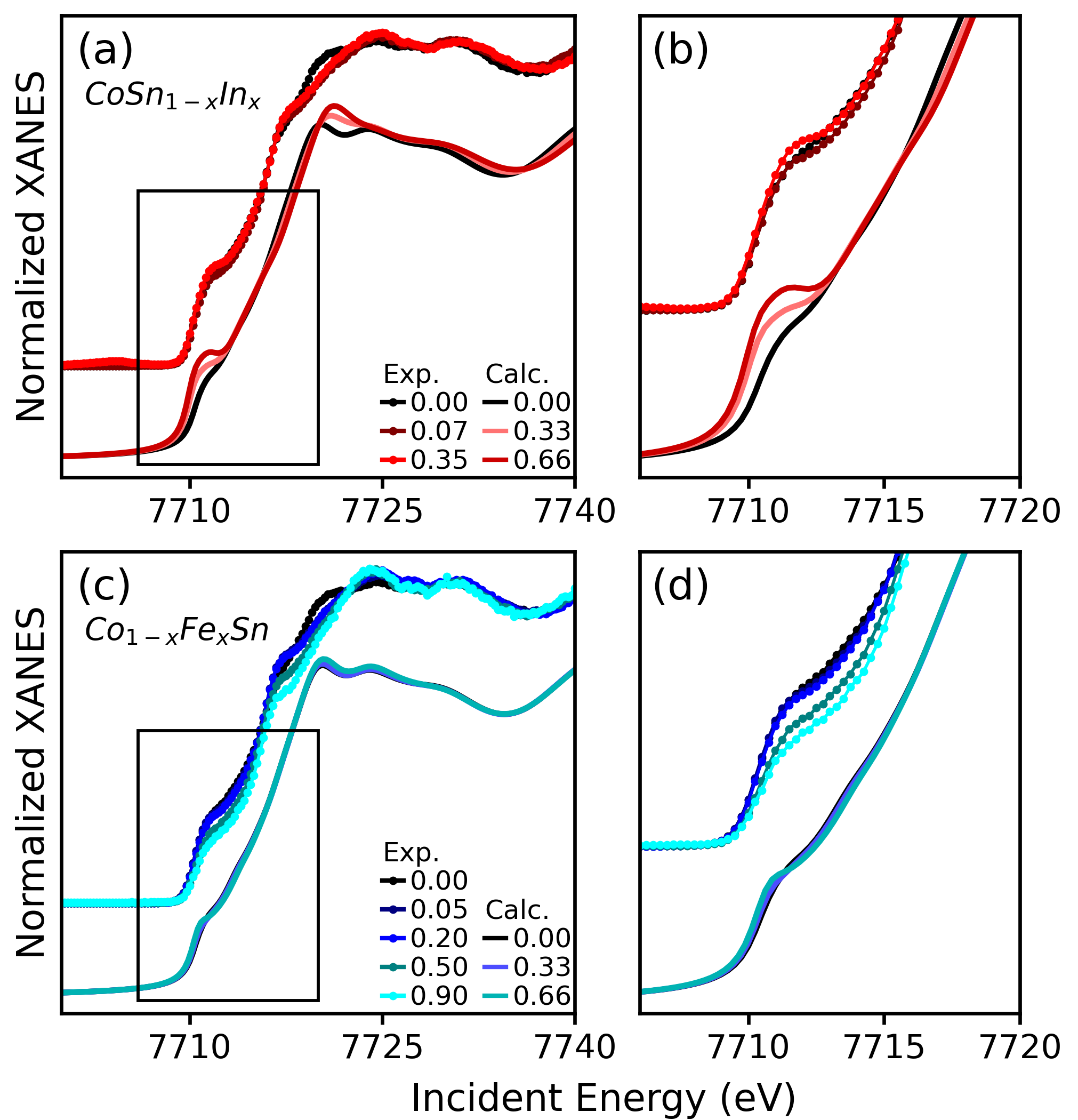}
	\caption{Co $K$-edge HERFD-XAS spectra (dots) and FDMNES simulations (solid lines) for CoSn$_{1-x}$In$_x$ (a,b) and Co$_{1-x}$Fe$_x$Sn (c,d) at the indicated dopings $x$, normalized to the edge jump. The experimental spectra are offset vertically for clarity. All simulations are performed in a non-magnetic $2\times2\times2$ supercell of the same volume, with the same convolution, and are aligned in energy on the calculated spectrum of pure CoSn of their own series \cite{sup}. (b) and (d) show the pre-edge region boxed in (a) and (c). The magnetic simulations are presented as Fig.~S17.} 
	\label{Fig_XAS}
\end{figure}

To get more information on the effect of these substitutions, we complemented this study with HERFD-XAS at the Co $K$-edge for several Fe and In dopings. XAS probes the unoccupied density of states and corresponds to a $1s\rightarrow 4p$ transition on Co, with sensitivity to the $3d$ states arising from $3d$-$4p$ hybridization and $1s\rightarrow3d$ transition allowed by the quadrupolar interaction, particularly visible in the pre-edge region. Interestingly, we observe in Fig.~\ref{Fig_XAS} an opposite evolution with In and Fe doping. The pre-edge intensity increases slightly in the case of In, consistent with a larger unoccupied density on Co, i.e. hole doping, but it decreases rather strongly in the case of Fe (note that the measurement extends to the high doping value of 90$\%$), indicating again a strong difference between the two substitutions. 

We simulated these spectra with the FDMNES code \cite{JolyPRB01,BunauJPCM09} and a $2\times2\times2$ supercell, where one Co atom out of three per kagome plane is replaced by Fe (or vice versa) or one of the two Sn sites by In. Every composition is simulated in the supercell of the same volume to be directly comparable (see supplementary for full details \cite{sup}). For In, the calculation reproduces well the measured evolution both in sign and in magnitude: the pre-edge area increases by 3.7$\%$ at $x_{In}$=1/3, to be compared with the 4.5$\%$ increase measured at $x_{In}$=0.35 [Fig. \ref{Fig_XAS}(a,b)]. This supports the conventional hole doping on the Co site inferred from ARPES when In substitutes for Sn.
 
For Fe, the simulation partially captures the different evolution compared to In. However, while the measured suppression of the pre-edge area grows steadily from 2$\%$ at $x_{Fe}$=0.05 to 19$\%$ at $x_{Fe}$=0.5 and 28$\%$ at $x_{Fe}$=0.9 [Fig. \ref{Fig_XAS}(c,d)], the calculated one is nearly independent of composition from the dilute limit up to $x_{Fe}$=2/3 and whether or not a local moment is allowed on Fe (the table in section IV-E of supplementary indicates changes from 3.9\%~to 8.5\%~\cite{sup}). We infer that a simulation of an ordered substitution of Fe for Co in a rigid CoSn lattice produces a local effect that saturates as soon as the first Fe is introduced, whereas the measured suppression accumulates with Fe content. This is the behavior expected if the Fe states near \Ef~are impurity-like and progressively reorganize the Co states as they overlap, a regime that an ordered ground-state calculation on a fixed lattice cannot describe.
 
The changes above the edge (above 7715~eV), which are more sensitive to the local structure, are on the contrary well captured by the simulations and can be interpreted as the lowering of symmetry produced by the ordered arrangement of the substituted atoms, from $P6/mmm$ to $Cmmm$ for Fe on the Co site \cite{sup}. A similar structural signature is observed and reproduced for In doping.

\vspace{0.2cm}

\begin{figure}[t]
	\center
	\includegraphics[width=0.99\linewidth]{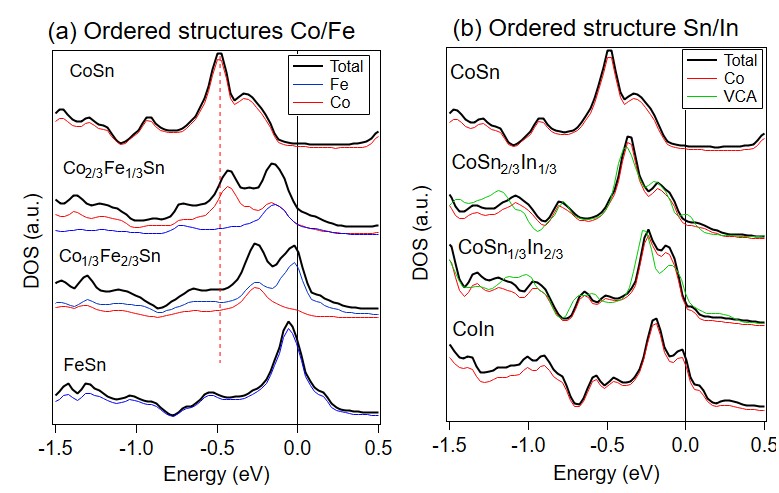}
	\caption{(a) DOS calculated for the pure compounds CoSn and FeSn and ordered supercells with one Fe and two Co or vice versa. The red and blue lines indicate the partial DOS of Co and Fe. All calculations are for a non-magnetic state. (b) DOS for a series of ordered supercells with Sn/In substitutions. The Co partial DOS is in red. The green line is obtained using VCA on Sn.}
	\label{Fig_DOS}
\end{figure}

To understand better the impact of substitutions on the electronic structure, we turn to DFT calculations, first in simple ordered cells, as for the FDMNES simulations. A spontaneous splitting between the Co and Fe contributions to the DOS emerges at $x_{Fe}=1/3$ and 2/3 [Fig.~\ref{Fig_DOS}(a)], the peak near \Ef~being predominantly of Fe character. This peak resembles that of non-magnetic FeSn [bottom of Fig.~\ref{Fig_DOS}(a)], where it originates from the FB (red, black, and blue markers in Fig.~\ref{Fig_ARPES}) lying near \Ef. Fe seems to inherit this electronic configuration, even as an impurity. In these non-magnetic calculations, the Co states however move toward \Ef~with increasing Fe content, opposite to the ARPES observations.  

For In, the peak simply shifts towards \Ef\ and keeps a dominant Co character [Fig.~\ref{Fig_DOS}(b)], very close to the VCA calculation on Sn (green line). In a hypothetical isostructural CoIn, with the same number of valence electrons as FeSn, the peak lies on average close to that of FeSn, although its shape differs slightly. This is the behavior expected for a regular hole doping.

To check whether this difference persists for lower and random substitutions, we performed complementary calculations in large supercells. We first consider the dilute limit, where a single Fe (or In) atom substitutes for Co (or Sn) in a relaxed $3\times3\times4$ Co$_{108}$Sn$_{108}$ supercell. Without spin polarization, the Fe partial DOS shows a sharp peak at \Ef\ (see Fig.~S11 \cite{sup}), absent in the Co partial DOS, quite similarly to Fig.~\ref{Fig_DOS}(a). This high DOS stabilizes a spin-polarized solution, both in the ordered cells and in the dilute limit, with a local moment larger than $1.5\mu_B$ on Fe \cite{sup}. Our calculations are thus consistent with attributing the Curie term in the magnetic susceptibility, roughly proportional to the number of Fe \cite{SalesPRM21}, to local moments on Fe. The moment keeps the hole on the impurity and makes the electrostatic potential on neighboring Co more attractive for electrons \cite{sup}, which shifts their states away from \Ef (see Fig. ~S12\cite{sup}). On the contrary, a single In introduces no states near \Ef~and remains non-magnetic. 

\begin{figure}[t]
	\center
	\includegraphics[width=0.99\linewidth]{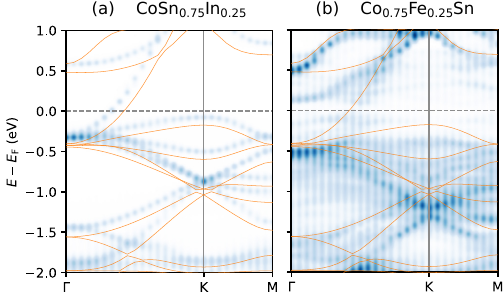}
	\caption{Unfolded band structure calculated in a supercell with 25\% In (a) or 25\% Fe (spin up channel, see SI\cite{sup} for spin down) (b). The CoSn band structure is recalled as thin orange lines. The intensity of panel (b) is enhanced by a factor of four relative to panel (a).}
	\label{Fig_Bands}
\end{figure}

To include the effect of disorder, we model Co$_{0.75}$Fe$_{0.25}$Sn and CoSn$_{0.75}$In$_{0.25}$ solid solutions with special quasirandom structures \cite{sup}. The Fe solid solution is spin-polarized, with the A-type antiferromagnetic arrangement of FeSn and Fe moments of about $2\mu_B$. In the unfolded band structures in Fig.~\ref{Fig_Bands}, indium shifts the bands upward almost rigidly, by 0.12~eV, whereas Fe broadens them strongly and shifts them downward non-rigidly, by 0.10~eV within 0.5~eV of \Ef\ and by about 0.2~eV deeper in the valence band. The diffuse character of the Fe bands reflects a stronger localization of the electronic states, since states localized in real space spread in reciprocal space (the degree of localization can be quantified by the higher inverse participation ratio \cite{sup}). Both the direction of these shifts and the stronger disorder for Fe agree extremely well with our observations in Fig.~\ref{Fig_ARPES}, validating this approach.

\vspace{0.2cm}

It is not rare that substituting transition metals for each other does not result in a simple doping. In Sr$_2$IrO$_4$, isovalent Rh substituting for Ir traps one electron and leads to hole doping \cite{ChikaraPRB15,LouatPRB19}, while Ru (one less electron than Ir) substitution leads to charge segregation at early dopings \cite{DhitalNatCom14,BrouetPRB21}, similarly to Fe in CoSn. In LaCoO$_3$, substituting non-magnetic Co by isovalent non-magnetic Rh induces magnetism \cite{StreltsovJPhysCondMat16}, confirming the complexity of these intricate interactions. Even in simple bcc Fe-Co alloys, the Fe moment increases when Fe is surrounded by Co \cite{SchwarzJPhysF84}. On the other hand, substituting Co for Fe in iron-based superconductors leads to an apparently simple electron doping, with a rigid upward shift of the bands in ARPES \cite{BrouetPRB09}, although the distribution between Fe and Co might remain slightly inhomogeneous \cite{LevyPRL12} and the ligand also plays an active role \cite{BaledentPRL15}. As substitutions are an essential tuning knob of the electronic structure, it is crucial to document and understand these behaviors.  

Finally, doping CoSn exhibits a richer phenomenology than previously anticipated. Our work suggests that the magnetism reported in Fe- and In-substituted CoSn \cite{SalesPRM21,SalesChemMat22} arises in two different ways, from magnetic impurities diluted in the metal for Fe and from the FB moved toward \Ef~for In. This should be an excellent opportunity to compare how features tentatively associated with the FB, like the large orbital susceptibility and the anisotropic resistivity \cite{HuangPRL22} or the magnetic fluctuations tracked by negative magnetoresistance \cite{SalesPRM19,ZhangCondMat21,SalesChemMat22}, evolve in the two different cases. The anomalies observed by STM at light Fe dopings (0.1<$x_{Fe}$<0.25) \cite{ChenPRX26} were naturally interpreted as resulting from a shift of the FBs to the Fermi level. Our work suggests a more complex picture, where the first FB forming at \Ef~upon Fe substitution derive from the minority-spin Fe levels, strongly hybridized with the Co bands. In fact, revisiting the STM $dI/dV$ spectra in light of our findings, it is interesting to note that they strongly resemble the DOS of Fig.~\ref{Fig_DOS} with two peaks developing with $x_{Fe}$ and shifting in opposite directions. Another open issue is how itinerant magnetism in FeSn emerges from the assembly of localized moments. All our calculations with Fe predict an induced moment on Co, suggesting a cooperative mechanism rather than simple percolation. Studying the spatial extent of the magnetic moment induced by Fe and how they are screened in the metal should reveal important features of the correlations present in this system, as the study of defects in many other correlated systems did before \cite{AlloulRMP09}.

\section{Acknowledgment}
We acknowledge the SOLEIL synchrotron for beamtime on the Cassiopée (proposals n° 20221212 and 20230669) and Galaxies beamlines (proposal n° 99260099), the use of the Ceres high
performance computer cluster to carry out the FDMNES
calculations, the Paris Ile-de-France Region in the framework of DIM MaTerRE (project DAC-VX), the France 2030 programme "ANR-11-IDEX-0003" via Integrative Institute of Materials from Paris-Saclay University - 2IM@UPSaclay and the MESURES PHYSIQUES and MICRO-NANO facilities of the Laboratoire de Physique des Solides, Université Paris-Saclay, for sample characterization. O.R. acknowledges computing resources provided by the Digital Research Alliance of Canada.

\bibliographystyle{apsrev4-2}
\bibliography{Biblio_kagome_CoSn,Biblio_Gal}

\begin{thebibliography}{55}%
\makeatletter
\providecommand \@ifxundefined [1]{%
 \@ifx{#1\undefined}
}%
\providecommand \@ifnum [1]{%
 \ifnum #1\expandafter \@firstoftwo
 \else \expandafter \@secondoftwo
 \fi
}%
\providecommand \@ifx [1]{%
 \ifx #1\expandafter \@firstoftwo
 \else \expandafter \@secondoftwo
 \fi
}%
\providecommand \natexlab [1]{#1}%
\providecommand \enquote  [1]{``#1''}%
\providecommand \bibnamefont  [1]{#1}%
\providecommand \bibfnamefont [1]{#1}%
\providecommand \citenamefont [1]{#1}%
\providecommand \href@noop [0]{\@secondoftwo}%
\providecommand \href [0]{\begingroup \@sanitize@url \@href}%
\providecommand \@href[1]{\@@startlink{#1}\@@href}%
\providecommand \@@href[1]{\endgroup#1\@@endlink}%
\providecommand \@sanitize@url [0]{\catcode `\\12\catcode `\$12\catcode
  `\&12\catcode `\#12\catcode `\^12\catcode `\_12\catcode `\%12\relax}%
\providecommand \@@startlink[1]{}%
\providecommand \@@endlink[0]{}%
\providecommand \url  [0]{\begingroup\@sanitize@url \@url }%
\providecommand \@url [1]{\endgroup\@href {#1}{\urlprefix }}%
\providecommand \urlprefix  [0]{URL }%
\providecommand \Eprint [0]{\href }%
\providecommand \doibase [0]{https://doi.org/}%
\providecommand \selectlanguage [0]{\@gobble}%
\providecommand \bibinfo  [0]{\@secondoftwo}%
\providecommand \bibfield  [0]{\@secondoftwo}%
\providecommand \translation [1]{[#1]}%
\providecommand \BibitemOpen [0]{}%
\providecommand \bibitemStop [0]{}%
\providecommand \bibitemNoStop [0]{.\EOS\space}%
\providecommand \EOS [0]{\spacefactor3000\relax}%
\providecommand \BibitemShut  [1]{\csname bibitem#1\endcsname}%
\let\auto@bib@innerbib\@empty
\bibitem [{\citenamefont {Yin}\ \emph {et~al.}(2022)\citenamefont {Yin},
  \citenamefont {Lian},\ and\ \citenamefont {Hasan}}]{YinHasanReview22}%
  \BibitemOpen
  \bibfield  {author} {\bibinfo {author} {\bibfnamefont {J.-X.}\ \bibnamefont
  {Yin}}, \bibinfo {author} {\bibfnamefont {B.}~\bibnamefont {Lian}},\ and\
  \bibinfo {author} {\bibfnamefont {M.~Z.}\ \bibnamefont {Hasan}},\ }\href
  {https://doi.org/10.1038/s41586-022-05516-0} {\bibfield  {journal} {\bibinfo
  {journal} {Nature}\ }\textbf {\bibinfo {volume} {612}},\ \bibinfo {pages}
  {647} (\bibinfo {year} {2022})}\BibitemShut {NoStop}%
\bibitem [{\citenamefont {Jiang}\ \emph {et~al.}(2019)\citenamefont {Jiang},
  \citenamefont {Kang}, \citenamefont {Huang}, \citenamefont {Xu},
  \citenamefont {Low},\ and\ \citenamefont {Liu}}]{JiangPRB19}%
  \BibitemOpen
  \bibfield  {author} {\bibinfo {author} {\bibfnamefont {W.}~\bibnamefont
  {Jiang}}, \bibinfo {author} {\bibfnamefont {M.}~\bibnamefont {Kang}},
  \bibinfo {author} {\bibfnamefont {H.}~\bibnamefont {Huang}}, \bibinfo
  {author} {\bibfnamefont {H.}~\bibnamefont {Xu}}, \bibinfo {author}
  {\bibfnamefont {T.}~\bibnamefont {Low}},\ and\ \bibinfo {author}
  {\bibfnamefont {F.}~\bibnamefont {Liu}},\ }\href
  {https://doi.org/10.1103/PhysRevB.99.125131} {\bibfield  {journal} {\bibinfo
  {journal} {Phys. Rev. B}\ }\textbf {\bibinfo {volume} {99}},\ \bibinfo
  {pages} {125131} (\bibinfo {year} {2019})}\BibitemShut {NoStop}%
\bibitem [{\citenamefont {Hu}\ \emph {et~al.}(2026)\citenamefont {Hu},
  \citenamefont {Vafek}, \citenamefont {Haule},\ and\ \citenamefont
  {Bernevig}}]{HuPRL26}%
  \BibitemOpen
  \bibfield  {author} {\bibinfo {author} {\bibfnamefont {H.}~\bibnamefont
  {Hu}}, \bibinfo {author} {\bibfnamefont {O.}~\bibnamefont {Vafek}}, \bibinfo
  {author} {\bibfnamefont {K.}~\bibnamefont {Haule}},\ and\ \bibinfo {author}
  {\bibfnamefont {B.~A.}\ \bibnamefont {Bernevig}},\ }\href
  {https://doi.org/10.1103/zdyq-3m9x} {\bibfield  {journal} {\bibinfo
  {journal} {Phys. Rev. Lett.}\ }\textbf {\bibinfo {volume} {136}},\ \bibinfo
  {pages} {256505} (\bibinfo {year} {2026})}\BibitemShut {NoStop}%
\bibitem [{\citenamefont {Chen}\ \emph {et~al.}(2025)\citenamefont {Chen},
  \citenamefont {Xie}, \citenamefont {Sur}, \citenamefont {Hu}, \citenamefont
  {Paschen}, \citenamefont {Cano},\ and\ \citenamefont {Si}}]{ChenCondMat25}%
  \BibitemOpen
  \bibfield  {author} {\bibinfo {author} {\bibfnamefont {L.}~\bibnamefont
  {Chen}}, \bibinfo {author} {\bibfnamefont {F.}~\bibnamefont {Xie}}, \bibinfo
  {author} {\bibfnamefont {S.}~\bibnamefont {Sur}}, \bibinfo {author}
  {\bibfnamefont {H.}~\bibnamefont {Hu}}, \bibinfo {author} {\bibfnamefont
  {S.}~\bibnamefont {Paschen}}, \bibinfo {author} {\bibfnamefont
  {J.}~\bibnamefont {Cano}},\ and\ \bibinfo {author} {\bibfnamefont
  {Q.}~\bibnamefont {Si}},\ }\href {https://arxiv.org/abs/2307.09431} {\bibinfo
  {title} {Metallic quantum criticality enabled by flat bands in a kagome
  lattice}} (\bibinfo {year} {2025}),\ \Eprint
  {https://arxiv.org/abs/2307.09431} {arXiv:2307.09431 [cond-mat.str-el]}
  \BibitemShut {NoStop}%
\bibitem [{\citenamefont {Guo}\ and\ \citenamefont {Franz}(2009)}]{GuoPRB09}%
  \BibitemOpen
  \bibfield  {author} {\bibinfo {author} {\bibfnamefont {H.-M.}\ \bibnamefont
  {Guo}}\ and\ \bibinfo {author} {\bibfnamefont {M.}~\bibnamefont {Franz}},\
  }\href {https://doi.org/10.1103/PhysRevB.80.113102} {\bibfield  {journal}
  {\bibinfo  {journal} {Phys. Rev. B}\ }\textbf {\bibinfo {volume} {80}},\
  \bibinfo {pages} {113102} (\bibinfo {year} {2009})}\BibitemShut {NoStop}%
\bibitem [{\citenamefont {Mazin}\ \emph {et~al.}(2014)\citenamefont {Mazin},
  \citenamefont {Jeschke}, \citenamefont {Lechermann}, \citenamefont {Lee},
  \citenamefont {Fink}, \citenamefont {Thomale},\ and\ \citenamefont
  {Valent{\'i}}}]{MazinNatCom14}%
  \BibitemOpen
  \bibfield  {author} {\bibinfo {author} {\bibfnamefont {I.~I.}\ \bibnamefont
  {Mazin}}, \bibinfo {author} {\bibfnamefont {H.~O.}\ \bibnamefont {Jeschke}},
  \bibinfo {author} {\bibfnamefont {F.}~\bibnamefont {Lechermann}}, \bibinfo
  {author} {\bibfnamefont {H.}~\bibnamefont {Lee}}, \bibinfo {author}
  {\bibfnamefont {M.}~\bibnamefont {Fink}}, \bibinfo {author} {\bibfnamefont
  {R.}~\bibnamefont {Thomale}},\ and\ \bibinfo {author} {\bibfnamefont
  {R.}~\bibnamefont {Valent{\'i}}},\ }\href
  {https://doi.org/10.1038/ncomms5261} {\bibfield  {journal} {\bibinfo
  {journal} {Nature Communications}\ }\textbf {\bibinfo {volume} {5}},\
  \bibinfo {pages} {4261} (\bibinfo {year} {2014})}\BibitemShut {NoStop}%
\bibitem [{\citenamefont {Yin}\ \emph {et~al.}(2019)\citenamefont {Yin},
  \citenamefont {Zhang}, \citenamefont {Chang}, \citenamefont {Wang},
  \citenamefont {Tsirkin}, \citenamefont {Guguchia}, \citenamefont {Lian},
  \citenamefont {Zhou}, \citenamefont {Jiang}, \citenamefont {Belopolski},
  \citenamefont {Shumiya}, \citenamefont {Multer}, \citenamefont {Litskevich},
  \citenamefont {Cochran}, \citenamefont {Lin}, \citenamefont {Wang},
  \citenamefont {Neupert}, \citenamefont {Jia}, \citenamefont {Lei},\ and\
  \citenamefont {Hasan}}]{YinNatPhys19}%
  \BibitemOpen
  \bibfield  {author} {\bibinfo {author} {\bibfnamefont {J.-X.}\ \bibnamefont
  {Yin}}, \bibinfo {author} {\bibfnamefont {S.~S.}\ \bibnamefont {Zhang}},
  \bibinfo {author} {\bibfnamefont {G.}~\bibnamefont {Chang}}, \bibinfo
  {author} {\bibfnamefont {Q.}~\bibnamefont {Wang}}, \bibinfo {author}
  {\bibfnamefont {S.~S.}\ \bibnamefont {Tsirkin}}, \bibinfo {author}
  {\bibfnamefont {Z.}~\bibnamefont {Guguchia}}, \bibinfo {author}
  {\bibfnamefont {B.}~\bibnamefont {Lian}}, \bibinfo {author} {\bibfnamefont
  {H.}~\bibnamefont {Zhou}}, \bibinfo {author} {\bibfnamefont {K.}~\bibnamefont
  {Jiang}}, \bibinfo {author} {\bibfnamefont {I.}~\bibnamefont {Belopolski}},
  \bibinfo {author} {\bibfnamefont {N.}~\bibnamefont {Shumiya}}, \bibinfo
  {author} {\bibfnamefont {D.}~\bibnamefont {Multer}}, \bibinfo {author}
  {\bibfnamefont {M.}~\bibnamefont {Litskevich}}, \bibinfo {author}
  {\bibfnamefont {T.~A.}\ \bibnamefont {Cochran}}, \bibinfo {author}
  {\bibfnamefont {H.}~\bibnamefont {Lin}}, \bibinfo {author} {\bibfnamefont
  {Z.}~\bibnamefont {Wang}}, \bibinfo {author} {\bibfnamefont {T.}~\bibnamefont
  {Neupert}}, \bibinfo {author} {\bibfnamefont {S.}~\bibnamefont {Jia}},
  \bibinfo {author} {\bibfnamefont {H.}~\bibnamefont {Lei}},\ and\ \bibinfo
  {author} {\bibfnamefont {M.~Z.}\ \bibnamefont {Hasan}},\ }\href
  {https://doi.org/10.1038/s41567-019-0426-7} {\bibfield  {journal} {\bibinfo
  {journal} {Nature Physics}\ }\textbf {\bibinfo {volume} {15}},\ \bibinfo
  {pages} {443} (\bibinfo {year} {2019})}\BibitemShut {NoStop}%
\bibitem [{\citenamefont {Han}\ \emph {et~al.}(2024)\citenamefont {Han},
  \citenamefont {John}, \citenamefont {Zheng}, \citenamefont {Fang},\ and\
  \citenamefont {Checkelsky}}]{HanPRB24}%
  \BibitemOpen
  \bibfield  {author} {\bibinfo {author} {\bibfnamefont {M.}~\bibnamefont
  {Han}}, \bibinfo {author} {\bibfnamefont {C.}~\bibnamefont {John}}, \bibinfo
  {author} {\bibfnamefont {J.}~\bibnamefont {Zheng}}, \bibinfo {author}
  {\bibfnamefont {S.}~\bibnamefont {Fang}},\ and\ \bibinfo {author}
  {\bibfnamefont {J.~G.}\ \bibnamefont {Checkelsky}},\ }\href
  {https://doi.org/10.1103/PhysRevB.109.L121112} {\bibfield  {journal}
  {\bibinfo  {journal} {Phys. Rev. B}\ }\textbf {\bibinfo {volume} {109}},\
  \bibinfo {pages} {L121112} (\bibinfo {year} {2024})}\BibitemShut {NoStop}%
\bibitem [{\citenamefont {Ye}\ \emph {et~al.}(2024)\citenamefont {Ye},
  \citenamefont {Fang}, \citenamefont {Kang}, \citenamefont {Kaufmann},
  \citenamefont {Lee}, \citenamefont {John}, \citenamefont {Neves},
  \citenamefont {Zhao}, \citenamefont {Denlinger}, \citenamefont {Jozwiak},
  \citenamefont {Bostwick}, \citenamefont {Rotenberg}, \citenamefont {Kaxiras},
  \citenamefont {Bell}, \citenamefont {Janson}, \citenamefont {Comin},\ and\
  \citenamefont {Checkelsky}}]{YeNatPhys24}%
  \BibitemOpen
  \bibfield  {author} {\bibinfo {author} {\bibfnamefont {L.}~\bibnamefont
  {Ye}}, \bibinfo {author} {\bibfnamefont {S.}~\bibnamefont {Fang}}, \bibinfo
  {author} {\bibfnamefont {M.}~\bibnamefont {Kang}}, \bibinfo {author}
  {\bibfnamefont {J.}~\bibnamefont {Kaufmann}}, \bibinfo {author}
  {\bibfnamefont {Y.}~\bibnamefont {Lee}}, \bibinfo {author} {\bibfnamefont
  {C.}~\bibnamefont {John}}, \bibinfo {author} {\bibfnamefont {P.~M.}\
  \bibnamefont {Neves}}, \bibinfo {author} {\bibfnamefont {S.~Y.~F.}\
  \bibnamefont {Zhao}}, \bibinfo {author} {\bibfnamefont {J.}~\bibnamefont
  {Denlinger}}, \bibinfo {author} {\bibfnamefont {C.}~\bibnamefont {Jozwiak}},
  \bibinfo {author} {\bibfnamefont {A.}~\bibnamefont {Bostwick}}, \bibinfo
  {author} {\bibfnamefont {E.}~\bibnamefont {Rotenberg}}, \bibinfo {author}
  {\bibfnamefont {E.}~\bibnamefont {Kaxiras}}, \bibinfo {author} {\bibfnamefont
  {D.~C.}\ \bibnamefont {Bell}}, \bibinfo {author} {\bibfnamefont
  {O.}~\bibnamefont {Janson}}, \bibinfo {author} {\bibfnamefont
  {R.}~\bibnamefont {Comin}},\ and\ \bibinfo {author} {\bibfnamefont {J.~G.}\
  \bibnamefont {Checkelsky}},\ }\href
  {https://doi.org/10.1038/s41567-023-02360-5} {\bibfield  {journal} {\bibinfo
  {journal} {Nature Physics}\ }\textbf {\bibinfo {volume} {20}},\ \bibinfo
  {pages} {610} (\bibinfo {year} {2024})}\BibitemShut {NoStop}%
\bibitem [{\citenamefont {Gao}\ \emph {et~al.}(2023)\citenamefont {Gao},
  \citenamefont {Zhang}, \citenamefont {Wang}, \citenamefont {Yan},
  \citenamefont {Han}, \citenamefont {Ji}, \citenamefont {Tao}, \citenamefont
  {Liu}, \citenamefont {Wang}, \citenamefont {Yuan}, \citenamefont {Qu},
  \citenamefont {Chen}, \citenamefont {Zhang}, \citenamefont {Huang},
  \citenamefont {Pan}, \citenamefont {Peng}, \citenamefont {Hu}, \citenamefont
  {Li}, \citenamefont {Huang}, \citenamefont {Zhou}, \citenamefont {Meng},
  \citenamefont {Yang}, \citenamefont {Wang}, \citenamefont {Yao},
  \citenamefont {Chen}, \citenamefont {Shi}, \citenamefont {Ding},
  \citenamefont {Yang}, \citenamefont {Jiang}, \citenamefont {Li},
  \citenamefont {Lei}, \citenamefont {Shi}, \citenamefont {Weng},\ and\
  \citenamefont {Qian}}]{GaoPRX23}%
  \BibitemOpen
  \bibfield  {author} {\bibinfo {author} {\bibfnamefont {S.}~\bibnamefont
  {Gao}}, \bibinfo {author} {\bibfnamefont {S.}~\bibnamefont {Zhang}}, \bibinfo
  {author} {\bibfnamefont {C.}~\bibnamefont {Wang}}, \bibinfo {author}
  {\bibfnamefont {S.}~\bibnamefont {Yan}}, \bibinfo {author} {\bibfnamefont
  {X.}~\bibnamefont {Han}}, \bibinfo {author} {\bibfnamefont {X.}~\bibnamefont
  {Ji}}, \bibinfo {author} {\bibfnamefont {W.}~\bibnamefont {Tao}}, \bibinfo
  {author} {\bibfnamefont {J.}~\bibnamefont {Liu}}, \bibinfo {author}
  {\bibfnamefont {T.}~\bibnamefont {Wang}}, \bibinfo {author} {\bibfnamefont
  {S.}~\bibnamefont {Yuan}}, \bibinfo {author} {\bibfnamefont {G.}~\bibnamefont
  {Qu}}, \bibinfo {author} {\bibfnamefont {Z.}~\bibnamefont {Chen}}, \bibinfo
  {author} {\bibfnamefont {Y.}~\bibnamefont {Zhang}}, \bibinfo {author}
  {\bibfnamefont {J.}~\bibnamefont {Huang}}, \bibinfo {author} {\bibfnamefont
  {M.}~\bibnamefont {Pan}}, \bibinfo {author} {\bibfnamefont {S.}~\bibnamefont
  {Peng}}, \bibinfo {author} {\bibfnamefont {Y.}~\bibnamefont {Hu}}, \bibinfo
  {author} {\bibfnamefont {H.}~\bibnamefont {Li}}, \bibinfo {author}
  {\bibfnamefont {Y.}~\bibnamefont {Huang}}, \bibinfo {author} {\bibfnamefont
  {H.}~\bibnamefont {Zhou}}, \bibinfo {author} {\bibfnamefont {S.}~\bibnamefont
  {Meng}}, \bibinfo {author} {\bibfnamefont {L.}~\bibnamefont {Yang}}, \bibinfo
  {author} {\bibfnamefont {Z.}~\bibnamefont {Wang}}, \bibinfo {author}
  {\bibfnamefont {Y.}~\bibnamefont {Yao}}, \bibinfo {author} {\bibfnamefont
  {Z.}~\bibnamefont {Chen}}, \bibinfo {author} {\bibfnamefont {M.}~\bibnamefont
  {Shi}}, \bibinfo {author} {\bibfnamefont {H.}~\bibnamefont {Ding}}, \bibinfo
  {author} {\bibfnamefont {H.}~\bibnamefont {Yang}}, \bibinfo {author}
  {\bibfnamefont {K.}~\bibnamefont {Jiang}}, \bibinfo {author} {\bibfnamefont
  {Y.}~\bibnamefont {Li}}, \bibinfo {author} {\bibfnamefont {H.}~\bibnamefont
  {Lei}}, \bibinfo {author} {\bibfnamefont {Y.}~\bibnamefont {Shi}}, \bibinfo
  {author} {\bibfnamefont {H.}~\bibnamefont {Weng}},\ and\ \bibinfo {author}
  {\bibfnamefont {T.}~\bibnamefont {Qian}},\ }\href
  {https://doi.org/10.1103/PhysRevX.13.041049} {\bibfield  {journal} {\bibinfo
  {journal} {Phys. Rev. X}\ }\textbf {\bibinfo {volume} {13}},\ \bibinfo
  {pages} {041049} (\bibinfo {year} {2023})}\BibitemShut {NoStop}%
\bibitem [{\citenamefont {Li}\ \emph {et~al.}(2022)\citenamefont {Li},
  \citenamefont {Wu}, \citenamefont {Fan}, \citenamefont {Zhang}, \citenamefont
  {Zhu}, \citenamefont {Zhu}, \citenamefont {Li},\ and\ \citenamefont
  {Wen}}]{LiPRB22}%
  \BibitemOpen
  \bibfield  {author} {\bibinfo {author} {\bibfnamefont {Q.}~\bibnamefont
  {Li}}, \bibinfo {author} {\bibfnamefont {Y.}~\bibnamefont {Wu}}, \bibinfo
  {author} {\bibfnamefont {X.}~\bibnamefont {Fan}}, \bibinfo {author}
  {\bibfnamefont {Y.-J.}\ \bibnamefont {Zhang}}, \bibinfo {author}
  {\bibfnamefont {X.}~\bibnamefont {Zhu}}, \bibinfo {author} {\bibfnamefont
  {Z.}~\bibnamefont {Zhu}}, \bibinfo {author} {\bibfnamefont {Y.}~\bibnamefont
  {Li}},\ and\ \bibinfo {author} {\bibfnamefont {H.-H.}\ \bibnamefont {Wen}},\
  }\href {https://doi.org/10.1103/PhysRevB.106.214501} {\bibfield  {journal}
  {\bibinfo  {journal} {Phys. Rev. B}\ }\textbf {\bibinfo {volume} {106}},\
  \bibinfo {pages} {214501} (\bibinfo {year} {2022})}\BibitemShut {NoStop}%
\bibitem [{\citenamefont {Liu}\ \emph {et~al.}(2024)\citenamefont {Liu},
  \citenamefont {Liu}, \citenamefont {Bao}, \citenamefont {Yang}, \citenamefont
  {Ji}, \citenamefont {Wu}, \citenamefont {Shen}, \citenamefont {Luo},
  \citenamefont {Yang}, \citenamefont {Liu}, \citenamefont {Xu}, \citenamefont
  {Yang}, \citenamefont {Chai}, \citenamefont {Lu}, \citenamefont {Liu},
  \citenamefont {Wang}, \citenamefont {Jiang}, \citenamefont {Tao},
  \citenamefont {Ren}, \citenamefont {Xu}, \citenamefont {Cao}, \citenamefont
  {Xu}, \citenamefont {Zhou}, \citenamefont {Cheng},\ and\ \citenamefont
  {Cao}}]{LiuNature24}%
  \BibitemOpen
  \bibfield  {author} {\bibinfo {author} {\bibfnamefont {Y.}~\bibnamefont
  {Liu}}, \bibinfo {author} {\bibfnamefont {Z.-Y.}\ \bibnamefont {Liu}},
  \bibinfo {author} {\bibfnamefont {J.-K.}\ \bibnamefont {Bao}}, \bibinfo
  {author} {\bibfnamefont {P.-T.}\ \bibnamefont {Yang}}, \bibinfo {author}
  {\bibfnamefont {L.-W.}\ \bibnamefont {Ji}}, \bibinfo {author} {\bibfnamefont
  {S.-Q.}\ \bibnamefont {Wu}}, \bibinfo {author} {\bibfnamefont {Q.-X.}\
  \bibnamefont {Shen}}, \bibinfo {author} {\bibfnamefont {J.}~\bibnamefont
  {Luo}}, \bibinfo {author} {\bibfnamefont {J.}~\bibnamefont {Yang}}, \bibinfo
  {author} {\bibfnamefont {J.-Y.}\ \bibnamefont {Liu}}, \bibinfo {author}
  {\bibfnamefont {C.-C.}\ \bibnamefont {Xu}}, \bibinfo {author} {\bibfnamefont
  {W.-Z.}\ \bibnamefont {Yang}}, \bibinfo {author} {\bibfnamefont {W.-L.}\
  \bibnamefont {Chai}}, \bibinfo {author} {\bibfnamefont {J.-Y.}\ \bibnamefont
  {Lu}}, \bibinfo {author} {\bibfnamefont {C.-C.}\ \bibnamefont {Liu}},
  \bibinfo {author} {\bibfnamefont {B.-S.}\ \bibnamefont {Wang}}, \bibinfo
  {author} {\bibfnamefont {H.}~\bibnamefont {Jiang}}, \bibinfo {author}
  {\bibfnamefont {Q.}~\bibnamefont {Tao}}, \bibinfo {author} {\bibfnamefont
  {Z.}~\bibnamefont {Ren}}, \bibinfo {author} {\bibfnamefont {X.-F.}\
  \bibnamefont {Xu}}, \bibinfo {author} {\bibfnamefont {C.}~\bibnamefont
  {Cao}}, \bibinfo {author} {\bibfnamefont {Z.-A.}\ \bibnamefont {Xu}},
  \bibinfo {author} {\bibfnamefont {R.}~\bibnamefont {Zhou}}, \bibinfo {author}
  {\bibfnamefont {J.-G.}\ \bibnamefont {Cheng}},\ and\ \bibinfo {author}
  {\bibfnamefont {G.-H.}\ \bibnamefont {Cao}},\ }\href
  {https://doi.org/10.1038/s41586-024-07761-x} {\bibfield  {journal} {\bibinfo
  {journal} {Nature}\ }\textbf {\bibinfo {volume} {632}},\ \bibinfo {pages}
  {1032} (\bibinfo {year} {2024})}\BibitemShut {NoStop}%
\bibitem [{\citenamefont {Zhang}\ \emph {et~al.}(2026)\citenamefont {Zhang},
  \citenamefont {Li}, \citenamefont {Song}, \citenamefont {Xie}, \citenamefont
  {Wang}, \citenamefont {Miao}, \citenamefont {Ye}, \citenamefont {Liu},
  \citenamefont {Liang}, \citenamefont {Cai}, \citenamefont {Chen},
  \citenamefont {Zhu}, \citenamefont {Xu}, \citenamefont {Li}, \citenamefont
  {Yu}, \citenamefont {Zhang}, \citenamefont {Zhang}, \citenamefont {Yang},
  \citenamefont {Wang}, \citenamefont {Peng}, \citenamefont {Mao},
  \citenamefont {Zhu}, \citenamefont {Liu}, \citenamefont {Xu}, \citenamefont
  {feng Yang}, \citenamefont {Ying}, \citenamefont {Zhao},\ and\ \citenamefont
  {Zhou}}]{ZhangCondMat2026}%
  \BibitemOpen
  \bibfield  {author} {\bibinfo {author} {\bibfnamefont {X.}~\bibnamefont
  {Zhang}}, \bibinfo {author} {\bibfnamefont {X.}~\bibnamefont {Li}}, \bibinfo
  {author} {\bibfnamefont {B.}~\bibnamefont {Song}}, \bibinfo {author}
  {\bibfnamefont {Y.}~\bibnamefont {Xie}}, \bibinfo {author} {\bibfnamefont
  {Q.}~\bibnamefont {Wang}}, \bibinfo {author} {\bibfnamefont {T.}~\bibnamefont
  {Miao}}, \bibinfo {author} {\bibfnamefont {S.}~\bibnamefont {Ye}}, \bibinfo
  {author} {\bibfnamefont {J.}~\bibnamefont {Liu}}, \bibinfo {author}
  {\bibfnamefont {B.}~\bibnamefont {Liang}}, \bibinfo {author} {\bibfnamefont
  {N.}~\bibnamefont {Cai}}, \bibinfo {author} {\bibfnamefont {H.}~\bibnamefont
  {Chen}}, \bibinfo {author} {\bibfnamefont {W.}~\bibnamefont {Zhu}}, \bibinfo
  {author} {\bibfnamefont {M.}~\bibnamefont {Xu}}, \bibinfo {author}
  {\bibfnamefont {W.-J.}\ \bibnamefont {Li}}, \bibinfo {author} {\bibfnamefont
  {S.-L.}\ \bibnamefont {Yu}}, \bibinfo {author} {\bibfnamefont
  {S.}~\bibnamefont {Zhang}}, \bibinfo {author} {\bibfnamefont
  {F.}~\bibnamefont {Zhang}}, \bibinfo {author} {\bibfnamefont
  {F.}~\bibnamefont {Yang}}, \bibinfo {author} {\bibfnamefont {Z.}~\bibnamefont
  {Wang}}, \bibinfo {author} {\bibfnamefont {Q.}~\bibnamefont {Peng}}, \bibinfo
  {author} {\bibfnamefont {H.}~\bibnamefont {Mao}}, \bibinfo {author}
  {\bibfnamefont {Z.}~\bibnamefont {Zhu}}, \bibinfo {author} {\bibfnamefont
  {G.}~\bibnamefont {Liu}}, \bibinfo {author} {\bibfnamefont {Z.}~\bibnamefont
  {Xu}}, \bibinfo {author} {\bibfnamefont {Y.}~\bibnamefont {feng Yang}},
  \bibinfo {author} {\bibfnamefont {T.}~\bibnamefont {Ying}}, \bibinfo {author}
  {\bibfnamefont {L.}~\bibnamefont {Zhao}},\ and\ \bibinfo {author}
  {\bibfnamefont {X.~J.}\ \bibnamefont {Zhou}},\ }\href
  {https://arxiv.org/abs/2604.02922} {\bibinfo {title} {Mott-derived local
  moments and kondo hybridization in a d-electron kagome lattice}} (\bibinfo
  {year} {2026}),\ \Eprint {https://arxiv.org/abs/2604.02922} {arXiv:2604.02922
  [cond-mat.str-el]} \BibitemShut {NoStop}%
\bibitem [{sup()}]{sup}%
  \BibitemOpen
  \href@noop {} {}\bibinfo {note} {See supplementary information}\BibitemShut
  {NoStop}%
\bibitem [{\citenamefont {Meier}\ \emph {et~al.}(2020)\citenamefont {Meier},
  \citenamefont {Du}, \citenamefont {Okamoto}, \citenamefont {Mohanta},
  \citenamefont {May}, \citenamefont {McGuire}, \citenamefont {Bridges},
  \citenamefont {Samolyuk},\ and\ \citenamefont {Sales}}]{MeierPRB20}%
  \BibitemOpen
  \bibfield  {author} {\bibinfo {author} {\bibfnamefont {W.~R.}\ \bibnamefont
  {Meier}}, \bibinfo {author} {\bibfnamefont {M.-H.}\ \bibnamefont {Du}},
  \bibinfo {author} {\bibfnamefont {S.}~\bibnamefont {Okamoto}}, \bibinfo
  {author} {\bibfnamefont {N.}~\bibnamefont {Mohanta}}, \bibinfo {author}
  {\bibfnamefont {A.~F.}\ \bibnamefont {May}}, \bibinfo {author} {\bibfnamefont
  {M.~A.}\ \bibnamefont {McGuire}}, \bibinfo {author} {\bibfnamefont {C.~A.}\
  \bibnamefont {Bridges}}, \bibinfo {author} {\bibfnamefont {G.~D.}\
  \bibnamefont {Samolyuk}},\ and\ \bibinfo {author} {\bibfnamefont {B.~C.}\
  \bibnamefont {Sales}},\ }\href {https://doi.org/10.1103/PhysRevB.102.075148}
  {\bibfield  {journal} {\bibinfo  {journal} {Phys. Rev. B}\ }\textbf {\bibinfo
  {volume} {102}},\ \bibinfo {pages} {075148} (\bibinfo {year}
  {2020})}\BibitemShut {NoStop}%
\bibitem [{\citenamefont {Liu}\ \emph {et~al.}(2020)\citenamefont {Liu},
  \citenamefont {Li}, \citenamefont {Wang}, \citenamefont {Wang}, \citenamefont
  {Wen}, \citenamefont {Jiang}, \citenamefont {Lu}, \citenamefont {Yan},
  \citenamefont {Huang}, \citenamefont {Shen}, \citenamefont {Yin},
  \citenamefont {Wang}, \citenamefont {Yin}, \citenamefont {Lei},\ and\
  \citenamefont {Wang}}]{LiuNatCom20}%
  \BibitemOpen
  \bibfield  {author} {\bibinfo {author} {\bibfnamefont {Z.}~\bibnamefont
  {Liu}}, \bibinfo {author} {\bibfnamefont {M.}~\bibnamefont {Li}}, \bibinfo
  {author} {\bibfnamefont {Q.}~\bibnamefont {Wang}}, \bibinfo {author}
  {\bibfnamefont {G.}~\bibnamefont {Wang}}, \bibinfo {author} {\bibfnamefont
  {C.}~\bibnamefont {Wen}}, \bibinfo {author} {\bibfnamefont {K.}~\bibnamefont
  {Jiang}}, \bibinfo {author} {\bibfnamefont {X.}~\bibnamefont {Lu}}, \bibinfo
  {author} {\bibfnamefont {S.}~\bibnamefont {Yan}}, \bibinfo {author}
  {\bibfnamefont {Y.}~\bibnamefont {Huang}}, \bibinfo {author} {\bibfnamefont
  {D.}~\bibnamefont {Shen}}, \bibinfo {author} {\bibfnamefont {J.-X.}\
  \bibnamefont {Yin}}, \bibinfo {author} {\bibfnamefont {Z.}~\bibnamefont
  {Wang}}, \bibinfo {author} {\bibfnamefont {Z.}~\bibnamefont {Yin}}, \bibinfo
  {author} {\bibfnamefont {H.}~\bibnamefont {Lei}},\ and\ \bibinfo {author}
  {\bibfnamefont {S.}~\bibnamefont {Wang}},\ }\href
  {https://doi.org/10.1038/s41467-020-17462-4} {\bibfield  {journal} {\bibinfo
  {journal} {Nature Communications}\ }\textbf {\bibinfo {volume} {11}},\
  \bibinfo {pages} {4002} (\bibinfo {year} {2020})}\BibitemShut {NoStop}%
\bibitem [{\citenamefont {Kang}\ \emph {et~al.}(2020)\citenamefont {Kang},
  \citenamefont {Fang}, \citenamefont {Ye}, \citenamefont {Po}, \citenamefont
  {Denlinger}, \citenamefont {Jozwiak}, \citenamefont {Bostwick}, \citenamefont
  {Rotenberg}, \citenamefont {Kaxiras}, \citenamefont {Checkelsky},\ and\
  \citenamefont {Comin}}]{KangNatCom20_CoSn}%
  \BibitemOpen
  \bibfield  {author} {\bibinfo {author} {\bibfnamefont {M.}~\bibnamefont
  {Kang}}, \bibinfo {author} {\bibfnamefont {S.}~\bibnamefont {Fang}}, \bibinfo
  {author} {\bibfnamefont {L.}~\bibnamefont {Ye}}, \bibinfo {author}
  {\bibfnamefont {H.~C.}\ \bibnamefont {Po}}, \bibinfo {author} {\bibfnamefont
  {J.}~\bibnamefont {Denlinger}}, \bibinfo {author} {\bibfnamefont
  {C.}~\bibnamefont {Jozwiak}}, \bibinfo {author} {\bibfnamefont
  {A.}~\bibnamefont {Bostwick}}, \bibinfo {author} {\bibfnamefont
  {E.}~\bibnamefont {Rotenberg}}, \bibinfo {author} {\bibfnamefont
  {E.}~\bibnamefont {Kaxiras}}, \bibinfo {author} {\bibfnamefont {J.~G.}\
  \bibnamefont {Checkelsky}},\ and\ \bibinfo {author} {\bibfnamefont
  {R.}~\bibnamefont {Comin}},\ }\href
  {https://doi.org/10.1038/s41467-020-17465-1} {\bibfield  {journal} {\bibinfo
  {journal} {Nature Communications}\ }\textbf {\bibinfo {volume} {11}},\
  \bibinfo {pages} {4004} (\bibinfo {year} {2020})}\BibitemShut {NoStop}%
\bibitem [{\citenamefont {Kang}\ \emph {et~al.}(2025)\citenamefont {Kang},
  \citenamefont {Kim}, \citenamefont {Qian}, \citenamefont {Neves},
  \citenamefont {Ye}, \citenamefont {Jung}, \citenamefont {Puntel},
  \citenamefont {Mazzola}, \citenamefont {Fang}, \citenamefont {Jozwiak},
  \citenamefont {Bostwick}, \citenamefont {Rotenberg}, \citenamefont {Fuji},
  \citenamefont {Vobornik}, \citenamefont {Park}, \citenamefont {Checkelsky},
  \citenamefont {Yang},\ and\ \citenamefont {Comin}}]{KangNatPhys25}%
  \BibitemOpen
  \bibfield  {author} {\bibinfo {author} {\bibfnamefont {M.}~\bibnamefont
  {Kang}}, \bibinfo {author} {\bibfnamefont {S.}~\bibnamefont {Kim}}, \bibinfo
  {author} {\bibfnamefont {Y.}~\bibnamefont {Qian}}, \bibinfo {author}
  {\bibfnamefont {P.~M.}\ \bibnamefont {Neves}}, \bibinfo {author}
  {\bibfnamefont {L.}~\bibnamefont {Ye}}, \bibinfo {author} {\bibfnamefont
  {J.}~\bibnamefont {Jung}}, \bibinfo {author} {\bibfnamefont {D.}~\bibnamefont
  {Puntel}}, \bibinfo {author} {\bibfnamefont {F.}~\bibnamefont {Mazzola}},
  \bibinfo {author} {\bibfnamefont {S.}~\bibnamefont {Fang}}, \bibinfo {author}
  {\bibfnamefont {C.}~\bibnamefont {Jozwiak}}, \bibinfo {author} {\bibfnamefont
  {A.}~\bibnamefont {Bostwick}}, \bibinfo {author} {\bibfnamefont
  {E.}~\bibnamefont {Rotenberg}}, \bibinfo {author} {\bibfnamefont
  {J.}~\bibnamefont {Fuji}}, \bibinfo {author} {\bibfnamefont {I.}~\bibnamefont
  {Vobornik}}, \bibinfo {author} {\bibfnamefont {J.-H.}\ \bibnamefont {Park}},
  \bibinfo {author} {\bibfnamefont {J.~G.}\ \bibnamefont {Checkelsky}},
  \bibinfo {author} {\bibfnamefont {B.-J.}\ \bibnamefont {Yang}},\ and\
  \bibinfo {author} {\bibfnamefont {R.}~\bibnamefont {Comin}},\ }\href
  {https://doi.org/10.1038/s41567-024-02678-8} {\bibfield  {journal} {\bibinfo
  {journal} {Nature Physics}\ }\textbf {\bibinfo {volume} {21}},\ \bibinfo
  {pages} {110} (\bibinfo {year} {2025})}\BibitemShut {NoStop}%
\bibitem [{\citenamefont {Brouet}\ \emph {et~al.}(2025)\citenamefont {Brouet},
  \citenamefont {Vedant}, \citenamefont {Bertran}, \citenamefont {Le~F\`evre},\
  and\ \citenamefont {Rubel}}]{BrouetPRB25}%
  \BibitemOpen
  \bibfield  {author} {\bibinfo {author} {\bibfnamefont {V.}~\bibnamefont
  {Brouet}}, \bibinfo {author} {\bibfnamefont {A.}~\bibnamefont {Vedant}},
  \bibinfo {author} {\bibfnamefont {F.}~\bibnamefont {Bertran}}, \bibinfo
  {author} {\bibfnamefont {P.}~\bibnamefont {Le~F\`evre}},\ and\ \bibinfo
  {author} {\bibfnamefont {O.}~\bibnamefont {Rubel}},\ }\href
  {https://doi.org/10.1103/5rpf-sd26} {\bibfield  {journal} {\bibinfo
  {journal} {Phys. Rev. B}\ }\textbf {\bibinfo {volume} {112}},\ \bibinfo
  {pages} {115135} (\bibinfo {year} {2025})}\BibitemShut {NoStop}%
\bibitem [{\citenamefont {Sales}\ \emph {et~al.}(2021)\citenamefont {Sales},
  \citenamefont {Meier}, \citenamefont {May}, \citenamefont {Xing},
  \citenamefont {Yan}, \citenamefont {Gao}, \citenamefont {Liu}, \citenamefont
  {Stone}, \citenamefont {Christianson}, \citenamefont {Zhang},\ and\
  \citenamefont {McGuire}}]{SalesPRM21}%
  \BibitemOpen
  \bibfield  {author} {\bibinfo {author} {\bibfnamefont {B.~C.}\ \bibnamefont
  {Sales}}, \bibinfo {author} {\bibfnamefont {W.~R.}\ \bibnamefont {Meier}},
  \bibinfo {author} {\bibfnamefont {A.~F.}\ \bibnamefont {May}}, \bibinfo
  {author} {\bibfnamefont {J.}~\bibnamefont {Xing}}, \bibinfo {author}
  {\bibfnamefont {J.-Q.}\ \bibnamefont {Yan}}, \bibinfo {author} {\bibfnamefont
  {S.}~\bibnamefont {Gao}}, \bibinfo {author} {\bibfnamefont {Y.~H.}\
  \bibnamefont {Liu}}, \bibinfo {author} {\bibfnamefont {M.~B.}\ \bibnamefont
  {Stone}}, \bibinfo {author} {\bibfnamefont {A.~D.}\ \bibnamefont
  {Christianson}}, \bibinfo {author} {\bibfnamefont {Q.}~\bibnamefont
  {Zhang}},\ and\ \bibinfo {author} {\bibfnamefont {M.~A.}\ \bibnamefont
  {McGuire}},\ }\href {https://doi.org/10.1103/PhysRevMaterials.5.044202}
  {\bibfield  {journal} {\bibinfo  {journal} {Phys. Rev. Mater.}\ }\textbf
  {\bibinfo {volume} {5}},\ \bibinfo {pages} {044202} (\bibinfo {year}
  {2021})}\BibitemShut {NoStop}%
\bibitem [{\citenamefont {Sales}\ \emph {et~al.}(2019)\citenamefont {Sales},
  \citenamefont {Yan}, \citenamefont {Meier}, \citenamefont {Christianson},
  \citenamefont {Okamoto},\ and\ \citenamefont {McGuire}}]{SalesPRM19}%
  \BibitemOpen
  \bibfield  {author} {\bibinfo {author} {\bibfnamefont {B.~C.}\ \bibnamefont
  {Sales}}, \bibinfo {author} {\bibfnamefont {J.}~\bibnamefont {Yan}}, \bibinfo
  {author} {\bibfnamefont {W.~R.}\ \bibnamefont {Meier}}, \bibinfo {author}
  {\bibfnamefont {A.~D.}\ \bibnamefont {Christianson}}, \bibinfo {author}
  {\bibfnamefont {S.}~\bibnamefont {Okamoto}},\ and\ \bibinfo {author}
  {\bibfnamefont {M.~A.}\ \bibnamefont {McGuire}},\ }\href
  {https://doi.org/10.1103/PhysRevMaterials.3.114203} {\bibfield  {journal}
  {\bibinfo  {journal} {Phys. Rev. Mater.}\ }\textbf {\bibinfo {volume} {3}},\
  \bibinfo {pages} {114203} (\bibinfo {year} {2019})}\BibitemShut {NoStop}%
\bibitem [{\citenamefont {Ren}\ \emph {et~al.}(2024)\citenamefont {Ren},
  \citenamefont {Huang}, \citenamefont {Tan}, \citenamefont {Biswas},
  \citenamefont {Pulkkinen}, \citenamefont {Zhang}, \citenamefont {Xie},
  \citenamefont {Yue}, \citenamefont {Chen}, \citenamefont {Xie}, \citenamefont
  {Allen}, \citenamefont {Wu}, \citenamefont {Ren}, \citenamefont
  {Rajapitamahuni}, \citenamefont {Kundu}, \citenamefont {Vescovo},
  \citenamefont {Kono}, \citenamefont {Morosan}, \citenamefont {Dai},
  \citenamefont {Zhu}, \citenamefont {Si}, \citenamefont {Min{\'a}r},
  \citenamefont {Yan},\ and\ \citenamefont {Yi}}]{RenNatCom24}%
  \BibitemOpen
  \bibfield  {author} {\bibinfo {author} {\bibfnamefont {Z.}~\bibnamefont
  {Ren}}, \bibinfo {author} {\bibfnamefont {J.}~\bibnamefont {Huang}}, \bibinfo
  {author} {\bibfnamefont {H.}~\bibnamefont {Tan}}, \bibinfo {author}
  {\bibfnamefont {A.}~\bibnamefont {Biswas}}, \bibinfo {author} {\bibfnamefont
  {A.}~\bibnamefont {Pulkkinen}}, \bibinfo {author} {\bibfnamefont
  {Y.}~\bibnamefont {Zhang}}, \bibinfo {author} {\bibfnamefont
  {Y.}~\bibnamefont {Xie}}, \bibinfo {author} {\bibfnamefont {Z.}~\bibnamefont
  {Yue}}, \bibinfo {author} {\bibfnamefont {L.}~\bibnamefont {Chen}}, \bibinfo
  {author} {\bibfnamefont {F.}~\bibnamefont {Xie}}, \bibinfo {author}
  {\bibfnamefont {K.}~\bibnamefont {Allen}}, \bibinfo {author} {\bibfnamefont
  {H.}~\bibnamefont {Wu}}, \bibinfo {author} {\bibfnamefont {Q.}~\bibnamefont
  {Ren}}, \bibinfo {author} {\bibfnamefont {A.}~\bibnamefont {Rajapitamahuni}},
  \bibinfo {author} {\bibfnamefont {A.~K.}\ \bibnamefont {Kundu}}, \bibinfo
  {author} {\bibfnamefont {E.}~\bibnamefont {Vescovo}}, \bibinfo {author}
  {\bibfnamefont {J.}~\bibnamefont {Kono}}, \bibinfo {author} {\bibfnamefont
  {E.}~\bibnamefont {Morosan}}, \bibinfo {author} {\bibfnamefont
  {P.}~\bibnamefont {Dai}}, \bibinfo {author} {\bibfnamefont {J.-X.}\
  \bibnamefont {Zhu}}, \bibinfo {author} {\bibfnamefont {Q.}~\bibnamefont
  {Si}}, \bibinfo {author} {\bibfnamefont {J.}~\bibnamefont {Min{\'a}r}},
  \bibinfo {author} {\bibfnamefont {B.}~\bibnamefont {Yan}},\ and\ \bibinfo
  {author} {\bibfnamefont {M.}~\bibnamefont {Yi}},\ }\href
  {https://doi.org/10.1038/s41467-024-53722-3} {\bibfield  {journal} {\bibinfo
  {journal} {Nature Communications}\ }\textbf {\bibinfo {volume} {15}},\
  \bibinfo {pages} {9376} (\bibinfo {year} {2024})}\BibitemShut {NoStop}%
\bibitem [{\citenamefont {Chen}\ \emph {et~al.}(2026)\citenamefont {Chen},
  \citenamefont {Zheng}, \citenamefont {He}, \citenamefont {Wu}, \citenamefont
  {Ying}, \citenamefont {Sankar}, \citenamefont {Li}, \citenamefont {Wei},
  \citenamefont {Dai}, \citenamefont {Po},\ and\ \citenamefont
  {J\"ack}}]{ChenPRX26}%
  \BibitemOpen
  \bibfield  {author} {\bibinfo {author} {\bibfnamefont {C.}~\bibnamefont
  {Chen}}, \bibinfo {author} {\bibfnamefont {J.}~\bibnamefont {Zheng}},
  \bibinfo {author} {\bibfnamefont {Y.}~\bibnamefont {He}}, \bibinfo {author}
  {\bibfnamefont {S.}~\bibnamefont {Wu}}, \bibinfo {author} {\bibfnamefont
  {X.}~\bibnamefont {Ying}}, \bibinfo {author} {\bibfnamefont {S.}~\bibnamefont
  {Sankar}}, \bibinfo {author} {\bibfnamefont {L.}~\bibnamefont {Li}}, \bibinfo
  {author} {\bibfnamefont {Y.}~\bibnamefont {Wei}}, \bibinfo {author}
  {\bibfnamefont {X.}~\bibnamefont {Dai}}, \bibinfo {author} {\bibfnamefont
  {H.~C.}\ \bibnamefont {Po}},\ and\ \bibinfo {author} {\bibfnamefont
  {B.}~\bibnamefont {J\"ack}},\ }\href {https://doi.org/10.1103/c93j-k53f}
  {\bibfield  {journal} {\bibinfo  {journal} {Phys. Rev. X}\ }\textbf {\bibinfo
  {volume} {16}},\ \bibinfo {pages} {011009} (\bibinfo {year}
  {2026})}\BibitemShut {NoStop}%
\bibitem [{\citenamefont {Sales}\ \emph {et~al.}(2022)\citenamefont {Sales},
  \citenamefont {Meier}, \citenamefont {Parker}, \citenamefont {Yin},
  \citenamefont {Yan}, \citenamefont {May}, \citenamefont {Calder},
  \citenamefont {Aczel}, \citenamefont {Zhang}, \citenamefont {Li},
  \citenamefont {Yilmaz}, \citenamefont {Vescovo}, \citenamefont {Miao},
  \citenamefont {Moseley}, \citenamefont {Hermann},\ and\ \citenamefont
  {McGuire}}]{SalesChemMat22}%
  \BibitemOpen
  \bibfield  {author} {\bibinfo {author} {\bibfnamefont {B.~C.}\ \bibnamefont
  {Sales}}, \bibinfo {author} {\bibfnamefont {W.~R.}\ \bibnamefont {Meier}},
  \bibinfo {author} {\bibfnamefont {D.~S.}\ \bibnamefont {Parker}}, \bibinfo
  {author} {\bibfnamefont {L.}~\bibnamefont {Yin}}, \bibinfo {author}
  {\bibfnamefont {J.}~\bibnamefont {Yan}}, \bibinfo {author} {\bibfnamefont
  {A.~F.}\ \bibnamefont {May}}, \bibinfo {author} {\bibfnamefont
  {S.}~\bibnamefont {Calder}}, \bibinfo {author} {\bibfnamefont {A.~A.}\
  \bibnamefont {Aczel}}, \bibinfo {author} {\bibfnamefont {Q.}~\bibnamefont
  {Zhang}}, \bibinfo {author} {\bibfnamefont {H.}~\bibnamefont {Li}}, \bibinfo
  {author} {\bibfnamefont {T.}~\bibnamefont {Yilmaz}}, \bibinfo {author}
  {\bibfnamefont {E.}~\bibnamefont {Vescovo}}, \bibinfo {author} {\bibfnamefont
  {H.}~\bibnamefont {Miao}}, \bibinfo {author} {\bibfnamefont {D.~H.}\
  \bibnamefont {Moseley}}, \bibinfo {author} {\bibfnamefont {R.~P.}\
  \bibnamefont {Hermann}},\ and\ \bibinfo {author} {\bibfnamefont {M.~A.}\
  \bibnamefont {McGuire}},\ }\href
  {https://doi.org/10.1021/acs.chemmater.2c01634} {\bibfield  {journal}
  {\bibinfo  {journal} {Chemistry of Materials}\ }\textbf {\bibinfo {volume}
  {34}},\ \bibinfo {pages} {7069} (\bibinfo {year} {2022})},\ \Eprint
  {https://arxiv.org/abs/https://doi.org/10.1021/acs.chemmater.2c01634}
  {https://doi.org/10.1021/acs.chemmater.2c01634} \BibitemShut {NoStop}%
\bibitem [{\citenamefont {H{\"a}m{\"a}l{\"a}inen}\ \emph
  {et~al.}(1991)\citenamefont {H{\"a}m{\"a}l{\"a}inen}, \citenamefont
  {Siddons}, \citenamefont {Hastings},\ and\ \citenamefont
  {Berman}}]{Hamalainen1991}%
  \BibitemOpen
  \bibfield  {author} {\bibinfo {author} {\bibfnamefont {K.}~\bibnamefont
  {H{\"a}m{\"a}l{\"a}inen}}, \bibinfo {author} {\bibfnamefont {D.~P.}\
  \bibnamefont {Siddons}}, \bibinfo {author} {\bibfnamefont {J.~B.}\
  \bibnamefont {Hastings}},\ and\ \bibinfo {author} {\bibfnamefont {L.~E.}\
  \bibnamefont {Berman}},\ }\href@noop {} {\bibfield  {journal} {\bibinfo
  {journal} {Physical Review Letters}\ }\textbf {\bibinfo {volume} {67}},\
  \bibinfo {pages} {2850} (\bibinfo {year} {1991})}\BibitemShut {NoStop}%
\bibitem [{\citenamefont {Glatzel}\ \emph {et~al.}(2013)\citenamefont
  {Glatzel}, \citenamefont {Weng}, \citenamefont {Kvashnina}, \citenamefont
  {Swarbrick}, \citenamefont {Sikora}, \citenamefont {Gallo}, \citenamefont
  {Smolentsev},\ and\ \citenamefont {Mori}}]{Glatzel2013}%
  \BibitemOpen
  \bibfield  {author} {\bibinfo {author} {\bibfnamefont {P.}~\bibnamefont
  {Glatzel}}, \bibinfo {author} {\bibfnamefont {T.-C.}\ \bibnamefont {Weng}},
  \bibinfo {author} {\bibfnamefont {K.}~\bibnamefont {Kvashnina}}, \bibinfo
  {author} {\bibfnamefont {J.}~\bibnamefont {Swarbrick}}, \bibinfo {author}
  {\bibfnamefont {M.}~\bibnamefont {Sikora}}, \bibinfo {author} {\bibfnamefont
  {E.}~\bibnamefont {Gallo}}, \bibinfo {author} {\bibfnamefont
  {N.}~\bibnamefont {Smolentsev}},\ and\ \bibinfo {author} {\bibfnamefont
  {R.~A.}\ \bibnamefont {Mori}},\ }\href@noop {} {\bibfield  {journal}
  {\bibinfo  {journal} {Journal of Electron Spectroscopy and Related
  Phenomena}\ }\textbf {\bibinfo {volume} {188}},\ \bibinfo {pages} {17}
  (\bibinfo {year} {2013})}\BibitemShut {NoStop}%
\bibitem [{\citenamefont {Lafuerza}\ \emph {et~al.}(2020)\citenamefont
  {Lafuerza}, \citenamefont {Retegan}, \citenamefont {Detlefs}, \citenamefont
  {Chatterjee}, \citenamefont {Yachandra}, \citenamefont {Yano},\ and\
  \citenamefont {Glatzel}}]{Lafuerza2020}%
  \BibitemOpen
  \bibfield  {author} {\bibinfo {author} {\bibfnamefont {S.}~\bibnamefont
  {Lafuerza}}, \bibinfo {author} {\bibfnamefont {M.}~\bibnamefont {Retegan}},
  \bibinfo {author} {\bibfnamefont {B.}~\bibnamefont {Detlefs}}, \bibinfo
  {author} {\bibfnamefont {R.}~\bibnamefont {Chatterjee}}, \bibinfo {author}
  {\bibfnamefont {V.}~\bibnamefont {Yachandra}}, \bibinfo {author}
  {\bibfnamefont {J.}~\bibnamefont {Yano}},\ and\ \bibinfo {author}
  {\bibfnamefont {P.}~\bibnamefont {Glatzel}},\ }\href@noop {} {\bibfield
  {journal} {\bibinfo  {journal} {Nanoscale}\ }\textbf {\bibinfo {volume}
  {12}},\ \bibinfo {pages} {16270} (\bibinfo {year} {2020})}\BibitemShut
  {NoStop}%
\bibitem [{\citenamefont {Rueff}\ \emph {et~al.}(2015)\citenamefont {Rueff},
  \citenamefont {Ablett}, \citenamefont {C{\'e}olin}, \citenamefont {Prieur},
  \citenamefont {Moreno}, \citenamefont {Bal{\'e}dent}, \citenamefont
  {Lassalle-Kaiser}, \citenamefont {Rault}, \citenamefont {Simon},\ and\
  \citenamefont {Shukla}}]{Rueff2015}%
  \BibitemOpen
  \bibfield  {author} {\bibinfo {author} {\bibfnamefont {J.-P.}\ \bibnamefont
  {Rueff}}, \bibinfo {author} {\bibfnamefont {J.~M.}\ \bibnamefont {Ablett}},
  \bibinfo {author} {\bibfnamefont {D.}~\bibnamefont {C{\'e}olin}}, \bibinfo
  {author} {\bibfnamefont {D.}~\bibnamefont {Prieur}}, \bibinfo {author}
  {\bibfnamefont {T.}~\bibnamefont {Moreno}}, \bibinfo {author} {\bibfnamefont
  {V.}~\bibnamefont {Bal{\'e}dent}}, \bibinfo {author} {\bibfnamefont
  {B.}~\bibnamefont {Lassalle-Kaiser}}, \bibinfo {author} {\bibfnamefont
  {J.~E.}\ \bibnamefont {Rault}}, \bibinfo {author} {\bibfnamefont
  {M.}~\bibnamefont {Simon}},\ and\ \bibinfo {author} {\bibfnamefont
  {A.}~\bibnamefont {Shukla}},\ }\href@noop {} {\bibfield  {journal} {\bibinfo
  {journal} {Journal of Synchrotron Radiation}\ }\textbf {\bibinfo {volume}
  {22}},\ \bibinfo {pages} {175} (\bibinfo {year} {2015})}\BibitemShut
  {NoStop}%
\bibitem [{\citenamefont {Ablett}\ \emph {et~al.}(2019)\citenamefont {Ablett},
  \citenamefont {Prieur}, \citenamefont {C{\'e}olin}, \citenamefont
  {Lassalle-Kaiser}, \citenamefont {Lebert}, \citenamefont {Sauvage},
  \citenamefont {Moreno}, \citenamefont {Bac}, \citenamefont {Bal{\'e}dent},
  \citenamefont {Ovono}, \citenamefont {Morand}, \citenamefont {G{\'e}lebart},
  \citenamefont {Shukla},\ and\ \citenamefont {Rueff}}]{Ablett2019}%
  \BibitemOpen
  \bibfield  {author} {\bibinfo {author} {\bibfnamefont {J.~M.}\ \bibnamefont
  {Ablett}}, \bibinfo {author} {\bibfnamefont {D.}~\bibnamefont {Prieur}},
  \bibinfo {author} {\bibfnamefont {D.}~\bibnamefont {C{\'e}olin}}, \bibinfo
  {author} {\bibfnamefont {B.}~\bibnamefont {Lassalle-Kaiser}}, \bibinfo
  {author} {\bibfnamefont {B.}~\bibnamefont {Lebert}}, \bibinfo {author}
  {\bibfnamefont {M.}~\bibnamefont {Sauvage}}, \bibinfo {author} {\bibfnamefont
  {T.~h.}\ \bibnamefont {Moreno}}, \bibinfo {author} {\bibfnamefont
  {S.}~\bibnamefont {Bac}}, \bibinfo {author} {\bibfnamefont {V.}~\bibnamefont
  {Bal{\'e}dent}}, \bibinfo {author} {\bibfnamefont {A.}~\bibnamefont {Ovono}},
  \bibinfo {author} {\bibfnamefont {M.}~\bibnamefont {Morand}}, \bibinfo
  {author} {\bibfnamefont {F.}~\bibnamefont {G{\'e}lebart}}, \bibinfo {author}
  {\bibfnamefont {A.}~\bibnamefont {Shukla}},\ and\ \bibinfo {author}
  {\bibfnamefont {J.-P.}\ \bibnamefont {Rueff}},\ }\href@noop {} {\bibfield
  {journal} {\bibinfo  {journal} {Journal of Synchrotron Radiation}\ }\textbf
  {\bibinfo {volume} {26}},\ \bibinfo {pages} {263} (\bibinfo {year}
  {2019})}\BibitemShut {NoStop}%
\bibitem [{\citenamefont {Blaha}\ \emph {et~al.}(2001)\citenamefont {Blaha},
  \citenamefont {Schwarz}, \citenamefont {Madsen}, \citenamefont {Kvasnicka},\
  and\ \citenamefont {Luitz}}]{blaha2001}%
  \BibitemOpen
  \bibfield  {author} {\bibinfo {author} {\bibfnamefont {P.}~\bibnamefont
  {Blaha}}, \bibinfo {author} {\bibfnamefont {K.}~\bibnamefont {Schwarz}},
  \bibinfo {author} {\bibfnamefont {G.~K.}\ \bibnamefont {Madsen}}, \bibinfo
  {author} {\bibfnamefont {D.}~\bibnamefont {Kvasnicka}},\ and\ \bibinfo
  {author} {\bibfnamefont {J.}~\bibnamefont {Luitz}},\ }\href@noop {}
  {\bibfield  {journal} {\bibinfo  {journal} {edited by {K}.Schwarz, Vienna
  University of Technology, Austria}\ } (\bibinfo {year} {2001})}\BibitemShut
  {NoStop}%
\bibitem [{\citenamefont {Kresse}\ and\ \citenamefont
  {Hafner}(1993)}]{Kresse_PRB_47_1993}%
  \BibitemOpen
  \bibfield  {author} {\bibinfo {author} {\bibfnamefont {G.}~\bibnamefont
  {Kresse}}\ and\ \bibinfo {author} {\bibfnamefont {J.}~\bibnamefont
  {Hafner}},\ }\href {https://doi.org/10.1103/PhysRevB.47.558} {\bibfield
  {journal} {\bibinfo  {journal} {Phys. Rev. B}\ }\textbf {\bibinfo {volume}
  {47}},\ \bibinfo {pages} {558} (\bibinfo {year} {1993})}\BibitemShut
  {NoStop}%
\bibitem [{\citenamefont {Kresse}\ and\ \citenamefont
  {Furthmüller}(1996)}]{Kresse_CMS_6_1996}%
  \BibitemOpen
  \bibfield  {author} {\bibinfo {author} {\bibfnamefont {G.}~\bibnamefont
  {Kresse}}\ and\ \bibinfo {author} {\bibfnamefont {J.}~\bibnamefont
  {Furthmüller}},\ }\href
  {https://doi.org/https://doi.org/10.1016/0927-0256(96)00008-0} {\bibfield
  {journal} {\bibinfo  {journal} {Computational Materials Science}\ }\textbf
  {\bibinfo {volume} {6}},\ \bibinfo {pages} {15} (\bibinfo {year}
  {1996})}\BibitemShut {NoStop}%
\bibitem [{\citenamefont {Kresse}\ and\ \citenamefont
  {Furthm\"uller}(1996)}]{Kresse_PRB_54_1996}%
  \BibitemOpen
  \bibfield  {author} {\bibinfo {author} {\bibfnamefont {G.}~\bibnamefont
  {Kresse}}\ and\ \bibinfo {author} {\bibfnamefont {J.}~\bibnamefont
  {Furthm\"uller}},\ }\href {https://doi.org/10.1103/PhysRevB.54.11169}
  {\bibfield  {journal} {\bibinfo  {journal} {Phys. Rev. B}\ }\textbf {\bibinfo
  {volume} {54}},\ \bibinfo {pages} {11169} (\bibinfo {year}
  {1996})}\BibitemShut {NoStop}%
\bibitem [{\citenamefont {Kresse}\ and\ \citenamefont
  {Joubert}(1999)}]{Kresse_PRB_59_1999}%
  \BibitemOpen
  \bibfield  {author} {\bibinfo {author} {\bibfnamefont {G.}~\bibnamefont
  {Kresse}}\ and\ \bibinfo {author} {\bibfnamefont {D.}~\bibnamefont
  {Joubert}},\ }\href {https://doi.org/10.1103/PhysRevB.59.1758} {\bibfield
  {journal} {\bibinfo  {journal} {Phys. Rev. B}\ }\textbf {\bibinfo {volume}
  {59}},\ \bibinfo {pages} {1758} (\bibinfo {year} {1999})}\BibitemShut
  {NoStop}%
\bibitem [{\citenamefont {Rubel}\ \emph {et~al.}(2014)\citenamefont {Rubel},
  \citenamefont {Bokhanchuk}, \citenamefont {Ahmed},\ and\ \citenamefont
  {Assmann}}]{RubelPRB14}%
  \BibitemOpen
  \bibfield  {author} {\bibinfo {author} {\bibfnamefont {O.}~\bibnamefont
  {Rubel}}, \bibinfo {author} {\bibfnamefont {A.}~\bibnamefont {Bokhanchuk}},
  \bibinfo {author} {\bibfnamefont {S.~J.}\ \bibnamefont {Ahmed}},\ and\
  \bibinfo {author} {\bibfnamefont {E.}~\bibnamefont {Assmann}},\ }\href
  {https://doi.org/10.1103/PhysRevB.90.115202} {\bibfield  {journal} {\bibinfo
  {journal} {Phys. Rev. B}\ }\textbf {\bibinfo {volume} {90}},\ \bibinfo
  {pages} {115202} (\bibinfo {year} {2014})}\BibitemShut {NoStop}%
\bibitem [{\citenamefont {Rubel}\ \emph {et~al.}(2023)\citenamefont {Rubel},
  \citenamefont {Moussy}, \citenamefont {Foulquier},\ and\ \citenamefont
  {Brouet}}]{Rubel23}%
  \BibitemOpen
  \bibfield  {author} {\bibinfo {author} {\bibfnamefont {O.}~\bibnamefont
  {Rubel}}, \bibinfo {author} {\bibfnamefont {J.-B.}\ \bibnamefont {Moussy}},
  \bibinfo {author} {\bibfnamefont {P.}~\bibnamefont {Foulquier}},\ and\
  \bibinfo {author} {\bibfnamefont {V.}~\bibnamefont {Brouet}},\ }\href
  {https://doi.org/https://doi.org/10.1016/j.cpc.2023.108800} {\bibfield
  {journal} {\bibinfo  {journal} {Computer Physics Communications}\ }\textbf
  {\bibinfo {volume} {291}},\ \bibinfo {pages} {108800} (\bibinfo {year}
  {2023})}\BibitemShut {NoStop}%
\bibitem [{\citenamefont {Zhang}\ \emph {et~al.}(2021)\citenamefont {Zhang},
  \citenamefont {Yilmaz}, \citenamefont {Meier}, \citenamefont {Pai},
  \citenamefont {Lapano}, \citenamefont {Li}, \citenamefont {Kaznatcheev},
  \citenamefont {Vescovo}, \citenamefont {Huon}, \citenamefont {Brahlek},
  \citenamefont {Ward}, \citenamefont {Lawrie}, \citenamefont {Moore},
  \citenamefont {Lee}, \citenamefont {Wang}, \citenamefont {Miao},\ and\
  \citenamefont {Sales}}]{ZhangCondMat21}%
  \BibitemOpen
  \bibfield  {author} {\bibinfo {author} {\bibfnamefont {J.}~\bibnamefont
  {Zhang}}, \bibinfo {author} {\bibfnamefont {T.}~\bibnamefont {Yilmaz}},
  \bibinfo {author} {\bibfnamefont {J.~W.~R.}\ \bibnamefont {Meier}}, \bibinfo
  {author} {\bibfnamefont {J.~Y.}\ \bibnamefont {Pai}}, \bibinfo {author}
  {\bibfnamefont {J.}~\bibnamefont {Lapano}}, \bibinfo {author} {\bibfnamefont
  {H.~X.}\ \bibnamefont {Li}}, \bibinfo {author} {\bibfnamefont
  {K.}~\bibnamefont {Kaznatcheev}}, \bibinfo {author} {\bibfnamefont
  {E.}~\bibnamefont {Vescovo}}, \bibinfo {author} {\bibfnamefont
  {A.}~\bibnamefont {Huon}}, \bibinfo {author} {\bibfnamefont {M.}~\bibnamefont
  {Brahlek}}, \bibinfo {author} {\bibfnamefont {T.~Z.}\ \bibnamefont {Ward}},
  \bibinfo {author} {\bibfnamefont {B.}~\bibnamefont {Lawrie}}, \bibinfo
  {author} {\bibfnamefont {R.~G.}\ \bibnamefont {Moore}}, \bibinfo {author}
  {\bibfnamefont {H.~N.}\ \bibnamefont {Lee}}, \bibinfo {author} {\bibfnamefont
  {Y.~L.}\ \bibnamefont {Wang}}, \bibinfo {author} {\bibfnamefont
  {H.}~\bibnamefont {Miao}},\ and\ \bibinfo {author} {\bibfnamefont
  {B.}~\bibnamefont {Sales}},\ }\href {https://arxiv.org/abs/2105.08888}
  {\bibinfo {title} {Flat band induced negative magnetoresistance in
  multi-orbital kagome metal}} (\bibinfo {year} {2021}),\ \Eprint
  {https://arxiv.org/abs/2105.08888} {arXiv:2105.08888 [cond-mat.str-el]}
  \BibitemShut {NoStop}%
\bibitem [{\citenamefont {Bellaiche}\ and\ \citenamefont
  {Vanderbilt}(2000)}]{BellaichePRB00}%
  \BibitemOpen
  \bibfield  {author} {\bibinfo {author} {\bibfnamefont {L.}~\bibnamefont
  {Bellaiche}}\ and\ \bibinfo {author} {\bibfnamefont {D.}~\bibnamefont
  {Vanderbilt}},\ }\href {https://doi.org/10.1103/PhysRevB.61.7877} {\bibfield
  {journal} {\bibinfo  {journal} {Phys. Rev. B}\ }\textbf {\bibinfo {volume}
  {61}},\ \bibinfo {pages} {7877} (\bibinfo {year} {2000})}\BibitemShut
  {NoStop}%
\bibitem [{\citenamefont {Joly}(2001)}]{JolyPRB01}%
  \BibitemOpen
  \bibfield  {author} {\bibinfo {author} {\bibfnamefont {Y.}~\bibnamefont
  {Joly}},\ }\href {https://doi.org/10.1103/PhysRevB.63.125120} {\bibfield
  {journal} {\bibinfo  {journal} {Phys. Rev. B}\ }\textbf {\bibinfo {volume}
  {63}},\ \bibinfo {pages} {125120} (\bibinfo {year} {2001})}\BibitemShut
  {NoStop}%
\bibitem [{\citenamefont {Bun{\u{a}}u}\ and\ \citenamefont
  {Joly}(2009)}]{BunauJPCM09}%
  \BibitemOpen
  \bibfield  {author} {\bibinfo {author} {\bibfnamefont {O.}~\bibnamefont
  {Bun{\u{a}}u}}\ and\ \bibinfo {author} {\bibfnamefont {Y.}~\bibnamefont
  {Joly}},\ }\href {https://doi.org/10.1088/0953-8984/21/34/345501} {\bibfield
  {journal} {\bibinfo  {journal} {J. Phys.: Condens. Matter}\ }\textbf
  {\bibinfo {volume} {21}},\ \bibinfo {pages} {345501} (\bibinfo {year}
  {2009})}\BibitemShut {NoStop}%
\bibitem [{\citenamefont {Chikara}\ \emph {et~al.}(2015)\citenamefont
  {Chikara}, \citenamefont {Haskel}, \citenamefont {Sim}, \citenamefont {Kim},
  \citenamefont {Chen}, \citenamefont {Fabbris}, \citenamefont {Veiga},
  \citenamefont {Souza-Neto}, \citenamefont {Terzic}, \citenamefont {Butrouna},
  \citenamefont {Cao}, \citenamefont {Han},\ and\ \citenamefont {van
  Veenendaal}}]{ChikaraPRB15}%
  \BibitemOpen
  \bibfield  {author} {\bibinfo {author} {\bibfnamefont {S.}~\bibnamefont
  {Chikara}}, \bibinfo {author} {\bibfnamefont {D.}~\bibnamefont {Haskel}},
  \bibinfo {author} {\bibfnamefont {J.-H.}\ \bibnamefont {Sim}}, \bibinfo
  {author} {\bibfnamefont {H.-S.}\ \bibnamefont {Kim}}, \bibinfo {author}
  {\bibfnamefont {C.-C.}\ \bibnamefont {Chen}}, \bibinfo {author}
  {\bibfnamefont {G.}~\bibnamefont {Fabbris}}, \bibinfo {author} {\bibfnamefont
  {L.~S.~I.}\ \bibnamefont {Veiga}}, \bibinfo {author} {\bibfnamefont {N.~M.}\
  \bibnamefont {Souza-Neto}}, \bibinfo {author} {\bibfnamefont
  {J.}~\bibnamefont {Terzic}}, \bibinfo {author} {\bibfnamefont
  {K.}~\bibnamefont {Butrouna}}, \bibinfo {author} {\bibfnamefont
  {G.}~\bibnamefont {Cao}}, \bibinfo {author} {\bibfnamefont {M.~J.}\
  \bibnamefont {Han}},\ and\ \bibinfo {author} {\bibfnamefont {M.}~\bibnamefont
  {van Veenendaal}},\ }\href@noop {} {\bibfield  {journal} {\bibinfo  {journal}
  {Physical Review B}\ }\textbf {\bibinfo {volume} {92}},\ \bibinfo {pages}
  {081114} (\bibinfo {year} {2015})}\BibitemShut {NoStop}%
\bibitem [{\citenamefont {Louat}\ \emph {et~al.}(2019)\citenamefont {Louat},
  \citenamefont {Lenz}, \citenamefont {Biermann}, \citenamefont {Martins},
  \citenamefont {Bertran}, \citenamefont {Le~F\`evre}, \citenamefont {Rault},
  \citenamefont {Bert},\ and\ \citenamefont {Brouet}}]{LouatPRB19}%
  \BibitemOpen
  \bibfield  {author} {\bibinfo {author} {\bibfnamefont {A.}~\bibnamefont
  {Louat}}, \bibinfo {author} {\bibfnamefont {B.}~\bibnamefont {Lenz}},
  \bibinfo {author} {\bibfnamefont {S.}~\bibnamefont {Biermann}}, \bibinfo
  {author} {\bibfnamefont {C.}~\bibnamefont {Martins}}, \bibinfo {author}
  {\bibfnamefont {F.}~\bibnamefont {Bertran}}, \bibinfo {author} {\bibfnamefont
  {P.}~\bibnamefont {Le~F\`evre}}, \bibinfo {author} {\bibfnamefont {J.~E.}\
  \bibnamefont {Rault}}, \bibinfo {author} {\bibfnamefont {F.}~\bibnamefont
  {Bert}},\ and\ \bibinfo {author} {\bibfnamefont {V.}~\bibnamefont {Brouet}},\
  }\href {https://doi.org/10.1103/PhysRevB.100.205135} {\bibfield  {journal}
  {\bibinfo  {journal} {Phys. Rev. B}\ }\textbf {\bibinfo {volume} {100}},\
  \bibinfo {pages} {205135} (\bibinfo {year} {2019})}\BibitemShut {NoStop}%
\bibitem [{\citenamefont {Dhital}\ \emph {et~al.}(2014)\citenamefont {Dhital},
  \citenamefont {Hogan}, \citenamefont {Zhou}, \citenamefont {Chen},
  \citenamefont {Ren}, \citenamefont {Pokharel}, \citenamefont {Okada},
  \citenamefont {Heine}, \citenamefont {Tian}, \citenamefont {Yamani},
  \citenamefont {Opeil}, \citenamefont {Helton}, \citenamefont {Lynn},
  \citenamefont {Wang}, \citenamefont {Madhavan},\ and\ \citenamefont
  {Wilson}}]{DhitalNatCom14}%
  \BibitemOpen
  \bibfield  {author} {\bibinfo {author} {\bibfnamefont {C.}~\bibnamefont
  {Dhital}}, \bibinfo {author} {\bibfnamefont {T.}~\bibnamefont {Hogan}},
  \bibinfo {author} {\bibfnamefont {W.}~\bibnamefont {Zhou}}, \bibinfo {author}
  {\bibfnamefont {X.}~\bibnamefont {Chen}}, \bibinfo {author} {\bibfnamefont
  {Z.}~\bibnamefont {Ren}}, \bibinfo {author} {\bibfnamefont {M.}~\bibnamefont
  {Pokharel}}, \bibinfo {author} {\bibfnamefont {Y.}~\bibnamefont {Okada}},
  \bibinfo {author} {\bibfnamefont {M.}~\bibnamefont {Heine}}, \bibinfo
  {author} {\bibfnamefont {W.}~\bibnamefont {Tian}}, \bibinfo {author}
  {\bibfnamefont {Z.}~\bibnamefont {Yamani}}, \bibinfo {author} {\bibfnamefont
  {C.}~\bibnamefont {Opeil}}, \bibinfo {author} {\bibfnamefont {J.~S.}\
  \bibnamefont {Helton}}, \bibinfo {author} {\bibfnamefont {J.~W.}\
  \bibnamefont {Lynn}}, \bibinfo {author} {\bibfnamefont {Z.}~\bibnamefont
  {Wang}}, \bibinfo {author} {\bibfnamefont {V.}~\bibnamefont {Madhavan}},\
  and\ \bibinfo {author} {\bibfnamefont {S.~D.}\ \bibnamefont {Wilson}},\
  }\href {https://doi.org/10.1038/ncomms4377} {\bibfield  {journal} {\bibinfo
  {journal} {Nature Communications}\ }\textbf {\bibinfo {volume} {5}},\
  \bibinfo {pages} {3377} (\bibinfo {year} {2014})}\BibitemShut {NoStop}%
\bibitem [{\citenamefont {Brouet}\ \emph {et~al.}(2021)\citenamefont {Brouet},
  \citenamefont {Foulquier}, \citenamefont {Louat}, \citenamefont {Bertran},
  \citenamefont {Le~F\`evre}, \citenamefont {Rault},\ and\ \citenamefont
  {Colson}}]{BrouetPRB21}%
  \BibitemOpen
  \bibfield  {author} {\bibinfo {author} {\bibfnamefont {V.}~\bibnamefont
  {Brouet}}, \bibinfo {author} {\bibfnamefont {P.}~\bibnamefont {Foulquier}},
  \bibinfo {author} {\bibfnamefont {A.}~\bibnamefont {Louat}}, \bibinfo
  {author} {\bibfnamefont {F.}~\bibnamefont {Bertran}}, \bibinfo {author}
  {\bibfnamefont {P.}~\bibnamefont {Le~F\`evre}}, \bibinfo {author}
  {\bibfnamefont {J.~E.}\ \bibnamefont {Rault}},\ and\ \bibinfo {author}
  {\bibfnamefont {D.}~\bibnamefont {Colson}},\ }\href
  {https://doi.org/10.1103/PhysRevB.104.L121104} {\bibfield  {journal}
  {\bibinfo  {journal} {Phys. Rev. B}\ }\textbf {\bibinfo {volume} {104}},\
  \bibinfo {pages} {L121104} (\bibinfo {year} {2021})}\BibitemShut {NoStop}%
\bibitem [{\citenamefont {Streltsov}\ \emph {et~al.}(2016)\citenamefont
  {Streltsov}, \citenamefont {Gapontsev},\ and\ \citenamefont
  {Khomskii}}]{StreltsovJPhysCondMat16}%
  \BibitemOpen
  \bibfield  {author} {\bibinfo {author} {\bibfnamefont {S.~V.}\ \bibnamefont
  {Streltsov}}, \bibinfo {author} {\bibfnamefont {V.~V.}\ \bibnamefont
  {Gapontsev}},\ and\ \bibinfo {author} {\bibfnamefont {D.~I.}\ \bibnamefont
  {Khomskii}},\ }\href {https://doi.org/10.1088/0953-8984/28/8/086005}
  {\bibfield  {journal} {\bibinfo  {journal} {Journal of Physics: Condensed
  Matter}\ }\textbf {\bibinfo {volume} {28}},\ \bibinfo {pages} {086005}
  (\bibinfo {year} {2016})}\BibitemShut {NoStop}%
\bibitem [{\citenamefont {Schwarz}\ \emph {et~al.}(1984)\citenamefont
  {Schwarz}, \citenamefont {Mohn}, \citenamefont {Blaha},\ and\ \citenamefont
  {Kubler}}]{SchwarzJPhysF84}%
  \BibitemOpen
  \bibfield  {author} {\bibinfo {author} {\bibfnamefont {K.}~\bibnamefont
  {Schwarz}}, \bibinfo {author} {\bibfnamefont {P.}~\bibnamefont {Mohn}},
  \bibinfo {author} {\bibfnamefont {P.}~\bibnamefont {Blaha}},\ and\ \bibinfo
  {author} {\bibfnamefont {J.}~\bibnamefont {Kubler}},\ }\href
  {https://doi.org/10.1088/0305-4608/14/11/021} {\bibfield  {journal} {\bibinfo
   {journal} {Journal of Physics F: Metal Physics}\ }\textbf {\bibinfo {volume}
  {14}},\ \bibinfo {pages} {2659} (\bibinfo {year} {1984})}\BibitemShut
  {NoStop}%
\bibitem [{\citenamefont {Brouet}\ \emph {et~al.}(2009)\citenamefont {Brouet},
  \citenamefont {Marsi}, \citenamefont {Mansart}, \citenamefont {Nicolaou},
  \citenamefont {Taleb-Ibrahimi}, \citenamefont {Le~F\`evre}, \citenamefont
  {Bertran}, \citenamefont {Rullier-Albenque}, \citenamefont {Forget},\ and\
  \citenamefont {Colson}}]{BrouetPRB09}%
  \BibitemOpen
  \bibfield  {author} {\bibinfo {author} {\bibfnamefont {V.}~\bibnamefont
  {Brouet}}, \bibinfo {author} {\bibfnamefont {M.}~\bibnamefont {Marsi}},
  \bibinfo {author} {\bibfnamefont {B.}~\bibnamefont {Mansart}}, \bibinfo
  {author} {\bibfnamefont {A.}~\bibnamefont {Nicolaou}}, \bibinfo {author}
  {\bibfnamefont {A.}~\bibnamefont {Taleb-Ibrahimi}}, \bibinfo {author}
  {\bibfnamefont {P.}~\bibnamefont {Le~F\`evre}}, \bibinfo {author}
  {\bibfnamefont {F.}~\bibnamefont {Bertran}}, \bibinfo {author} {\bibfnamefont
  {F.}~\bibnamefont {Rullier-Albenque}}, \bibinfo {author} {\bibfnamefont
  {A.}~\bibnamefont {Forget}},\ and\ \bibinfo {author} {\bibfnamefont
  {D.}~\bibnamefont {Colson}},\ }\href
  {https://doi.org/10.1103/PhysRevB.80.165115} {\bibfield  {journal} {\bibinfo
  {journal} {Phys. Rev. B}\ }\textbf {\bibinfo {volume} {80}},\ \bibinfo
  {pages} {165115} (\bibinfo {year} {2009})}\BibitemShut {NoStop}%
\bibitem [{\citenamefont {Levy}\ \emph {et~al.}(2012)\citenamefont {Levy},
  \citenamefont {Sutarto}, \citenamefont {Chevrier}, \citenamefont {Regier},
  \citenamefont {Blyth}, \citenamefont {Geck}, \citenamefont {Wurmehl},
  \citenamefont {Harnagea}, \citenamefont {Wadati}, \citenamefont {Mizokawa},
  \citenamefont {Elfimov}, \citenamefont {Damascelli},\ and\ \citenamefont
  {Sawatzky}}]{LevyPRL12}%
  \BibitemOpen
  \bibfield  {author} {\bibinfo {author} {\bibfnamefont {G.}~\bibnamefont
  {Levy}}, \bibinfo {author} {\bibfnamefont {R.}~\bibnamefont {Sutarto}},
  \bibinfo {author} {\bibfnamefont {D.}~\bibnamefont {Chevrier}}, \bibinfo
  {author} {\bibfnamefont {T.}~\bibnamefont {Regier}}, \bibinfo {author}
  {\bibfnamefont {R.}~\bibnamefont {Blyth}}, \bibinfo {author} {\bibfnamefont
  {J.}~\bibnamefont {Geck}}, \bibinfo {author} {\bibfnamefont {S.}~\bibnamefont
  {Wurmehl}}, \bibinfo {author} {\bibfnamefont {L.}~\bibnamefont {Harnagea}},
  \bibinfo {author} {\bibfnamefont {H.}~\bibnamefont {Wadati}}, \bibinfo
  {author} {\bibfnamefont {T.}~\bibnamefont {Mizokawa}}, \bibinfo {author}
  {\bibfnamefont {I.~S.}\ \bibnamefont {Elfimov}}, \bibinfo {author}
  {\bibfnamefont {A.}~\bibnamefont {Damascelli}},\ and\ \bibinfo {author}
  {\bibfnamefont {G.~A.}\ \bibnamefont {Sawatzky}},\ }\href
  {https://doi.org/10.1103/PhysRevLett.109.077001} {\bibfield  {journal}
  {\bibinfo  {journal} {Phys. Rev. Lett.}\ }\textbf {\bibinfo {volume} {109}},\
  \bibinfo {pages} {077001} (\bibinfo {year} {2012})}\BibitemShut {NoStop}%
\bibitem [{\citenamefont {Bal\'edent}\ \emph {et~al.}(2015)\citenamefont
  {Bal\'edent}, \citenamefont {Rullier-Albenque}, \citenamefont {Colson},
  \citenamefont {Ablett},\ and\ \citenamefont {Rueff}}]{BaledentPRL15}%
  \BibitemOpen
  \bibfield  {author} {\bibinfo {author} {\bibfnamefont {V.}~\bibnamefont
  {Bal\'edent}}, \bibinfo {author} {\bibfnamefont {F.}~\bibnamefont
  {Rullier-Albenque}}, \bibinfo {author} {\bibfnamefont {D.}~\bibnamefont
  {Colson}}, \bibinfo {author} {\bibfnamefont {J.~M.}\ \bibnamefont {Ablett}},\
  and\ \bibinfo {author} {\bibfnamefont {J.-P.}\ \bibnamefont {Rueff}},\ }\href
  {https://doi.org/10.1103/PhysRevLett.114.177001} {\bibfield  {journal}
  {\bibinfo  {journal} {Phys. Rev. Lett.}\ }\textbf {\bibinfo {volume} {114}},\
  \bibinfo {pages} {177001} (\bibinfo {year} {2015})}\BibitemShut {NoStop}%
\bibitem [{\citenamefont {Huang}\ \emph {et~al.}(2022)\citenamefont {Huang},
  \citenamefont {Zheng}, \citenamefont {Lin}, \citenamefont {Guo},
  \citenamefont {Wang}, \citenamefont {Zhang}, \citenamefont {Zhang},
  \citenamefont {Sun}, \citenamefont {Wang}, \citenamefont {Weng},
  \citenamefont {Li}, \citenamefont {Wu}, \citenamefont {Chen},\ and\
  \citenamefont {Zeng}}]{HuangPRL22}%
  \BibitemOpen
  \bibfield  {author} {\bibinfo {author} {\bibfnamefont {H.}~\bibnamefont
  {Huang}}, \bibinfo {author} {\bibfnamefont {L.}~\bibnamefont {Zheng}},
  \bibinfo {author} {\bibfnamefont {Z.}~\bibnamefont {Lin}}, \bibinfo {author}
  {\bibfnamefont {X.}~\bibnamefont {Guo}}, \bibinfo {author} {\bibfnamefont
  {S.}~\bibnamefont {Wang}}, \bibinfo {author} {\bibfnamefont {S.}~\bibnamefont
  {Zhang}}, \bibinfo {author} {\bibfnamefont {C.}~\bibnamefont {Zhang}},
  \bibinfo {author} {\bibfnamefont {Z.}~\bibnamefont {Sun}}, \bibinfo {author}
  {\bibfnamefont {Z.}~\bibnamefont {Wang}}, \bibinfo {author} {\bibfnamefont
  {H.}~\bibnamefont {Weng}}, \bibinfo {author} {\bibfnamefont {L.}~\bibnamefont
  {Li}}, \bibinfo {author} {\bibfnamefont {T.}~\bibnamefont {Wu}}, \bibinfo
  {author} {\bibfnamefont {X.}~\bibnamefont {Chen}},\ and\ \bibinfo {author}
  {\bibfnamefont {C.}~\bibnamefont {Zeng}},\ }\href
  {https://doi.org/10.1103/PhysRevLett.128.096601} {\bibfield  {journal}
  {\bibinfo  {journal} {Phys. Rev. Lett.}\ }\textbf {\bibinfo {volume} {128}},\
  \bibinfo {pages} {096601} (\bibinfo {year} {2022})}\BibitemShut {NoStop}%
\bibitem [{\citenamefont {Alloul}\ \emph {et~al.}(2009)\citenamefont {Alloul},
  \citenamefont {Bobroff}, \citenamefont {Gabay},\ and\ \citenamefont
  {Hirschfeld}}]{AlloulRMP09}%
  \BibitemOpen
  \bibfield  {author} {\bibinfo {author} {\bibfnamefont {H.}~\bibnamefont
  {Alloul}}, \bibinfo {author} {\bibfnamefont {J.}~\bibnamefont {Bobroff}},
  \bibinfo {author} {\bibfnamefont {M.}~\bibnamefont {Gabay}},\ and\ \bibinfo
  {author} {\bibfnamefont {P.~J.}\ \bibnamefont {Hirschfeld}},\ }\href
  {https://doi.org/10.1103/RevModPhys.81.45} {\bibfield  {journal} {\bibinfo
  {journal} {Rev. Mod. Phys.}\ }\textbf {\bibinfo {volume} {81}},\ \bibinfo
  {pages} {45} (\bibinfo {year} {2009})}\BibitemShut {NoStop}%
\bibitem [{\citenamefont {{Rigaku Oxford
  Diffraction}}(2022)}]{CrysAlisPRO2022}%
  \BibitemOpen
  \bibfield  {author} {\bibinfo {author} {\bibnamefont {{Rigaku Oxford
  Diffraction}}},\ }\href@noop {} {\bibinfo {title} {{CrysAlis PRO}, software
  system, version 1.171...}} (\bibinfo {year} {2022})\BibitemShut {NoStop}%
\bibitem [{\citenamefont {Perdew}\ \emph {et~al.}(1996)\citenamefont {Perdew},
  \citenamefont {Burke},\ and\ \citenamefont {Ernzerhof}}]{Perdew_PRL_77_1996}%
  \BibitemOpen
  \bibfield  {author} {\bibinfo {author} {\bibfnamefont {J.~P.}\ \bibnamefont
  {Perdew}}, \bibinfo {author} {\bibfnamefont {K.}~\bibnamefont {Burke}},\ and\
  \bibinfo {author} {\bibfnamefont {M.}~\bibnamefont {Ernzerhof}},\ }\href
  {https://doi.org/10.1103/PhysRevLett.77.3865} {\bibfield  {journal} {\bibinfo
   {journal} {Phys. Rev. Lett.}\ }\textbf {\bibinfo {volume} {77}},\ \bibinfo
  {pages} {3865} (\bibinfo {year} {1996})}\BibitemShut {NoStop}%
\bibitem [{\citenamefont {Pashartis}\ and\ \citenamefont
  {Rubel}(2017)}]{Pashartis_PRA_7_2017}%
  \BibitemOpen
  \bibfield  {author} {\bibinfo {author} {\bibfnamefont {C.}~\bibnamefont
  {Pashartis}}\ and\ \bibinfo {author} {\bibfnamefont {O.}~\bibnamefont
  {Rubel}},\ }\href {https://doi.org/10.1103/PhysRevApplied.7.064011}
  {\bibfield  {journal} {\bibinfo  {journal} {Phys. Rev. Appl.}\ }\textbf
  {\bibinfo {volume} {7}},\ \bibinfo {pages} {064011} (\bibinfo {year}
  {2017})}\BibitemShut {NoStop}%
\bibitem [{\citenamefont {Zunger}\ \emph {et~al.}(1990)\citenamefont {Zunger},
  \citenamefont {Wei}, \citenamefont {Ferreira},\ and\ \citenamefont
  {Bernard}}]{Zunger_PRL_65_1990}%
  \BibitemOpen
  \bibfield  {author} {\bibinfo {author} {\bibfnamefont {A.}~\bibnamefont
  {Zunger}}, \bibinfo {author} {\bibfnamefont {S.-H.}\ \bibnamefont {Wei}},
  \bibinfo {author} {\bibfnamefont {L.~G.}\ \bibnamefont {Ferreira}},\ and\
  \bibinfo {author} {\bibfnamefont {J.~E.}\ \bibnamefont {Bernard}},\ }\href
  {https://doi.org/10.1103/PhysRevLett.65.353} {\bibfield  {journal} {\bibinfo
  {journal} {Phys. Rev. Lett.}\ }\textbf {\bibinfo {volume} {65}},\ \bibinfo
  {pages} {353} (\bibinfo {year} {1990})}\BibitemShut {NoStop}%
\end{thebibliography}%


\clearpage

\onecolumngrid

\setcounter{secnumdepth}{3}
\renewcommand{\thefigure}{S\arabic{figure}}
\setcounter{figure}{0}

{\huge \textbf{Supplementary Material:}}


\vspace{0.5cm}

\section{Samples}

\begin{figure}[h]
	\includegraphics[width=0.8\linewidth]{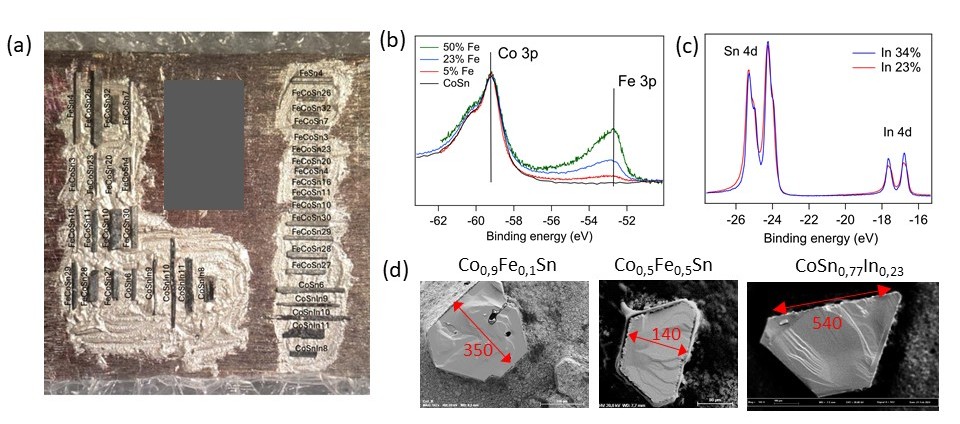}
	\caption{(a) Collection of samples used for XAS measurements. (b-c) Core-levels of cleaved samples measured at 126eV for Co and Fe 3p (b) and Sn and In 4d (c). d) Images obtained by scanning electron microscope of some of the ARPES cleaved samples. The number indicate dimensions in microns.}
	\label{Sup_CleavedSamples}
\end{figure}

Fig.~\ref{Sup_CleavedSamples}(a) shows a collection of samples used for XAS measurements. They are needles-like and can be mm long along c-axis with a 100 to 500 $\mu$m diameter. All sample compositions were established from EDX measurements with a precision of about 2\% and no measurable inhomogeneity. The composition can be confirmed on cleaved samples by the ratio of Co and Fe core-levels or Sn and In core levels, see Fig. \ref{Sup_CleavedSamples}(b-c). Examples of cleaved surfaces are shown in Fig. \ref{Sup_CleavedSamples}(d).

\begin{figure}[b]
	\centering
	\includegraphics[width=0.4\linewidth]{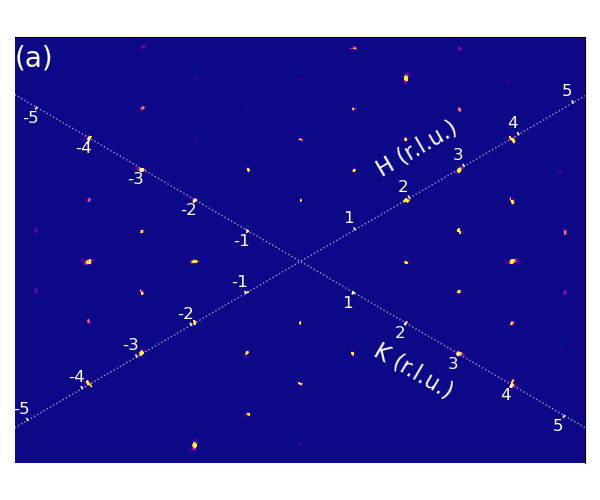}
	\includegraphics[width=0.4\linewidth]{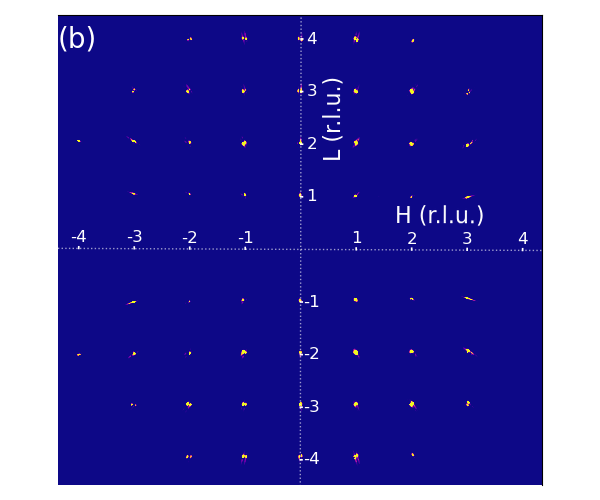}
	\caption{Reconstructed sections of the reciprocal space measured by
		single-crystal X-ray diffraction on $Co_{0.94}Fe_{0.06}Sn$: (a) $HK0$ plane, showing the
		sixfold symmetry of the kagome plane, and (b) $H0L$ plane. All 804
		measured reflections are indexed by a single hexagonal lattice.}
	\label{Sup_XRD}
\end{figure}

The crystalline quality and the absence of secondary phases were checked by single-crystal X-ray diffraction. Diffractograms of x=0.06 $Co_{1-x}Fe_xSn$ single crystal were recorded on the high-pressure diffraction setup developed at the Laboratoire de Physique des Solides (Orsay, France). The diffractometer uses a microfocus Mo X-ray tube (Incoatec GmbH) equipped with Montel multilayer focusing optics, delivering a monochromatic beam ($\lambda = 0.711$~\AA) with a 100~$\mu$m spot size and a 0.5~mrad divergence at the sample position. The diffracted intensity is collected on a large-area MAR345 image-plate detector (MarXperts GmbH). Data were acquired in $1^{\circ}$ steps over a $340^{\circ}$ rotation range with a 60~s exposure per frame, for a total acquisition time of about 6~hours. The reciprocal space was then reconstructed from the 340 frames using CrysAlisPro \cite{CrysAlisPRO2022}. Reconstructed sections of the reciprocal space are shown in Fig. \ref{Sup_XRD} for the $HK0$ (a) and $H0L$ (b) planes. All 804 reflections detected over the full data set are indexed by a single hexagonal lattice (space group $P6/mmm$), with refined parameters $a = b = $~5.27(1)~\AA\ and $c = $~4.26(1)~\AA. No unindexed reflection, no splitting of the Bragg spots and no diffuse streaking are observed, excluding the presence of secondary crystallites, of twin domains or of an intergrown competing phase. The substituted crystals therefore consist of a single crystallographic domain retaining the parent CoSn structure, the dopants being incorporated substitutionally on the Co or Sn sublattices without structural phase separation.

\section{ARPES analysis}    

\begin{figure}[thb]
	\includegraphics[width=1\linewidth]{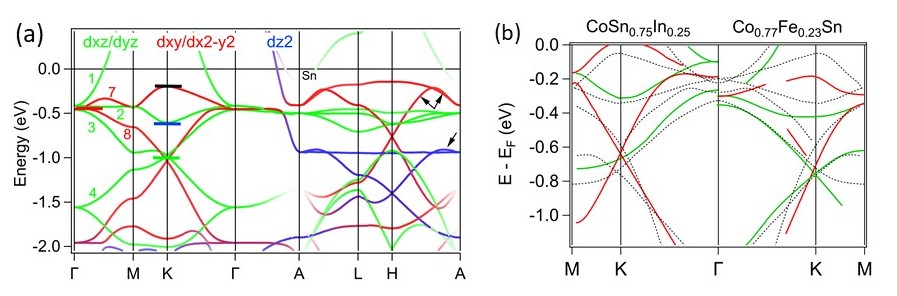}
	\caption{(a) DFT calculation of CoSn showing with colors the main orbital character. Horizontal markers indicate the points used to track the evolution of the band structure with doping in Figure 1. Arrows indicate the position of bands detected at \kz=1 (see Fig. \ref{CoSnVsKz}). (b) Model dispersions extracted from ARPES images in Fig. \ref{Sup_Doping25}, green for \dxz/\dyz~and red for \dxy/\dx2my2. Dotted black lines give comparison to models fitted to CoSn \cite{BrouetPRB25}. }
	\label{Sup_Bands}
\end{figure}

In Figure \ref{Sup_Bands}(a), we recall the band structure of CoSn calculated using Wien2k, as we published in ref. \onlinecite{BrouetPRB25} and similar to previous reports. The extrema of dispersion that we chose in Fig. 1 to track the evolution of the electronic structure are indicated as thick horizontal markers. In Figure \ref{Sup_Bands}(b), we show the model dispersion extracted from the ARPES data with 25\% Fe or In doping, compared to those of CoSn, as will be further discussed in the next section. 

\subsection{EDC and MDC corresponding to images of Figure 1}

All raw images used for Fig. 1 are shown in the attached file "Additional Data", together with complementary measurements at other photon energies or compositions.

\begin{figure}[thb]
	\includegraphics[width=1\linewidth]{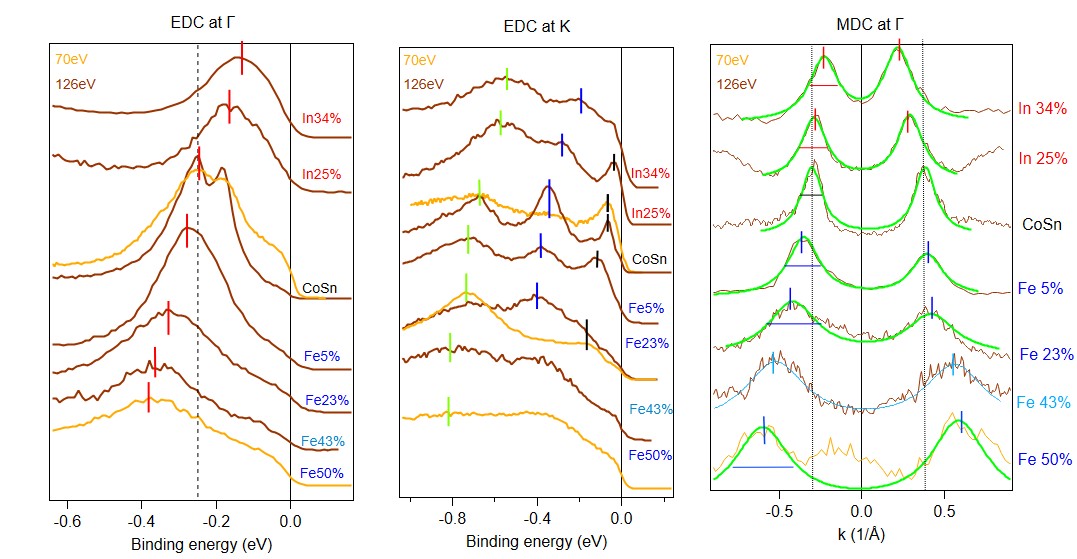}
	\caption{EDC at $\Gamma$ and K extracted from the images of Fig.1 and additional measurements shown in attached supplementary file. Different photon energies are indicated by different colors. The markers indicate position reported in Fig.~1(g). Right : MDC spectra of Fig.~1(h) fitted by two lorentzians (green line).}
	\label{Sup_EDC_MDC}
\end{figure}

In Figure \ref{Sup_EDC_MDC}(a), we show the EDC spectra at $\Gamma$ that allow to track the position of the red marker, indicating the bottom of the \dxz/\dyz~band crossing the Fermi level and four nearly degenerate bands labelled (2,3,7,8) in Fig. \ref{Sup_Bands}(a). This includes the two nearly Flat Bands 7 and 2. Note that this point is not dispersive at all along $\Gamma$A, which means that its position is nearly independent of \kz. The splitting between these bands, mainly due to spin-orbit coupling, is visible only in CoSn. We chose the EDC maxima to represent this position.

\begin{figure}[thb]
	\includegraphics[width=1\linewidth]{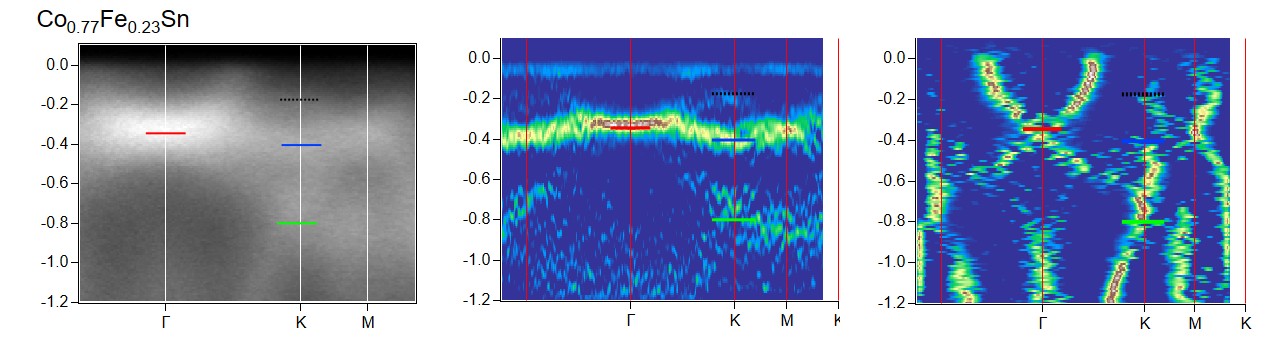}
	\caption{Energy-momentum image for $x_{Fe}$=0.23 (Fig. 1e) with second derivative taken as a function of energy (middle) or momentum (right). As expected, the derivation as a function of energy enhances weakly dispersive bands (red, blue and black markers) and the one as a function of momentum the fast dispersive ones (Dirac cone, with green marker). Despite the fact that the FB marked by black is weak at 126~eV, the second derivative reveals a trace of it.}
	\label{Sup_2ndDeriv}
\end{figure}

In Figure \ref{Sup_EDC_MDC}(b), we show EDC at K, where the top of Flat Band 7 (black marker), the bottom of Flat Band 2 (blue marker) and the two degenerate Dirac cones (green marker) are located. We systematically observe a larger intensity for 7 at 70eV and for 2 at 126eV, so that we used both EDC to determine the positions more precisely. The Dirac cone position is not very precise if extracted from the position at K alone due to the background and its width. However, it becomes rather clear when the full dispersion is considered, as can be observed, for example, in Fig. \ref{Sup_Doping25} at 25\% doping or in additional data. Sometimes, images from the second derivative can help visualize the underlying bands, as shown in Fig. \ref{Sup_2ndDeriv} for $x_{Fe}$ = 0.23. For similar reasons, we did not assign the blue marker to the broad peak near -0.4eV in the EDC of x$_{Fe}$ = 0.43 and 0.5, although it appears close to the expected position, because we never clearly observed a dispersion of the corresponding band.

Finally, we show in Figure \ref{Sup_EDC_MDC}(c) the MDC spectra of Fig.~1(h) together with the fits used to extract the \kf~position and the full width at half maximum in Fig.~1(i-j). The spectra were fitted with two lorentzians constrained to have the same width. For x$_{Fe}$=0.5, we excluded the points near \kz=0 from the fit where residual intensity is found. 

\subsection{ARPES in C\lowercase{o}S\lowercase{n} at 25$\%$ hole doping}

\begin{figure}
	\includegraphics[width=1\linewidth]{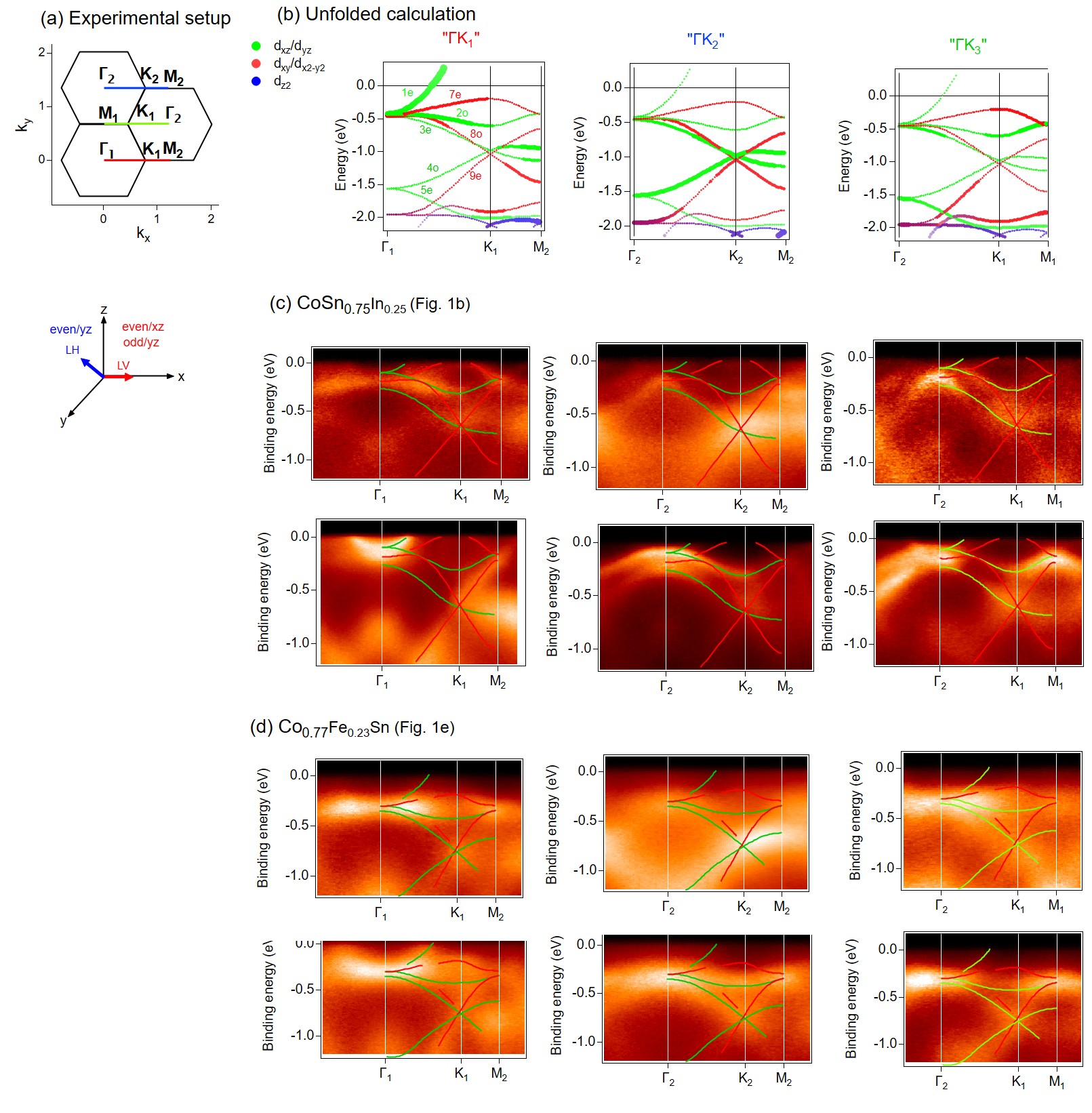}
	\caption{(a) Sketch of the three inequivalent $\Gamma$KM paths with respect to the kagome BZ (black hexagon). See ref. \onlinecite{BrouetPRB25} for more details. Below : Direction of polarization for Linear Horizontal (LH) and Linear Horizontal (LV) in our ARPES setup. (b) Unfolded band structure in CoSn at $k_z$=0 for each inequivalent $\Gamma$KM path. They were calculated with WIEN2k and the fold2Bloch procedure. Spin-orbit coupling is not included. Colors indicate the main orbital character. Size of the markers indicate the unfolded spectral weight along the particular path. The letters indicate the parity, even or odd with respect to the perpendicular plane containing $\Gamma$K. (c) Energy-momentum images obtained by ARPES at 126eV ($k_z=0$) in CoSn$_{0.75}$In$_{0.25}$. The six panels are obtained along $\Gamma$KM, but with LH (LV) polarization for the first (second) rows and the paths shown in panel (a) as red, green and blue. Thin lines are models extracted from the data, allowing to compare the bands between different panels. They are compared in Fig. \ref{Sup_Bands}(b) to the band structure of CoSn. (d) Same as (c) for Co$_{0.77}$Fe$_{0.23}$Sn. }
	\label{Sup_Doping25}
\end{figure}

In Figure \ref{Sup_Doping25}, we compare ARPES data taken in compounds where about 0.25 hole per Co were introduced to CoSn, either by Sn/In substitutions [Fig. \ref{Sup_Doping25}(c)] or Co/Fe ones [Fig. \ref{Sup_Doping25}(d)]. It is clear that the data are markedly different, despite the similar hole counting. 

Although all images are taken along a $\Gamma$KM path at $k_z$=0, the different polarizations or Brillouin Zones [see Fig. \ref{Sup_Doping25}(a)] highlight different bands. As explained in our previous work \cite{BrouetPRB25}, interferences between the three atoms of the kagome motif modulate the spectral weight of bands in neighboring BZ in a way that can be captured by unfolded calculations, such as those in Fig. \ref{Sup_Doping25}(b). This explains very well most of the observed intensity modulations, once the selection between even and odd bands  is included [see letters on Fig. \ref{Sup_Doping25}(b) for band parity]. Band\#1 is for example observed strongly around $\G_1$ in LV polarization but is weak everywhere else. On the contrary, Band\#3 is clearer around $\G_2$ with LV. This is why many images are needed to obtain a complete view of the electronic structure. In the main text, we present for each doping data obtained by adding four images for the red and blue paths and the two polarizations. This allows us to visualize most bands. All raw data are given in the supplementary file.

We may then extract models to the experimental data following the high intensity contours. They are shown as color lines in Fig. \ref{Sup_Doping25} (in an additional supplementary file, the same data are shown without these lines). The models are compared with those of CoSn in Fig. \ref{Sup_Bands}(b). There is a global shift of 0.2eV between the two compounds, the bands in \In25~shifting by +0.12eV towards \Ef~compared to CoSn (dotted black line \cite{BrouetPRB25}) and by -0.08eV in \Fe23. These shift values are close to those found in the calculation at 25\% doping (+0.12 and -0.1eV respectively) shown in Fig. 4. While the band structures are similar in the two cases, there are subtle differences. The EDCs at $\Gamma$ shown in Fig. \ref{Sup_EDC_MDC}(a) evidence the shifts between the two compounds, but also a clearer splitting between the hole bands\#3 and \#8 for In (-0.24~eV and -0.09~eV in the In case, while only a broad peak around -0.34~eV is observed in the Fe case). This splitting depends both on the spin-orbit coupling and the relative position of the orbitals, suggesting an orbital dependent shift. The FB dispersion also appears slightly larger for In than Fe, a feature that is also reproduced in the calculation in Fig. 4, where the Flat Bands become quite diffuse and overlap with lower lying bands. The striking difference between the two is however the electron-like shift of the Fe doped sample, opposite to the naive expectation of hole doping and discussed in the main text.

\clearpage

\subsection{Three dimensional electronic structure}

\begin{figure*}[b]
	\includegraphics[width=0.85\linewidth]{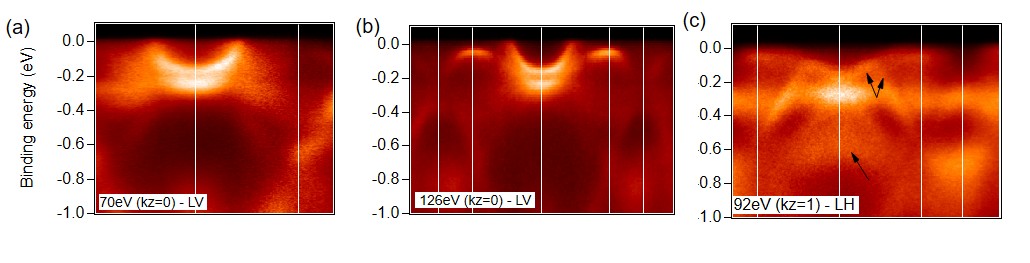}
	\caption{Bands measured in CoSn along $\Gamma$KM at different photon energies [70eV (a), 126eV (b) and 92eV (c)] and with polarization either horizontal (LH) or vertical (LV). }
	\label{CoSnVsKz}
\end{figure*}

\begin{figure*}[b]
	\includegraphics[width=0.95\linewidth]{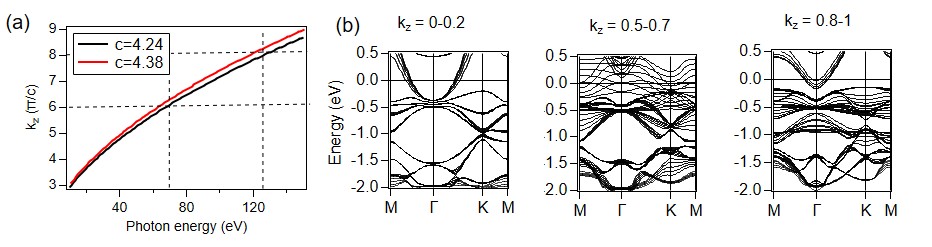}
	\caption{(a) Variation of \kz in units of $\pi$/c for two extreme values of c : 4.24\AA~for CoSn and 4.38\AA~for Co$_{0.5}$Fe$_{0.5}$Sn. Dotted lines indicate the two values of kz=0[2$\pi$/c], namely 70 and 126eV. (b) DFT calculation of CoSn plotted along the direction M$\Gamma$KM for different ranges of \kz.}
	\label{Sup_BandsVsKz}
\end{figure*}

In the main text, we only report measurements performed at \kz=0, acquired with photon energy 70 or 126eV. In Fig. \ref{CoSnVsKz}(a-b) we show images acquired in CoSn at these two photon energies to confirm the identical band structure. Other examples can be found in supplementary additional file. We use the \kz~scale shown in Fig. \ref{Sup_BandsVsKz}(a) obtained by adjusting the periodicity detected in experiments as a function of \kz with the formula : 
\begin{equation}
	k_z=\frac{\sqrt{2m*(E_{ph}-W+V_0)}/\hbar}{\pi/c}
\end{equation}
and using $V_0$=8eV. These values are in excellent agreement with previous estimations \cite{LiuNatCom20,KangNatCom20_CoSn}.

The complete electronic structure is however quite strongly three dimensional, as shown by the band calculation displayed in Fig. \ref{Sup_Bands}(a). Notably, the \dz2~band is strongly dispersive, not visible at \kz=0 in a 2eV window below \Ef, but visible at \kz=1. Fig. \ref{CoSnVsKz}(c) displays the electronic structure for CoSn at 92eV corresponding to \kz=1, showing the characteristic inverted W shape expected for the \dx2my2~band at \kz=1 [see arrows in Fig. \ref{Sup_Bands}(a)], and a flat band around -0.6eV corresponding well to \dz2. This gives confidence in our \kz~scale and on the good agreement between the ARPES data and the DFT calculation.

The two bands with strong \dz2~character cross the Fermi level near \kz=0.7 along $\Gamma$A. While it would be in principle important to also image these bands, it is difficult due to the broadening associated with their strong \kz~dispersion. It is well known that \kz~is not a well defined quantity in the photoemission process, $\delta$\kz~is of the order of the photoelectron escape depth (5-10\AA~at these photon energies). This leads to a significant broadening when the three-dimensional dispersion is large. In Fig. \ref{Sup_BandsVsKz}(b), we show the calculated band structure for the indicated range of \kz, corresponding to a realistic broadening in ARPES. Near \kz=0, this broadening is modest, preserving high quality ARPES data. However, it becomes quite severe at other \kz~and particularly for the two \dz2~bands crossing \Ef~near \kz=0.7, whose dispersion along \kz~is very large. This makes them impossible to detect clearly by ARPES. Near \kz=1, the \dx2my2~is sharp again, allowing good visualization, while the \dz2~is broader, but still visible for the flat part at $\Gamma$. 

Although the \kz~scale should in principle change with doping, as the c-axis expands, the largest expected change is $\delta$\kz=0.2$\pi$/c, as shown in Fig.~\ref{Sup_BandsVsKz}(a), of the order of the broadening presented on the right panels. At \kz=0, the broadening on this \kz window is $\delta$k=0.1\AA$^{-1}$, smaller than the sharpest MDC width for CoSn ($\delta$k=0.15\ \AA$^{-1}$), so that it can hardly be detected.

\section{DFT calculations}
\subsection{VCA calculations}
\begin{figure*}[h]
	\includegraphics[width=0.95\linewidth]{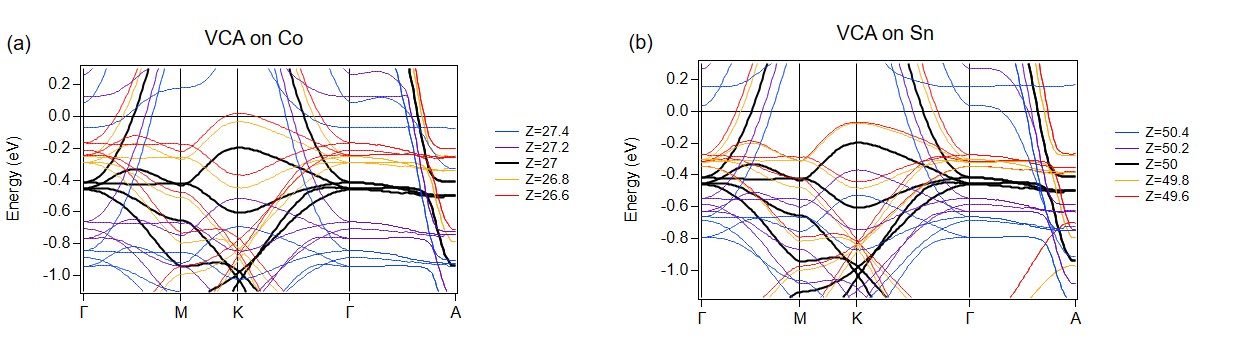}
	\caption{Evolution of band structure at \kz=0 and along $\Gamma$A when the Z value of either Co (a) or Sn (b) is changed to simulate holes/electron doping by a lower/higher value of Z, as indicated. }
	\label{Sup_VCA}
\end{figure*}

To evaluate the number of carriers in absence of the full 3D information from ARPES, we rely on VCA calculations to get a relation between the value of \kf~at \kz=0 the total the number of carriers. The calculation at \kz=0 and along $\Gamma$A are presented in Fig. \ref{Sup_VCA} with carriers introduced either on Co (a) or Sn (b), by changing their respective Z value. In both cases, the bands move up (down) when holes (electrons) are introduced, as expected from a rigid shift and contrary to ARPES for Fe doping. The resulting variation of \kf~at \kz=0 is indeed strikingly similar in the two cases, as presented in Fig.~1(i). 

However, it is interesting to note that the shift is more pronounced for VCA on Co compared to Sn. In the case of Co, the flat band at K crosses the Fermi level for x=-0.4 (Z=26.6), while it is still below \Ef~for Sn. In our experiment, the flat band clearly crosses \Ef~near K for x$_{In}$=0.25, which would not be expected in the VCA calculation on Sn. As there is a significant renormalization of the electron band at $\Gamma$ (its band bottom is E=-0.24eV in ARPES (red marker) compared to E=-0.43eV in the calculation), this deviation could be expected. On the other hand, the Luttinger theorem states that electron correlations do not change the Fermi surface volume, so that reasoning on the \kf~values remain meaningful. In a multi-band system, electrons could be transferred between orbitals. The calculation along $\Gamma$A includes the two other bands crossing the Fermi level. They do not indicate a major three-dimensional redistribution of carriers.  

\subsection{Ordered unit-cell calculations}

\begin{figure*}[h]
	\includegraphics[width=1\linewidth]{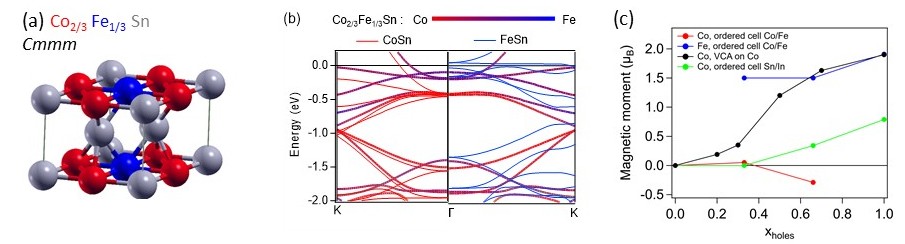}
	\caption{(a) Sketch of an ordered structure corresponding to Co$_{2/3}$Fe$_{1/3}$Sn. (b) Bands obtained along $\Gamma$K at $k_z$=0 for Co$_{2/3}$Fe$_{1/3}$Sn. The Co/Fe spectral weights are coded by a red/blue scale. Thin lines are the bands of pure CoSn (left) and non-magnetic FeSn (right). (c) Magnetic moment obtained in a ferromagnetic calculations for the different cases indicated in caption.}
	\label{Sup_OrderedCells}
\end{figure*}

We show in Fig.~\ref{Sup_OrderedCells} an ordered structure with one Co atom out of three replaced by Fe, corresponding to Co$_{2/3}$Fe$_{1/3}$Sn. The unit cell is unchanged but the symmetry is lowered to space group \#65 ($Cmmm$). This structure was used to calculate the partial DOS shown in Fig.~3. We further show in Fig. \ref{Sup_OrderedCells}(b) how the bands reconstruct at $x_{Fe}$=1/3. The 5 bands close to the Fermi level split into two groups, dominated by Fe (blue) close to \Ef~and Co (red) further away, even though the mutual hybridization remains sizable (according to the DOS of Fig.~3(a), it is as strong as 50\% for the states near \Ef). Interestingly, each group remains close to the energy positions they had in the pure compounds, either CoSn for Co on the left or FeSn for Fe the right, as if these positions act as an anchor for the Fe and Co weight distribution. This suggests that there is a persisting tendency to segregate charges on Fe and Co, with Fe states closer to \Ef. 

Finally, we let the calculation find a ferromagnetic ground state to evaluate the tendencies to magnetism. We display in  Fig. \ref{Sup_OrderedCells}(c) the optimized magnetic moment. We find that Fe always develops a rather large moment M>1.5$\mu_B$, while the Co moment is smaller and opposite to Fe at $x_{Fe}$=2/3. In the VCA calculation, the magnetic moment is growing more continuously, as can be expected from the forced homogeneous distribution of charges. For the ordered cells with Sn/In substitutions, magnetism also develops, but reaches a smaller magnetic moment.

\subsection{Supercell calculations: dilute limit}

Calculations were performed using density functional theory (DFT) and the projector augmented-wave (PAW) method \cite{Kresse_PRB_59_1999} as implemented in the VASP code \cite{Kresse_PRB_47_1993, Kresse_CMS_6_1996, Kresse_PRB_54_1996}. Exchange and correlation were treated within the generalized gradient approximation using the Perdew-Burke-Ernzerhof (PBE) functional \cite{Perdew_PRL_77_1996}. We used the Co\_pv and Fe\_pv PAW potentials, which include the semicore $3p$ states in the valence (15 and 14 valence electrons, respectively), and the standard Sn and In potentials with 4 and 3 valence electrons ($5s$ and $5p$ states). The structure of pristine CoSn (space group $P6/mmm$) was first optimized in the primitive cell, starting from the experimental structure, with a plane-wave cutoff energy of 339~eV, which is 25\% above the highest cutoff recommended for the PAW potentials, to reduce the Pulay stress. The Brillouin zone was sampled with a $\Gamma$-centered $8\times8\times8$ $k$-mesh, generated automatically with a length parameter $R_k=35$~\AA. A Gaussian smearing of 10~meV was used, and the total energy was converged to $10^{-7}$~eV. All atoms occupy Wyckoff positions without free coordinates, so the forces vanish by symmetry and only the lattice parameters were optimized. The residual stress after optimization did not exceed 1~kbar. The resulting lattice parameters, $a=5.2925$~\AA\ and $c=4.2190$~\AA, were kept fixed in subsequent calculations with impurities. Compared with experiment [SpringerMaterials (entry SD1500002)], PBE overestimates $a$ by 0.3\% and underestimates $c$ by 1.0\%.

Using the optimized structure of CoSn, we built a $3\times3\times4$ supercell containing 216 atoms (Co$_{108}$Sn$_{108}$). The same supercell was used to study isolated impurities and solid solutions. To model isolated impurities, we substituted a single Co atom with Fe or a single Sn atom with In. The structure contains two inequivalent Sn sites, Sn1 located within the kagome plane and Sn2 located in the stanene layer, and both sites were considered for In. After the substitution, the atomic positions were optimized while the lattice vectors were kept fixed at the values determined for pristine CoSn. In these calculations, the plane-wave cutoff energy was set to the highest value recommended for the PAW potentials involved, which is 293~eV in the presence of Fe and 271~eV otherwise, and the residual forces did not exceed 0.01~eV/\AA. The Brillouin zone was sampled with a $\Gamma$-centered $3\times3\times3$ $k$-mesh, generated with a $k$-point spacing of 0.18~\AA$^{-1}$ (including the factor $2\pi$), which gives approximately the same $k$-point density as for the primitive cell. For each impurity, both non-magnetic and spin-polarized solutions were considered. In the spin-polarized calculations, the impurity atom was initialized with a magnetic moment of $2\mu_B$ and all other atoms with zero moment. The In impurity converged to a non-magnetic solution, with no magnetic moment on In or on any other atom of the supercell. For an isolated In impurity, the Sn1 site is energetically favored over the Sn2 site, with a total energy difference of approximately 0.1~eV. The Fe impurity converged to a magnetic solution with a moment of $2.14\mu_B$ within the Fe PAW sphere (radius 1.16~\AA), accompanied by small induced moments on the neighboring Co and more distant atoms. This magnetic solution is lower in total energy than the non-magnetic one by 0.30~eV. We also optimized the structure with the Fe impurity separately without spin polarization. The resulting local distortions are small and depend on the magnetic state of Fe. Without spin polarization, all Fe-neighbor distances remain within about 0.01~\AA\ of those in pristine CoSn. With spin polarization, the four nearest Co atoms move towards Fe by 0.07~\AA\ ($-2.7\%$), while the two Sn1 and four Sn2 neighbors move away by 0.09 and 0.04~\AA\ ($+3.3\%$ and $+1.7\%$). The six nearest Co atoms move away from In by 0.05~\AA\ ($+1.8\%$) for the Sn1 site and by 0.03~\AA\ ($+1.1\%$) for the Sn2 site. Beyond 6~\AA\ from the impurity, all displacements are below 0.01~\AA.

\begin{figure*}[h]
	\includegraphics{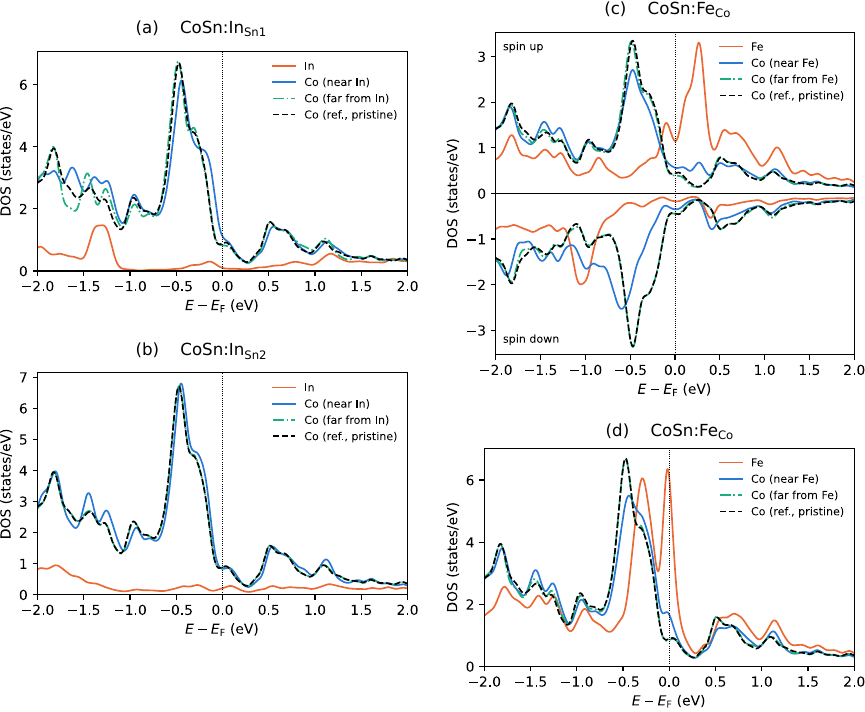}
	\caption{Single atom-projected density of states in a CoSn supercell hosting Fe or In impurities in the dilute limit: (a) 
		In on Sn1 site, (b) In on Sn2 site, (c,d) Fe on Co site. 
		Panels (a,b,d) correspond to non-spin-polarized calculations. 
		Panel (c) shows results of a spin-polarized calculation. 
		`Ref. pristine' labels the pure CoSn structure.}
	\label{figSI-PDOS-dilute}
\end{figure*}

Figure~\ref{figSI-PDOS-dilute} shows the density of states projected on the impurity and on Co atoms near and far from it, compared with Co in pristine CoSn. In the non-magnetic calculation, the Fe partial DOS shows a sharp peak at the Fermi level [Fig.~\ref{figSI-PDOS-dilute}(d)], which is absent for Co in pristine CoSn. This high DOS at \Ef~fulfills the Stoner criterion on Fe, which explains why the Fe impurity prefers the magnetic solution. In the spin-polarized calculation [Fig.~\ref{figSI-PDOS-dilute}(c)], the Fe peak splits, with the minority-spin Fe states moving to about 0.3~eV above \Ef~and the majority-spin Fe states moving below \Ef. The Fe moment of $2.14\mu_B$ is accompanied by moments of $0.21\mu_B$ on each of the four nearest Co atoms, aligned parallel to Fe, while more distant atoms carry small moments that are antiparallel to Fe on average, so that the total moment of the supercell is reduced to $1.58\mu_B$. Unlike Fe, the In impurity does not produce a peak near \Ef~[Fig.~\ref{figSI-PDOS-dilute}(a,b)], and the DOS of Co near In remains close to that of pristine CoSn.

\begin{figure*}[h]
	\includegraphics{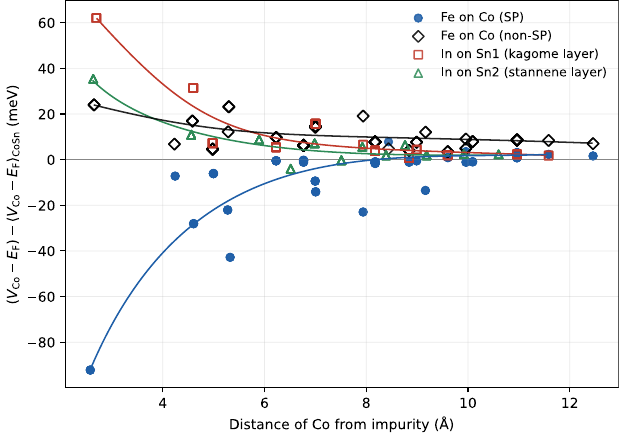}
	\caption{Electrostatic potential at the core of Co atoms as a function of the distance from the impurity, relative to that in pristine CoSn. The Fe impurity is shown with and without spin polarization (SP). Lines are guides to the eye.}
	\label{figSI-Co-core-potential-dilute}
\end{figure*}

The impurities also create a local electrostatic potential that acts on the surrounding Co atoms. Figure~\ref{figSI-Co-core-potential-dilute} shows the electrostatic potential at the Co cores, measured relative to \Ef~and referenced to pristine CoSn, as a function of the distance from the impurity. In the magnetic ground state, Fe makes the potential on the neighboring Co atoms more attractive for electrons, whereas In makes it less attractive, more strongly on the Sn1 site than on the Sn2 site. In both cases the perturbation is local, although individual Co atoms near Fe deviate from a monotonic decay. The charge density is also perturbed only locally. In supercells with unrelaxed positions and the same plane-wave cutoff of 300~eV, the net charge enclosed within a sphere centred on the impurity, obtained from the difference between the charge densities of the doped and pristine supercells, drops below 0.03~e in magnitude at a radius of 3~\AA\ and to 0.01~e or less beyond 5~\AA, for both magnetic Fe and In. The charge of each nearest Co atom changes by less than 0.02~e. Therefore, the downward shift of the bands is more subtle than a simple charge transfer from Fe to Co but it is captured by the change of the local potential shown in Fig.~\ref{figSI-Co-core-potential-dilute}. The local moment on Fe is essential for this result. Without spin polarization, the potential near Fe changes sign and becomes less attractive for electrons, much as near In, and the change on the nearest Co atoms amounts to about 0.12~eV. The potential also remains raised far from the impurity, which reflects a partial release of the hole into the host.

\begin{figure*}[h]
	\includegraphics{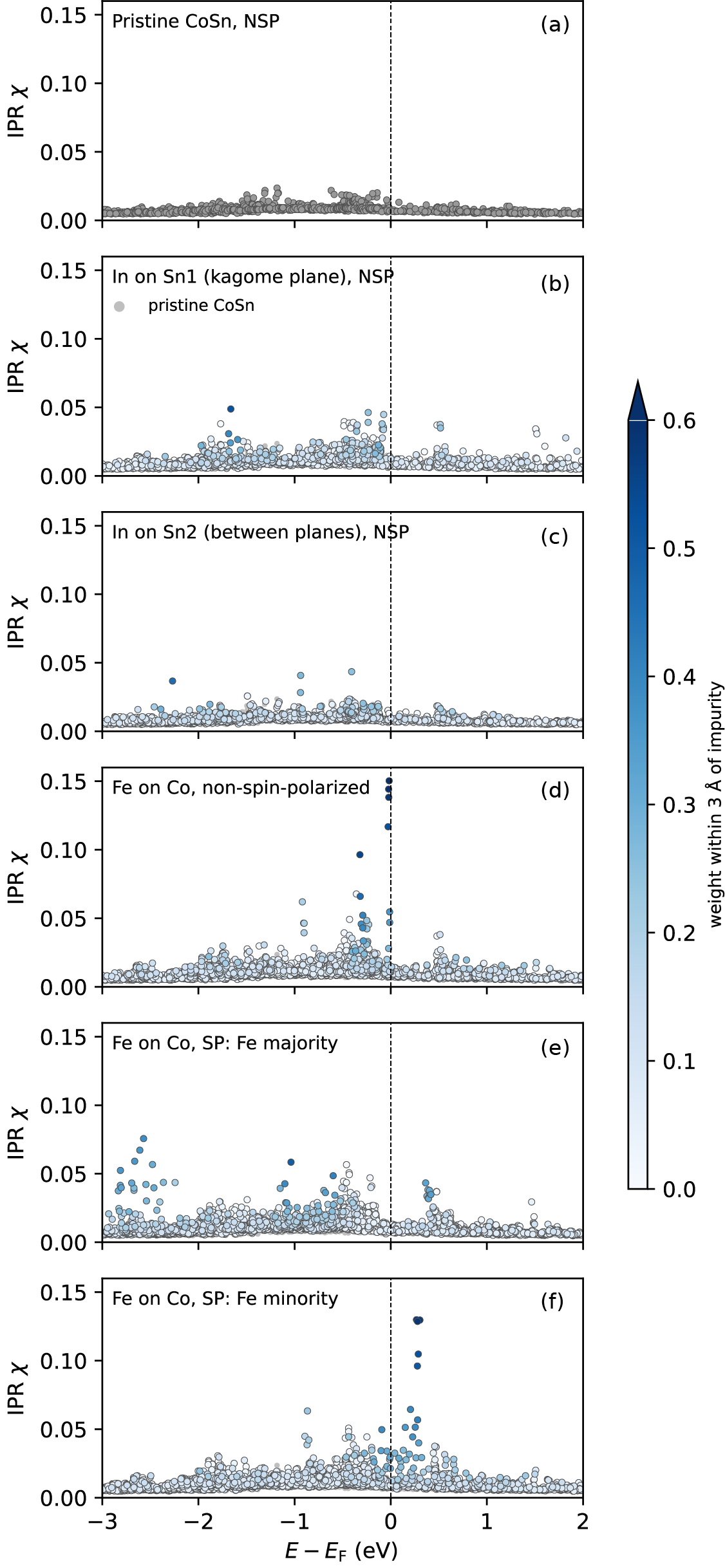}
	\caption{Inverse participation ratio $\chi$ of the Kohn-Sham states in the $3\times3\times4$ CoSn supercell: (a) pristine, (b) In on the Sn1 site, (c) In on the Sn2 site, (d) Fe on the Co site without spin polarization, and (e,f) Fe on the Co site with spin polarization for the Fe majority and minority spin, respectively. The color shows the fraction of the state weight within 3~\AA~of the impurity. Grey points in (b-f) show pristine CoSn for reference. The majority spin is `down' for CoSn:Fe$_{\text{Co}}$ in Fig.~\ref{figSI-PDOS-dilute}(c).}
	\label{figSI-IPR-dilute}
\end{figure*}

Finally, we examined the spatial localization of electronic states caused by the impurities using the inverse participation ratio (IPR) $\chi$ \cite{Pashartis_PRA_7_2017}, evaluated from the projections of the Kohn-Sham states on the atomic PAW spheres. The IPR is small for delocalized states and reaches its upper limit of 1 when a state is fully localized within the sphere of a single atom. Its lower limit is $1/N$, where $N=216$ is the number of atoms in the supercell. In practice, even the states of pristine CoSn are not equally distributed among the atoms, and part of their weight lies in the interstitial region, so the IPR of a delocalized state exceeds this limit. We therefore use the pristine CoSn supercell as a baseline [Fig.~\ref{figSI-IPR-dilute}(a)] and compare it with the supercells containing an impurity, which have the same number of atoms. All impurities produce some states that are more localized than in pristine CoSn, but they differ in how strongly these states are localized and where they lie. For In, the localization is modest. The states closest to \Ef~carry most of their weight on atoms away from the impurity, while the states centered on In lie deeper in the valence band. In on the Sn2 site perturbs the states even less than on the Sn1 site. Fe produces a much stronger localization, centered on the impurity. Without spin polarization, the most localized states lie at \Ef~[Fig.~\ref{figSI-IPR-dilute}(d)], as the counterpart of the Fe peak in the partial DOS that drives the magnetic instability. Spin polarization removes these states from \Ef. The localized minority-spin states move above \Ef~[Fig.~\ref{figSI-IPR-dilute}(f)], and the majority-spin ones spread over the valence band [Fig.~\ref{figSI-IPR-dilute}(e)], in line with the exchange splitting of the Fe states in the partial DOS [Fig.~\ref{figSI-PDOS-dilute}(c)].

The data that support the findings of this subsection are openly available in the Zenodo repository (DOI: 10.5281/zenodo.23539503).

\subsection{Supercell calculations: solid solutions}

The solid solutions were modeled with the same PAW potentials and computational settings as described in the previous subsection. We considered a 25\% substitution on the corresponding sublattice, Co$_{81}$Fe$_{27}$Sn$_{108}$ and Co$_{108}$Sn$_{81}$In$_{27}$, in the same $3\times3\times4$ supercell. Since In can occupy both Sn1 and Sn2 sites and the energy difference between them is relatively small (0.11~eV for an isolated In), we did not assume that In occupies only the Sn1 sites. Instead, the In populations of the two sublattices were estimated from a thermal equilibrium between them, assuming ideal mixing on each sublattice, at 1030~K (760~$^\circ$C). This temperature lies in the middle of the range between 620 and 900~$^\circ$C over which the crystals grow from the flux \cite{MeierPRB20, SalesPRM21}. This gives 15 In atoms on the 36 Sn1 sites and 12 In atoms on the 72 Sn2 sites. The Fe and In atoms were then distributed over their sublattices using special quasirandom structures (SQS) \cite{Zunger_PRL_65_1990}. The SQS were generated by simulated annealing, with atom swaps restricted to the same sublattice and layer, followed by a search over all single swaps. The objective was to reproduce the correlation functions of a random alloy for pair clusters up to 7.1~{\AA} and for triplet clusters with all sides up to 5.4~{\AA}, where the random targets for In were set by the In concentration on each Sn sublattice. The number of dopant atoms was fixed in each layer, with 7, 7, 7 and 6 Fe atoms in the four kagome planes, and 4, 4, 4, 3 In atoms on the Sn1 sites and 3 In atoms on the Sn2 sites of each layer. The sum of absolute deviations of the correlation functions from the random-alloy values is 0.18 for the Fe and 0.24 for the In solid solution, compared with $1.09\pm0.26$ and $1.39\pm0.31$ for random configurations with the same layer occupations, and all pair Warren-Cowley parameters are below 0.035 in magnitude. Both the atomic positions and the lattice parameters were optimized while preserving the hexagonal metric of the supercell ($a=b$, $\alpha=\beta=90^\circ$, $\gamma=120^\circ$). The plane-wave cutoff energy was set to the highest value recommended for the PAW potentials involved, the residual forces did not exceed 0.01~eV/{\AA}, and the $k$-mesh, smearing and energy convergence criterion were the same as for the isolated impurities. The Fe solid solution was calculated with spin polarization, starting from an A-type antiferromagnetic arrangement with ferromagnetic kagome planes coupled antiferromagnetically along $c$ and initial moments of $\pm2\mu_B$ on Fe. All 27 Fe atoms retained the sign of their initial moment, so the converged solution preserves the A-type arrangement. The Fe moments within the PAW spheres are $1.93\pm0.16\mu_B$ (mean and standard deviation over the 27 sites, ranging from 1.60 to 2.19~$\mu_B$), which is on average slightly smaller than the $2.14\mu_B$ of an isolated Fe impurity. The In solid solution was calculated without spin polarization.

\begin{figure*}[h]
	\includegraphics{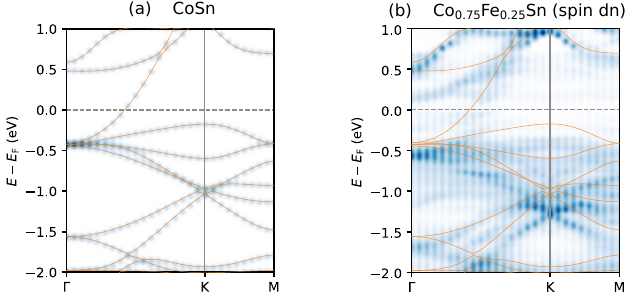}
	\caption{Band structures unfolded onto the Brillouin zone of primitive CoSn along $\Gamma$-K-M at $k_z=0$: (a) pristine CoSn supercell and (b) Co$_{0.75}$Fe$_{0.25}$Sn solid solution, spin-down channel. Each state is shown as a Gaussian spot with $\sigma=30$~meV. The band structure of the primitive CoSn cell is shown as thin orange lines. The intensity in panel (b) is enhanced by the same factor as in Fig.~4(b) of the main text.}
	\label{figSI-CoSn-unfolded-25-perc-Fe}
\end{figure*}

Figure~\ref{figSI-CoSn-unfolded-25-perc-Fe} shows band structures unfolded with the \texttt{fold2Bloch} code \cite{RubelPRB14, Rubel23} from the $3\times3\times4$ supercells onto the Brillouin zone of the primitive CoSn cell along $\Gamma$-K-M at $k_z=0$. Each unfolded state is represented by a Gaussian spot with a width of $\sigma=30$~meV in energy and an equivalent width in $k$. The spectral weights of overlapping spots are added on a common pixel grid, so that the intensity increases near degeneracies, which would not be visible if each band were drawn with its own spectral weight only. For the pristine CoSn supercell [Fig.~\ref{figSI-CoSn-unfolded-25-perc-Fe}(a)], which serves as a benchmark, the unfolded bands coincide with the band structure of the primitive cell, which confirms the validity of the unfolding procedure. The enhanced intensity near degeneracies is visible for the flat bands at $\Gamma$ and for the Dirac-like crossings at K. Figure~\ref{figSI-CoSn-unfolded-25-perc-Fe}(b) shows the spin-down channel of the Co$_{0.75}$Fe$_{0.25}$Sn solid solution, which complements the spin-up channel shown in Fig.~4(b) of the main text. In the antiferromagnetic arrangement used here, the numbers of Fe atoms with opposite moments differ by only one (14 and 13), so the two spin channels are very similar. As in the spin-up channel, the bands shift away from \Ef~in a non-rigid way, more strongly at K than at $\Gamma$, and the flat band near K is pushed below its position in pristine CoSn. In addition, the unfolded bands of the Fe solid solution have a pronounced diffuse character, which reflects the disorder introduced by Fe. Since states localized in real space are spread in reciprocal space, this diffuse character is consistent with the increased localization of states in the Fe solid solution discussed below.

\begin{figure*}[h]
	\includegraphics[width=0.5\linewidth]{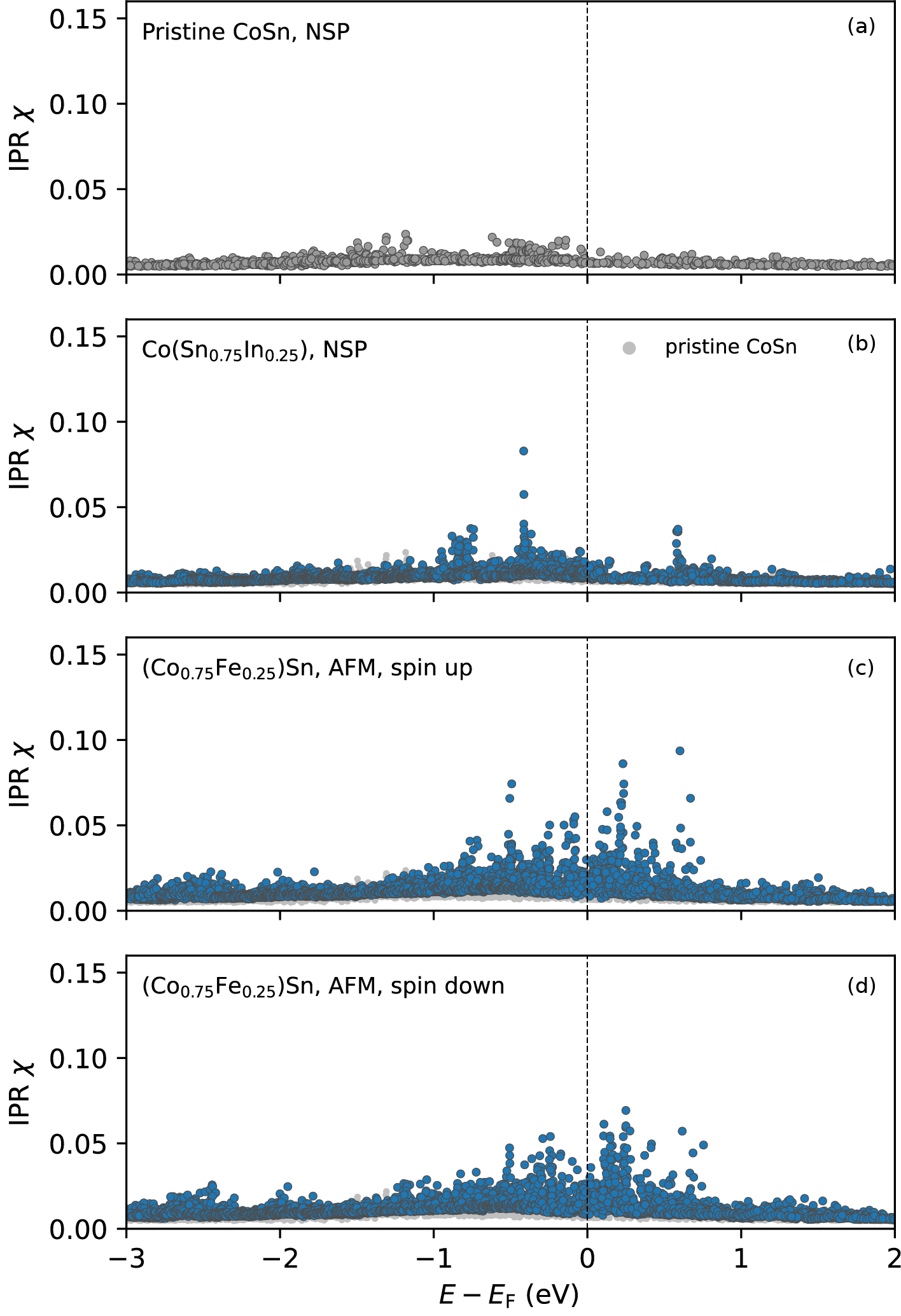}
	\caption{Inverse participation ratio $\chi$ in $3\times3\times4$ supercells: (a) pristine CoSn without spin polarization, (b) CoSn$_{0.75}$In$_{0.25}$ solid solution without spin polarization, and (c,d) Co$_{0.75}$Fe$_{0.25}$Sn solid solution with antiferromagnetic order, spin up and spin down, respectively. Grey points in (b-d) show pristine CoSn for reference.}
	\label{figSI-IPR-solid-solution}
\end{figure*}

Figure~\ref{figSI-IPR-solid-solution} shows that the two substitutions affect the spatial character of the electronic states in very different ways. In substitution leaves the deep valence band essentially as in pristine CoSn and produces only a few localized states, concentrated in a narrow energy window below \Ef, where an isolated In on the Sn2 site also produces its most localized state [Fig.~\ref{figSI-IPR-dilute}(c)]. Fe substitution, in contrast, raises the IPR over the whole valence band, and its localized states are spread from about $-1$~eV to above \Ef~rather than concentrated at particular energies. The localized states above \Ef~are reminiscent of the minority-spin states of an isolated Fe impurity [Fig.~\ref{figSI-IPR-dilute}(f)]. Averaged between $-1$~eV and \Ef, the IPR increases by about 30\% for In and by about 60\% for Fe with respect to pristine CoSn. The two spin channels of the Fe solid solution behave alike, as expected for the nearly compensated antiferromagnetic arrangement. Even the most localized states extend over more than about ten atoms ($1/\chi>10$), so the localization remains partial. The stronger localization for Fe is consistent with the diffuse character of its unfolded bands [Fig.~\ref{figSI-CoSn-unfolded-25-perc-Fe}(b)] and with the larger broadening of the ARPES bands observed for Fe than for In substitution (Fig.~1 of the main text).

The data that support the findings of this subsection are openly available in the Zenodo repository (DOI: 10.5281/zenodo.23539503).

\clearpage

\section{FDMNES simulations}

Co $K$-edge XANES spectra were simulated with the FDMNES code
\cite{JolyPRB01, BunauJPCM09}, which performs real-space \textit{ab initio} calculations of X-ray absorption spectra within a one-electron (independent-particle) framework: the cross-section is a matrix element towards a one-electron final state computed in the presence of the screened core hole, so that multiplet structure and configuration interaction in the final state are not included. At the $K$ pre-edge of a $3d$ metal the $1s$ hole is spherical and its exchange coupling to the $3d$ shell is weak, so this description is expected to be adequate, in contrast with the $L_{2,3}$ edges. All simulations use the finite difference method (FDM), which makes no muffin-tin approximation on the shape of the potential and is therefore better suited to the pre-edge region, together with a cluster radius of 7~\AA\ (101 atoms; the potential is built over 9.5~\AA, 207 atoms) and the \texttt{Quadrupole} keyword, which includes the $1s\rightarrow3d$ quadrupolar channel contributing to the pre-edge. The \texttt{SCF} keyword activates a self-consistent determination of the electronic charge density within the cluster prior to the final spectrum calculation: the Kohn-Sham equations are solved iteratively with a Hedin-Lundqvist exchange-correlation potential until convergence of the electron density, so that the effective potential felt by the photoelectron is self-consistently determined rather than built from a superposition of atomic densities \cite{BunauJPCM09}. Spectra are broadened with the internal convolution of FDMNES, which uses the self-consistent Fermi level $E_{\mathrm{cut}}$ of each calculation; the same convention is applied to every composition, so that intensities can be compared from one calculation to the next.

\subsection{Unit cells}
All compositions are described in $2\times2\times2$ supercells of the parent CoSn cell ($a=b=5.27$~\AA, $c=4.25$~\AA), so that every calculation of this work is performed at the same volume per atom (17.04~\AA$^3$) and with the same number of atoms (48 per primitive cell, 817.77~\AA$^3$). Two settings of this supercell are used:

\begin{itemize}
	\item Co/Fe substitutions require an orthohexagonal, $C$-centred setting ($Cmmm$, No.~65) with $a=10.540$~\AA, $b=18.256$~\AA, $c=8.500$~\AA\ (96 atoms in the conventional cell); 
	\item Sn/In substitutions are described in the hexagonal setting ($P6/mmm$, No.~191) with $a=b=10.540$~\AA, $c=8.500$~\AA\ (48 atoms).
\end{itemize}

\noindent The supercell contains 24 Co and 24 Sn sites. The compositions simulated are $x_{\mathrm{Fe}} = 1/24$, $1/3$, $2/3$ and $23/24$, obtained by replacing the corresponding number of Co sites by Fe, and $x_{\mathrm{In}} = 1/24$, $1/3$ and $2/3$, obtained by replacing Sn sites by In. For each composition and each series, the $x=0$ reference is recalculated in the \emph{same} supercell, so that a doped spectrum is always compared to an undoped one treated identically.

\subsection{Consistency of the description}
The quantity that controls the calculated Co $K$ spectrum is not the size of the unit cell but the self-consistent interstitial potential $V_0$ resulting from the SCF cycle. We verified this explicitly. At $x_{\mathrm{Fe}}=1/3$, the same ferromagnetic structure described in a 12-atom $Cmmm$ cell, in the 24-atom cell required by the A-type antiferromagnetic stacking, and in the 48-atom $2\times2\times2$ supercell yields $V_0 = -9.388$~eV and $E_{\mathrm{cut}} = -0.663$~eV in the three cases, and three normalized spectra identical to within $2\times10^{-5}$. The same holds at $x_{\mathrm{Fe}}=2/3$ ($V_0 = -9.566$~eV). Cell size and setting are therefore immaterial once $V_0$ is converged, and the orthohexagonal and hexagonal descriptions of pure CoSn, which have the same volume and the same number of atoms, differ by only 1.5\% in pre-edge area and by $-0.10$~eV in position. The Fe and In series are consequently directly comparable.

This also sets the reason for using supercells throughout. Pure CoSn described in the primitive 3-atom hexagonal cell converges to $V_0 = -6.950$~eV, i.e.\ 2.3 to 2.4~eV above \emph{every} other calculation of this work (which all lie between $-9.08$ and $-9.57$~eV), although the cluster (101 atoms within 7~\AA), the point group ($mmm$) and $E_{\mathrm{cut}}$ are identical to those of the supercell. Its pre-edge is 10\% more intense and displays a sharp peak at 7711~eV absent from the supercell calculations. Taking this small-cell result as the $x=0$ reference would place the undoped compound above all substituted ones for purely numerical reasons; all references quoted below are therefore supercell calculations.

Because $V_0$ varies slightly from one composition to the next, part of the calculated change upon substitution is a rigid translation of the whole spectrum rather than a change of shape. Since no shift of the Co $K$ edge is detected experimentally, each simulated spectrum is realigned on the $x=0$ reference of its own series by minimizing the residual over 7705--7735~eV, which isolates the change in shape. The shifts applied are small for Fe ($+0.02$~eV at $x=1/3$, $+0.13$~eV at $x=2/3$) and larger for In ($+0.40$ and $+0.53$~eV).

\subsection{Diluted limit}
\begin{figure}[h]
	\centering
	\includegraphics[width=\columnwidth]{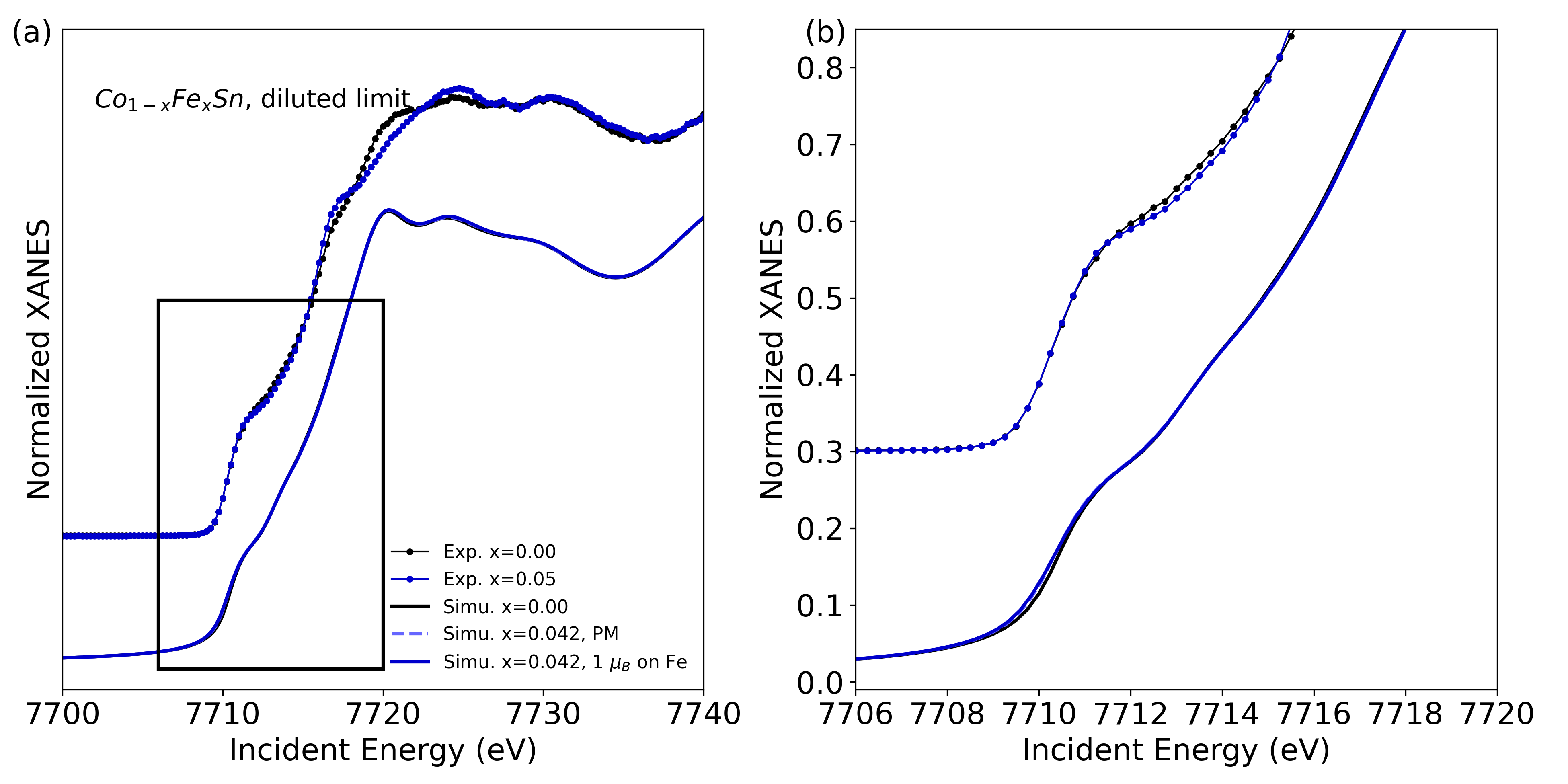}
	\caption{Diluted limit of Fe substitution. (a) Measured Co $K$-edge spectra at $x=0$ and $x=0.05$ (dots, shifted vertically for clarity) and FDMNES simulations for pure CoSn (black) and for one Fe atom among 24 Co sites ($x_{\mathrm{Fe}}=0.042$), either non-magnetic (dashed) or with a self-consistent 1~$\mu_B$ moment on Fe (solid blue). All calculations are performed in the same $2\times2\times2$ supercell with the same convolution. (b) Enlarged view of the pre-edge region. Once realigned in energy, the two doped calculations differ from the undoped one by 7.3\% (non-magnetic) and 6.0\% (with a moment on Fe) in pre-edge area, already as much as at $x_{\mathrm{Fe}}=2/3$.}
	\label{Sup_FDMNES_diluted}
\end{figure}

Figure \ref{Sup_FDMNES_diluted} shows the diluted limit, one Fe atom among 24 Co sites ($x_{\mathrm{Fe}} = 0.042$), compared to pure CoSn calculated in the same supercell and to the measured spectra at $x=0$ and $x=0.05$. Most of the raw difference between the two calculated spectra is a rigid translation: realigning the doped spectrum on the undoped one reduces the residual from 0.013 to 0.003 in the non-magnetic case, so that 77\% of the difference is a shift of $+0.22$~eV and not a redistribution of spectral weight; the proportion reaches 85\% ($+0.41$~eV) when a local moment is allowed on Fe. Once this shift is removed, a genuine change of shape remains, corresponding to a 7.3\% reduction of the pre-edge area (6.0\% with the moment). This is already as large as the reduction obtained at $x_{\mathrm{Fe}}=2/3$, whereas the measured suppression at the closest available composition, $x_{\mathrm{Fe}}=0.05$, is only 1.8\%. The calculated effect is therefore saturated from the very first substitution, which is the clearest expression of its local character.

Allowing spin polarization, the Fe moment converges to 0.99~$\mu_B$ starting from a 2~$\mu_B$ input, with induced moments of 0.01 to 0.08~$\mu_B$ on the surrounding Co, parallel to the Fe moment. The corresponding spectrum is shown as the solid blue line in Fig.~\ref{Sup_FDMNES_diluted}.

\subsection{Ordered configurations and magnetism}
At $x_{\mathrm{Fe}} = 1/3$ and $2/3$, the Fe atoms are placed on an ordered sublattice of the supercell. Both non-magnetic and spin-polarized calculations were performed, the latter with ferromagnetic and with A-type antiferromagnetic stacking of the kagome planes. The converged Fe moment grows with Fe content, from 0.99~$\mu_B$ in the diluted limit to 1.15~$\mu_B$ at $x=1/3$, 1.57~$\mu_B$ at $x=2/3$ and 1.67--1.92~$\mu_B$ at $x=23/24$, the latter comparable to the 1.85~$\mu_B$ measured in FeSn \cite{SalesPRM19}. The moment obtained in the diluted limit is however appreciably smaller than the $2\mu_B$ given by the supercell DFT calculations of Sec.~V~C and than the value suggested by the Curie term of the susceptibility \cite{SalesPRM21}; the cluster self-consistency with a Hedin-Lundqvist exchange-correlation potential is expected to underestimate moments in itinerant magnets, and the discrepancy is largest where the moment is least stable. Ferromagnetic and antiferromagnetic stackings give identical Co $K$ spectra, which shows that the pre-edge is insensitive to the relative orientation of the moments between planes and probes only the local Fe moment.

The spin-polarized calculations also produce a sizeable induced moment on Co, parallel to that of Fe: $0.01$ to $0.08~\mu_B$ in the diluted limit, then $0.34~\mu_B$ at $x=1/3$, $0.49~\mu_B$ at $x=2/3$ and $0.30~\mu_B$ at $x=23/24$. The parallel orientation on the Co neighbours agrees with the supercell calculations of Sec.~V~C. Co is therefore not a spectator of the magnetism carried by Fe, although the Co moment remains three to four times smaller than the Fe one.

We deliberately refrain from quoting $d$-orbital populations as a measure of charge transfer between Co and Fe. These populations are integrated inside atomic spheres whose radii are neither equal for the two species nor constant from one calculation to the next: switching on spin polarization reduces the Fe radius from 1.43 to 1.29~\AA, which removes about 0.3 electron from the Fe sphere for purely geometrical reasons. At fixed radius, the Co $d$ count varies by at most 0.055 electron (0.7\%) over the whole series, from 7.906 in pure CoSn to 7.851--7.894 in the substituted compounds, which is consistent with the few-percent changes of the calculated pre-edge.

For Sn/In substitutions, In may occupy either the in-plane Sn site ($1a$ in the parent cell) or the between-plane site ($2d$). Each composition is therefore reported as the average of the two configurations. Contrary to what a calculation in the primitive cell suggests, the two choices are not equivalent in the supercell: the pre-edge areas obtained for the two configurations differ by 2.2\% at $x_{\mathrm{In}}=1/3$ and by 6.4\% at $2/3$. This spread should be regarded as a systematic uncertainty on the simulated In spectra.

\subsection{Comparison with the measured pre-edge}
Throughout this work the pre-edge intensity is quantified by its area between 7710 and 7714~eV, after subtraction of a linear background fitted between 7700 and 7707~eV, the spectra being normalized to the edge jump. Simulated spectra are first realigned in energy on the $x=0$ reference of their own series, as described above, so that the quoted areas measure a change of shape and not a residual translation. All variations below are relative to that $x=0$ reference. Table \ref{tab_preedge} compares the relative change of the pre-edge area (7710--7714~eV, linear background subtracted) between experiment and simulation, each referred to its own $x=0$.

\begin{table}[h]
	\begin{tabular}{lrrr}
		\hline
		& experiment & \multicolumn{2}{c}{simulation} \\
		&            & non-magn. & with moment on Fe \\
		\hline
		Co$_{1-x}$Fe$_x$Sn & & & \\
		\quad $x=0.05$ / $1/24$   & $-1.8$\%  & $-7.3$\% & $-6.0$\% \\
		\quad $x=0.20$ / $1/3$    & $-4.5$\%  & $-3.9$\% & $-7.6$\% \\
		\quad $x=0.50$ / $2/3$    & $-19.5$\% & $-6.6$\% & $-8.5$\% \\
		\quad $x=0.90$ / $23/24$  & $-28.2$\% & $-0.6$\% & $+4.6$\% \\
		CoSn$_{1-x}$In$_x$ & & & \\
		\quad $x=0.07$ / $1/24$   & $-3.7$\%  & $-1.7$\% & -- \\
		\quad $x=0.35$ / $1/3$    & $+4.5$\%  & $+3.7$\% & -- \\
		\quad\phantom{$x=0.35$} / $2/3$ & --  & $+12.7$\% & -- \\
		\hline
	\end{tabular}
	\caption{Relative change of the Co $K$ pre-edge area upon substitution, for
	experiment and for the FDMNES simulations. The pre-edge area is integrated
	between 7710 and 7714~eV after subtraction of a linear background fitted
	between 7700 and 7707~eV, on spectra normalized to the edge jump, and each
	value is referred to the $x=0$ reference of its own series. Simulated spectra
	are first realigned in energy on that reference, so that the quoted changes
	measure a change of shape and not a rigid displacement of the edge. Each row
	pairs the measured composition with the closest composition accessible in the
	$2\times2\times2$ supercell. The last column gives the spin-polarized
	calculations, with a local moment on Fe; the In substitution does not support
	a magnetic solution, and no measurement is available at $x_{\mathrm{In}}=2/3$.}

	\label{tab_preedge}
\end{table}

The In series is reproduced quantitatively at the single composition where measurement and calculation can be compared. The Fe series is reproduced in sign but not in amplitude: the calculated suppression stays between 4 and 8\%\ over the whole range $1/24 \le x \le 2/3$, while the measured one grows by more than a factor of ten over the same interval. The departure is largest at $x_{Fe}=23/24$, which is also the composition for which neglecting the lattice relaxation is most severe: all calculations use the lattice parameters of CoSn ($a=5.27$~\AA, $c=4.25$~\AA), whereas the end member FeSn has $c=4.48$~\AA, 5.4\%\ larger. The pre-edge being sensitive to the local geometry, this alone may account for part of the discrepancy at high Fe content.

\begin{figure*}[h]
	\centering
	\includegraphics[width=0.95\linewidth]{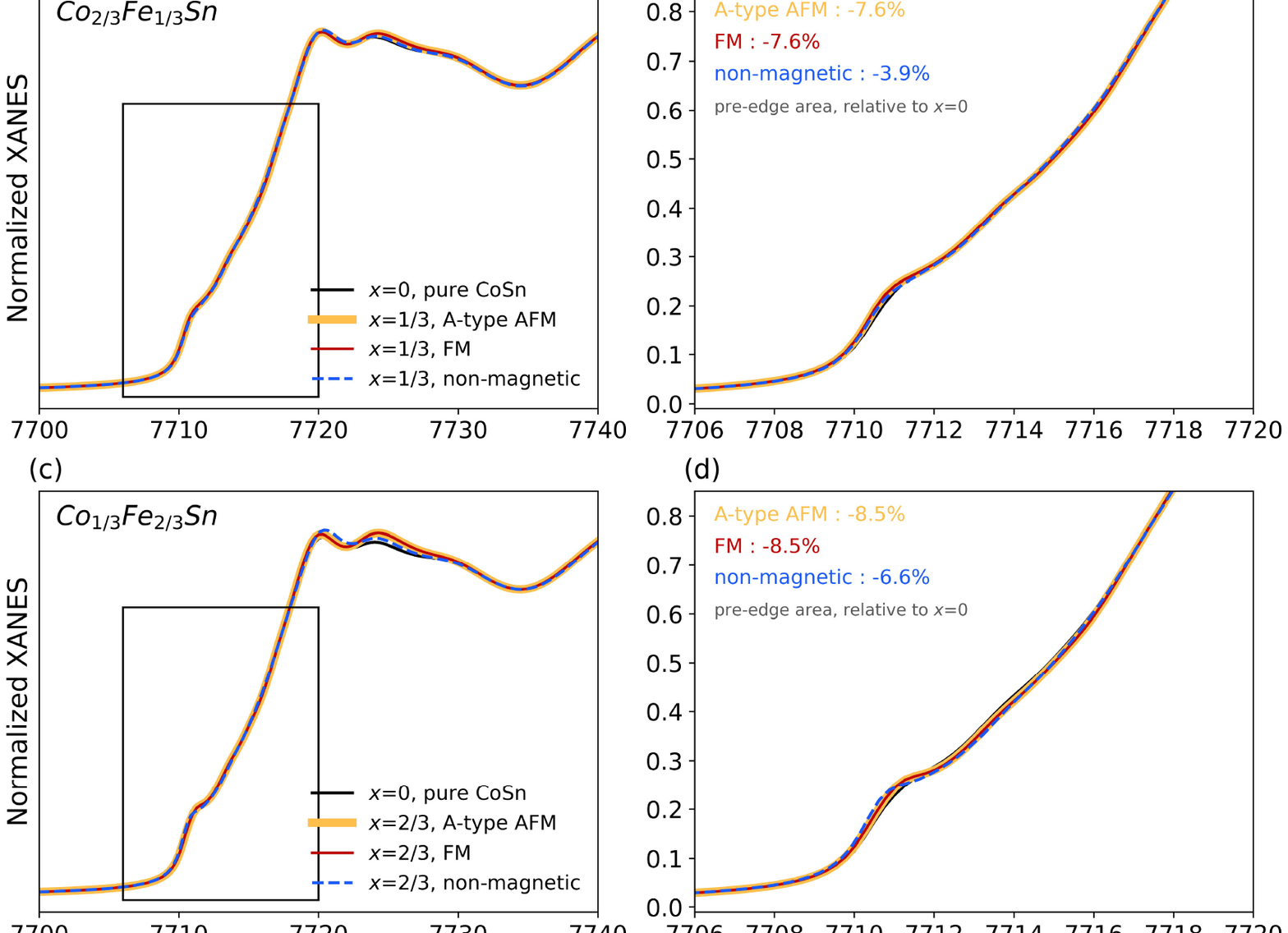}
	\caption{Effect of spin polarization on the simulated Co $K$-edge
		spectrum of Co$_{1-x}$Fe$_x$Sn, for $x=1/3$ (a,b) and $x=2/3$ (c,d). Each panel compares, for a single composition, the non-magnetic calculation (blue, dashed), the ferromagnetic one (dark red, thin) and the A-type antiferromagnetic one (orange, thick), together with pure CoSn (black) as the $x=0$ reference. All four calculations use the same $2\times2\times2$ supercell, the same cluster radius and the same internal convolution, and the substituted spectra are realigned in energy on the $x=0$ reference. (b) and (d) show the pre-edge region boxed in (a) and (c), with the corresponding change in pre-edge area relative to $x=0$. The ferromagnetic curve lies entirely within the thick antiferromagnetic one: the pre-edge does not distinguish the two magnetic stackings. At fixed composition, however, the onset of the local moment on Fe deepens the calculated suppression, bringing it closer to the measured one.}
	\label{Sup_FDMNES_mag}
\end{figure*}

Ferromagnetic and A-type antiferromagnetic stackings give strictly identical Co $K$ spectra (same $V_0$, same $E_{\mathrm{cut}}$, spectra identical to within $2\times10^{-5}$), as shown in Fig.~\ref{Sup_FDMNES_mag}: the Co pre-edge is insensitive to the relative orientation of the Fe moments between kagome planes, and probes only the local moment. Allowing that moment to form does change the spectrum, and in the direction of the experiment: at fixed composition, the calculated suppression of the pre-edge area goes from 3.9 to 7.6\%\ at $x_{\mathrm{Fe}}=1/3$, and from 6.6 to 8.5\%\ at $2/3$. With the moment included the suppression also becomes monotonic in $x$ up to $2/3$, which it is not in the non-magnetic series. The magnetic calculation is thus closer to the measurement than the paramagnetic one, which supports independently, through the Co $K$ pre-edge, the local-moment picture developed in the main text -- although it still accounts for less than half of the measured suppression at $x_{\mathrm{Fe}}=0.5$.

\end{document}